\input harvmac.tex

\def\hA{{{\hat A}}}
\def\cF{{{\cal F}}}
\def\CF{{{\cal F}}}
\def\CM{{{\cal M}}}
\def\Det{{\rm Det}}
\def\Z{{{\bf Z}}}
\def\im{{\rm Im}}
\def\re{{\rm Re}}

\def\half{{{1\over 2}}}

\def\ch{{\rm ch}}

\def\dg{{\dagger}}
\def\wdg{{\wedge}}
\def\a{{\alpha}}

\def\b{{\beta}}
\def\c{{\gamma}}
\def\d{{\delta}}
\def\one{{\bf 1}}
\def\l{{\lambda}}
\def\vph{{\varphi}}
\def\L_R{{\Lambda}}

\def\Z{{\bf Z}}

\def\s{{\sigma}}

\def\x{{\xi}}

\def\tht{{\theta}}

\def\tG{{{\tilde G}}}

\def\tX{{{\tilde X}}}
\def\hA{{{\hat A}}}

\def\cA{{{\cal A}}}

\def\hB{{{\hat B}}}
\def\hD{{{\hat D}}}
\def\hC{{{\hat C}}}
\def\hE{{{\hat E}}}
\def\hF{{{\hat F}}}
\def\hN{{{\hat N}}}
\def\hM{{{\hat M}}}

\def\cF{{{\cal F}}}
\def\cR{{\cal R}}

\def\bpsi{{{\bar \psi}}}
\def\bPsi{{{\overline {\bf \Psi}}}}
\def\bW{{{\overline {\bf W}}}}
\def\bSigma{{{\overline \Sigma}}}
\def\bLambda{{{\overline \Lambda}}}
\def\bartl{{{\bar {\tilde l}}}}
\def\bartmu{{{\bar {\tilde \mu}}}}
\def\bareta{{{\bar \eta}}}
\def\barl{{{\bar l}}}
\def\barmu{{{\bar \mu}}}

\def\bGamma{{{\bar \Gamma}}}

\def\p{{\partial}}
\lref\DMW{D.E.Diaconescu, G.Moore, E. Witten,
''E8 Gauge Theory, and a Derivation of K-Theory from M-Theory'',
hep-th/005090 }
\lref\DMWUN{D.E.Diaconescu, G.Moore, E. Witten, unpublished.}
\lref\DM{D.E. Diaconescu and G. Moore, unpublished. } 
\lref\MW{ G.Moore, E. Witten, ``Integration over the u-plane in
Donaldson theory'',hep-th/9709193}
\lref\freed{D. Freed,
Dirac Charge Quantization and Generalized Differential Cohomology,''
hep-th/0011220 } 
\lref\Hull{C.M. Hull,''Massive String Theories From M-Theory and F-Theory'',
JHEP 9811 (1998) 027,hep-th/
9811021 }
\lref\Town{C.Hull,P.Townsend,''Unity of Superstring Dualities,''
Nucl.Phys. B438 (1995) 109-137,hep-th/9410167}
\lref\GPR{ A. Giveon, M. Porrati, E. Rabinovici,
'' Target Space Duality in String Theory'', hep-th/9401139 }
\lref\BRG{E.Bergshoeff,M.de Roo,M.B.Green, G.Papadopoulos,
P.K.Townsend,''Duality of II 7-branes and 8-branes'', hep-th/9601150}
\lref\Lavr{I.Lavrinenko, H.Lu, C.Pope,
T.Tran,``U-duality as general Coordinate Transformations, and Spacetime
Geometry'', hep-th/9807006 }
\lref\Cvet{ M. Cvetic, H. Lu, C.N. Pope, K.S. Stelle,''
T-Duality in the Green-Schwarz Formalism, 
and the Massless/Massive IIA Duality Map'',Nucl.Phys. B573 (2000) 149-176,
 hep-th/9907202}
\lref\Piol{E.Kiritsis and B. Pioline, ``On $R^4$ threshhold corrections
in IIB string theory and (p,q) string instantons,'' 
Nucl. Phys. {\bf B508}(1997)509; 
B. Pioline, H. Nicolai, J. Plefka, A. Waldron,
 $R^4$ couplings, the fundamental membrane and 
exceptional theta correspondences, hep-th/0102123 }  
\lref\mm{R. Minasian and G. Moore,``K Theory and Ramond-Ramond Charge,''
JHEP {\bf 9711}:002, 1997; hep-th/9710230.}

\lref\mw{G. Moore and E. Witten, ``Self-Duality, Ramond-Ramond Fields, and K-Theory,''
hep-th/9912279;JHEP 0005 (2000) 032}
\lref\Warn{C.Isham,C.Pope,N.Warner,
''Nowhere-vanishing spinors and triality rotations in 8-manifolds,''
Class.Quantum Grav.5(1988)1297}
\lref\fourflux{E. Witten, ``On Flux Quantization in $M$-Theory
and the Effective Action,'' hep-th/9609122; Journal of
Geometry and Physics, {\bf 22} (1997) 1.}
\lref\wittenk{E. Witten, ``$D$-Branes And $K$-Theory,''
JHEP {\bf 9812}:019, 1998; hep-th/9810188.}      
\lref\wittenvbr{E.Witten,'' Five-brane effective action in M-theory,''
J.Geom.Phys.,22 (1997)103,hep-th/9610234}
\lref\wittenduality{E. Witten, ``Duality Relations Among Topological Effects In String Theory,''
hep-th/9912086;JHEP 0005 (2000) 031} 
\lref\wittenstrings{E. Witten, ``Overview of K-theory applied to
strings,'' hep-th/0007175.}          
\lref\Sethi{ S. Sethi, C. Vafa, E. Witten,
``Constraints on Low-Dimensional String Compactifications'',
hep-th/9606122, Nucl.Phys. B480 (1996) 213-224 }
\lref\Wits{E.Witten,''On S-Duality in Abelian Gauge Theory'',
hep-th/9505186}
\lref\Verl{E. Verlinde,''Global Aspects of Electric-Magnetic Duality'',
Nucl.Phys. B455 (1995) 211-228,hep-th/9506011}
\lref\Romans{L. Romans,''Massive IIA Supergravity in Ten Dimensions',
Phys.Let.169B(1986)374}
\lref\Myers{P. Meessen, T. Ortin,
''An Sl(2,Z) Multiplet of Nine-Dimensional Type II Supergravity Theories'',
Nucl.Phys. B541 (1999) 195-245, hep-th/9806120 }
\lref\Siegel{W.Siegel,''Hidden ghosts'', Phys.Lett.B93(1980),170 }
\lref\Deser{S.Deser,P.van Nieuwenhuizen,''One-loop
divergences of quantized Einstein-maxwell fields'',Phys.Rev.D(1974),
v.10,p.401}
\lref\Hooft{G.'t Hooft, M.Veltman,''One loop divergences in
the theory of gravitation'', Annales 
Inst.Henri Poincare ,vol.20,(1974),69}
\lref\CJS{E.Cremmer, B.Julia, J.Scherk, 
``Supergravity theory in 11 dimensions'',
Phys.Lett. B76(1978)409}
\lref\Nielsen{N.Nielsen, 
 ``Ghost counting  in supergravity'', NPB 140(1978)499}
\lref\Kallosh{R.Kallosh,''Modified  Feynman
rules in supergravity'',NPB 141(1978)141}
\lref\Fuj{K.Fujikawa,''
Path  Integral for gauge theories with fermions'',
Phys. Rev.D.21(1980) 2848}
\lref\Ibusa{J.Igusa, {\it Theta-functions}, Berlin, New York, 
Springer-Verlag, 1972.}
\lref\Phong{E.D'Hoker,D.Phong,''Conformal scalar fields 
and chiral splitting on SuperRiemann
surfaces'',Comm.Math.Phys.125
(1989)469;

K.Aoki,E.D'Hoker,D.Phong,
''Unitarity of closed superstring Perturbation Theory'',
NPB 342(1990)149-230 
}
\lref\Geg{J.Gegenborg, G.Kunstatter,''The partition function for
topological field theories``,hep-th/9304016}
\lref\bs{G.Moore,''Some comments on Branes, G-flux, and K-theory'',
Int.J.Mod.Phys.A16   (2001)936-944,
hep-th/0012007} 
\lref\Mathai{V. Mathai, D. Stevenson, 
``Chern character in twisted K-theory: equivariant and holomorphic cases''
hep-th/0201010}
\lref\MathaiB{V. Mathai, R.B. Melrose, and I.M. Singer, 
``The index of projective families of elliptic operators,'' 
math.DG/0206002}
\lref\Rabin{J.Rabin, M.Bergvelt,''Super curves, their Jacobians,
and super KP equations'',alg-geom/9601012}
\lref\Sch{S.Fredenhagen, V.Schomerus, Non-commutative geometry,
gluon condensates, and K-theory'',JHEP 0004(2001)007,
hep-th/0012164}
\lref\Stan{S.Stanciu,''An illustrated guide to D-branes in SU(3)'',
hep-th/0111221}
\lref\MMS{J.Maldacena,G.Moore, N.Seiberg, ``D-brane instantons
and K-theory charges'', JHEP 0111(2001)062; hep-th/0108100 }
\lref\mmsiii{J. Maldacena, G. Moore, and N. Seiberg, 
``D-brane charges in five-brane backgrounds,'' JHEP 0110 (2001)
005; hep-th/0108152} 

\lref\Mathaii{P.Bouwknegt, V.Mathai,``D-branes, B-fields and 
twisted K-theory'', ~~
JHEP 0003
(2000)007, hep-th/0002023}
\lref\APS{M.Atiyah,V.Patodi, I.Singer,
Math.Proc.Cambridge Phil.Soc.77(1975)43;405.}
\lref\fh{D. S. Freed, M. J. Hopkins, 
``On Ramond-Ramond fields and K-theory,'' JHEP 0005 (2000) 044;
hep-th/0002027}
\lref\birreldavies{N.D. Birrell and P.C.W. Davies, 
{\it Quantum Fields in Curved Space}, Cambridge Univ. Press, 1982}
 
\lref\stong{R. Stong, ``Calculation of
$\Omega_{11}^{spin}(K({\bf Z},4))$'' in
{\it Unified String Theories}, 1985 Santa Barbara
Proceedings, M. Green and D. Gross, eds. World Scientific 1986.}

\lref\asv{M. F. Atiyah, I. M. Singer, ``The index of
elliptic operators: V'' Ann. Math. {\bf 93} (1971) 139.}

\lref\szabo{K. Olsen and R.J. Szabo,
``Constructing $D$-Branes from $K$-Theory,''
hep-th/9907140.  }

\lref\hori{K. Hori, ``D-branes, T-duality, and Index Theory,''
Adv.Theor.Math.Phys. 3 (1999) 281-342; hep-th/9902102  }

\lref\evslin{A.Adams,J.Evslin,''The Loop Group of $E_8$ and 
K-Theory from 11d,''  hep-th/0203218}

%%%%%%%%%%%%%%%%%%%%%%%%%%%%%%%%%%%%%%%%%%%%%%%%%%%%%%%%%%%%%%  
% paper starts here !!!  

%%%%%%%%%%%%%%%%%%%%%%%%%%%%%%%%%%%%%%%%%%%%%%%%%%%%%%%%%%%%%%%%%%%%%%  
%%%%%%%%%%%%%%%%%%%%%%%%%%%%%%%%%%%%%%%%%%%%%%%%%%%%%%%%%%%%%%%%%%%%%% 
%\rightline{ }  
\rightline{RUNHETC-2002-15; NI 02013-MTH }  
\Title{  
\rightline{hep-th/0206092}}  
{\vbox{
\centerline{T-Duality, and the  }
\smallskip
\centerline{K-Theoretic Partition Function  }
\smallskip
\centerline{of TypeIIA Superstring Theory } }}  
\smallskip  
\centerline{ Gregory Moore$^1$ and Natalia Saulina$^2$}  
\smallskip  
\centerline{$^1$ {\it Department of Physics, Rutgers University}}  
\centerline{\it Piscataway, NJ 08855-0849, USA}  
\smallskip  
\centerline{$^2$ {\it Department of Physics, Princeton University}}  
\centerline{\it Princeton, NJ 08544, USA}

\bigskip  
\noindent  
We study the partition function of type IIA string theory 
on 10-manifolds of the form $T^2 \times X$ where $X$ is 
8-dimensional, compact, and spin. We pay particular 
attention to the effects of the topological phases in 
the supergravity action implied by the K-theoretic 
formulation of RR fields, and we use these to 
check the $T$-duality 
invariance of the partition function. We find that 
the partition function is only $T$-duality invariant
 when we take into 
account the $T$-duality anomalies in the 
RR sector, the fermionic path integral 
(including 4-fermi interaction terms),  and 1-loop 
corrections including worldsheet instantons. We comment 
on applications of our computation to speculations 
about the role of the Romans mass in $M$-theory. 
We also discuss some issues which arise when one 
attempts to extend these considerations to checking 
the full $U$-duality invariance of the theory. 

\bigskip
  
\vfill  
  
\Date{June 10, 2002}

%%%%%%%%%%%%%%%%%%%%%%%%%%%%%%%%%%%%%%%%%%%%%%%%%%%%%%%%%%%%%%%%%%%%%%  
%%%%%%%%%%%%%%%%%%%%%%%%%%%%%%%%%%%%%%%%%%%%%%%%%%%%%%%%%%%%%%%%%%%%%%  

\newsec{Introduction \& Summary }

Duality symmetries, such as the $U$-duality symmetry 
of toroidally compactified $M$-theory,
 have been of central importance in the 
  definition of string theory and M-theory. 
Topologically nontrivial effects associated with 
the RR sector have also played a crucial role in defining 
the theory.  
It is currently believed that RR fieldstrengths (and their 
D-brane charge sources) are 
classified topologically using K-theory 
\refs{\mm,\wittenk,\szabo,\wittenduality,\mw,\fh,\DMW,\freed}. Unfortunately, 
this classification is not $U$-duality invariant. Finding 
a U-duality invariant formulation of M-theory which
at the same time naturally incorporates the K-theoretic 
formulation of RR fields remains an outstanding open problem. 

With this problem as motivation, 
the present paper investigates the interplay between the 
K-theoretic formulation of RR fields and the T-duality 
group, an important subgroup of the full U-duality group. 
While T-duality invariance of the theory was one of the 
guiding principles in the definition of the K-theoretic 
theta function \wittenduality\DMW\  we will see that the 
full implementation of T-duality invariance of the 
low energy effective action of type II string theory 
is in fact surprisingly subtle, 
even on backgrounds as simple as $T^2 \times X$, 
where $T^2$ is a two-dimensional torus, and $X$ is an 
8-dimensional compact spin manifold. 
We will show that, in fact,  in the RR sector there is a T-duality 
anomaly. This anomaly is cancelled by a compensating anomaly from
 fermion determinants 
together with quantum corrections to the 8D effective action. 
A by-product of our computation is a complete analysis of the 1-loop
determinants of IIA supergravity on $X\times T^2.$

As an application of our discussion, we re-examine a 
proposal of C. Hull \Hull\ for interpreting the 
Romans mass of IIA supergravity in the framework of 
M-theory. We will show that, while the interpretation 
cannot hold at the level of classical field 
theory, it might well hold as a quantum-mechanical 
equivalence. In section 10 we comment on some of the issues
which arise in extending our computation to a fully
U-duality invariant partition function. This includes the
role of twisted K-theory in formulating the partition
sum.

This paper is long and technical. Therefore we
have attempted to write a readable summary of our results
in the remainder of the introduction. 

\subsec{The effective eight-dimensional supergravity, and its 
partition function} 

 Previous studies of the  partition function in type II string theory 
 \wittenduality\DMW\ considered  
 the limit of a large 10-manifold. One chose 
a  family of  Riemannian 
metrics $g = t^2 g_0$ with $t\to \infty$ and $g_0$ fixed. 
Simultaneously, one took the string coupling to zero. 
The focus of these works was on   the sum over classical 
field configurations of  the RR fields.  In 
 this paper we consider the limit where only 8 of the dimensions are
large.  
The 
metric has the form 
\eqn\introi{
ds^2 = ds^2_{T^2} + t^2 ds^2_{X}
}
where $ds^2_{T^2}$ is flat when pulled back to $T^2.$
  The background dilaton $g_{\rm string}^2 = e^{2\xi}$ is constant. 
We will work in the limit
\eqn\limit{
\eqalign{
t & \to \infty \cr
e^{-2\xi }& := e^{-2\phi} V \to \infty \cr}
}
where $V$ is the volume of $T^2$ and $\phi$ is the  10-dimensional
dilaton. Finally - 
and this is important -until section 10  we assume the background
NSNS 3-form flux, ${\hat H},$ is identically zero. In particular, the 
2-form potential,
$ {\hat B},$ is a globally well-defined
harmonic form on $X\times T^2.$

As is well-known the 
background data for the toroidal compactification \introi\ 
include a pair of points 
$(\tau,\rho)\in \CH \times \CH $ where $\CH$ is the upper half complex
plane. $\tau$ is the 
Teichmuller parameter of the torus and $\rho:=B_0 + i V$, 
where   $B_0d\s^8\wdg d\s^9$ is an harmonic
2-form  on $T^2$. While we work in the limit \limit, within this approximation 
we work with exact expressions in 
the geometrical data $(\tau, \rho).$
In this way we go beyond \DMW.
 
It is extremely well-known that the 
low energy effective 8D supergravity theory obtained by 
Kaluza-Klein reduction of type II supergravity on $T^2$ has 
a  ``$U$-duality symmetry'' which is classically 
$SL(3,R) \times SL(2,R)$, and is broken to 
$\CD := SL(3,Z) \times SL(2,Z)$ by quantum effects
\refs{\Town,\GPR,\BRG,
\Lavr,\Cvet}.
These are symmetries of the equations of motion and not of 
the action. (The implementation of these symmetries 
at the level of the action involves
 a Legendre transformation of the fields.)  What is 
perhaps less well-known is that the K-theoretic formulation 
of RR fields leads to an extra term in the supergravity 
action which is nonvanishing in the presence of nontrivial 
flux configurations. Indeed, the proper formulation of 
this term is unknown for arbitrary flux configurations
with $[\hat H_3]\not=0$, but for  
topologically trivial NSNS flux  the extra 
term is known \DMW\ and is recalled in 
equations (1.14) and (1.15) below. 
This term breaks naive duality invariance 
of the classical supergravity theory already for 
the T-duality subgroup of the U-duality group, and 
makes the discussion of T-duality nontrivial. 

Let us now summarize the fields and T-duality transformation 
laws in the conventional description of the 
eight-dimensional   effective supergravity theory  on $X.$ 
The T-duality group is 
$\CD_T = SL(2,Z)_{\tau} \times SL(2,Z)_{\rho}$. The 
theory has the following bosonic fields. From the NSNS 
sector there is a scalar $t$, characterizing the size of $X,$
a unit volume metric  $g_{MN}$, a 2-form potential
\foot{We will always indicate by the subscript $(p)$
the degree $p$ of a differential $p$-form on $X$} $B_{(2)}$, with fieldstrength $H_{(3)},$ 
and 
a dilaton $\xi$, all of which are invariant under  $\CD_T.$ In addition, 
there is a multiplet of 1-form potentials ${\bf A}_{(1)}^{ m \a}$
transforming in the $({\bf 2},{\bf 2})$ of $\CD_T$. 
Finally, the pair of  scalars $(\tau,\rho)$, 
transform under $(\gamma_1,\gamma_2)\in \CD_T$ as 
$(\tau,\rho) \to (\gamma_1\cdot \tau, \gamma_2\cdot \rho)$ where 
$\gamma\cdot$ is the action by a fractional linear transformation. 
We therefore call the factors $SL(2,Z)_\tau, SL(2,Z)_\rho$, respectively. 

The fieldstrengths from the RR sector include a 0-form and a
 2-form, $ g_{(p)}^{\a}$, $p=0,2,\a =1,2$ transforming in the 
$({\bf 1}, {\bf 2})$ of $\CD_T$, and a 1-form and 
3-form  $g_{(p)m}$, $p=1,3,m=8,9$ transforming in the 
$({\bf 2'},{\bf 1} )$ of $\CD_T$.
Finally there is a 4-form fieldstrength 
$g_{(4)}$ on $X$. This field does not transform locally under $T$-duality, 
rather its equation of motion mixes with its Bianchi identity \Cvet. 
The fermionic partners are described in section 7 below.

The real part of the standard bosonic  supergravity action takes the form
\eqn\standact{
Re\left (S_{\rm boson}^{(8D)}\right ) =
 S_{NSNS} + \sum_{p=0}^3S_p\left ( g_{(p)}\right)+
S_4\left (g_{(4)}\right) 
}

In the action \standact\   all of the terms except for
 the last term are 
manifestly T-duality invariant. 
The detailed forms of the actions are: 
\eqn\sns{
\eqalign{
S_{NSNS} & = {1 \over 2\pi}
\int_X e^{-2\xi}\Biggl\{ t^6\bigl(\CR(g) +4  d\xi \wedge * d\xi
+ 28 t^{-2}dt \wdg *dt \bigr ) +
 \half t^2 H_{(3)}\wedge * H_{(3)}\cr
& + \half t^6{d\tau \wedge * d\tau \over (\im \tau)^2 } +
\half t^6 {d\rho \wedge * d\rho \over (\im \rho)^2 } 
+ \half t^4 g_{mn}{\cal G}_{\a \b}{\bf F}_{(2)}^{m \a}
\wdg * {\bf F}_{(2)}^{n \b} \Biggr \} \cr}}
where  $*$ stands for the
 Hodge dual with the metric $g_{MN}$, we also denote  
\eqn\snsiii{{\bf F}_{(2)}^{m \a}=d{\bf A}_{(1)}^{m \a},\quad
H_{(3)}=dB_{(2)}-\half \epsilon_{mn}{\cal E}_{\a \b}
{\bf A}_{(1)}^{m \a} {\bf F}_{(2)}^{n \b},}
$\epsilon_{mn}$ and ${\cal E}_{\a \b}$ are
invariant antisymmetric  tensors for $SL(2,\Z)_{\tau}$ and $SL(2,\Z)_{\rho}$
respectively,
\eqn\mrho{g_{mn}=\CM(\tau),\quad g^{mn}=\CM(\tau)^{-1},\quad
{\cal G}_{\a \b}=\CM(\rho)}
and finally
\eqn\matrdef{
\CM(z):= {1\over  \im z} 
\pmatrix{1 & \re z \cr \re z & \vert z\vert^2\cr}.
}
The real part of the RR sector action is given by
\eqn\rri{
\sum_{p=0}^3 S_p\left ( g_{(p)}\right ) =
\pi \int_X \Biggl\{
 t^8 {\cal G}_{\a \b} g^{\a}_{(0)}\wdg *g^{\b}_{(0)} +
 t^6 g^{mn} g_{(1)m}\wdg * g_{(1)n} +}
$$  t^4 {\cal G}_{\a \b} g^{\a}_{(2)}\wdg *g^{\b}_{(2)} + 
 t^2 g^{mn} g_{(3)m}\wdg * g_{(3)n} \Biggr \}
$$
together with 

\eqn\gfouract{
S_{4}\left (g_{(4)}\right) =  \pi \int_{X} \im(\rho) g_{(4)} \wedge * g_{(4)} .}

\subsec{The semiclassical expansion}

The vevs of the two fields $t$ and $e^{-2\xi}$ (the 8-dimensional 
length scale of $X$ and the 
inverse-square 8D string 
coupling) define semiclassical expansions when they become large. 
We will expand around a solution of the equations of 
motion on $X$. To leading order in our expansion this 
means $X$ admits  a  Ricci flat
 metric\foot{Almost nothing in what follows relies on the Ricci flatness 
of the metric. We avoid using this condition since a $T$-duality 
anomaly on non-Ricci flat manifolds would signal an important 
inconsistency in formulating string theory on 
manifolds of topology $X \times T^2$.}  
 $g_{MN}.$  We also have constant 
scalars $t,\xi,\tau,\rho$, and ${\bf F}_{(2)}^{m\a}=0,\quad
H_{(3)}=0,$ so the background action $S_{NSNS}$ is zero. 
Finally, we expand around a classical field configuration 
for the RR fluxes, and to leading order these fluxes $g_{(p)}$ are
harmonic forms. Nonzero fluxes contribute terms 
to the partition function going like  $\CO(e^{-t^{8-2p}}) $. 

Let us consider the leading order contribution to the 
partition function. There are several sources of contributions 
even at leading order, but, since we are interested in 
questions of T-duality, most of these
 can be neglected. 
\foot{In particular we are negelecting determinants of KK and string modes, 
 and perturbative corrections  
${\cal O}(g_{\rm string}^2)$. These are all $T$-duality invariant. The 
 backreaction of nonzero RR fluxes on the NSNS action simply renormalizes
$V$ to $V_{eff},$ where   $\rho=B_0+iV_{eff}$ is 
the variable  on which $SL(2,\Z)_{\rho}$ acts by fractional linear
transformations.}
The volume of $X$ suppresses the contribution of 
fluxes  ${ g}_{(p)}, p=0,1,2,3$, 
and, to leading order in the 
$t \to \infty$ expansion these can be set to their classical values. 
Note, however, that neither the string coupling, nor the volume of $X$, 
suppress the action for $g_{(4)}$, and thus we must work in a fully quantum 
mechanical way with this field. This is just as well, since (not 
coincidentally) this is the term in the action which is not manifestly 
T-duality invariant. Fortunately, in our approximation, $g_{(4)}$ is a 
free, nonchiral  field and hence quantization is straightforward
(after the $K$-theory subtleties are taken into account). 
Including subleading 
terms in the expansion parameter $t$ involves (among other things) 
summing  over 
the RR fluxes ${ g}_{(p)}$, $p=0,1,2,3$. 

Finally, in order to be consistent with our approximation scheme we must 
allow the possibility of {\it flat} potentials
in the background. 
\foot{By ``flat'' we mean the DeRham representative of 
the relevant fieldstrength is zero. }
These contribute nontrivially to the partition function through important
phases and accordingly, we will generalize our background to include these. 
The real part of the
action for the flat configurations vanishes, of course, and hence
in the physical partition function one must integrate over  
these flat configurations. In the RR sector the flat potentials 
are thought to be classified by $K^1(X_{10};U(1))$ \mw. These contribute 
no phase to the action and we will henceforth ignore them. 
\foot{If treated as differential forms, RR zero modes
 do contribute to the overall
dependence of the partition sum on $\tilde t=te^{-\xi/3}$.
See  eq.(7.39) below.) In the K-theoretic treatment they also
give a factor of $\vert K^0_{tors}(X\times T^2) \vert$.} 
The space of flat NSNS potentials 
is  
$$H^2(X;U(1)) \times (H^1(X;U(1)))^4. $$
In this paper we will   work only with the identity 
component of this torus. 
 Accordingly, we will identify the space of flat NSNS 
potentials with the torus 
\eqn\flatpot{
{{\cal H}^2(X)\over {\cal H}^2_{\Z}(X)}\times
 \left({{\cal H}^1(X)\over {\cal H}^1_{\Z}(X)}\right )^4
}
where ${\cal H}^p(X)$ is a space of harmonic p-forms on $X$
 and ${\cal H}^p_{\Z}(X)$
is the lattice of integrally normalized harmonic p-forms on $X.$
The first factor is for $B_{(2)}$ and the second factor for the fields
${\bf A}^{m \a}_{(1)}$ transforming in the ${\bf (2,2)}$ of $\CD_T.$

Putting all these ingredients together the partition function 
we wish to study can be schematically written as 
\eqn\schemstruct{
Z(t,g_{MN},\xi,\tau,\rho)
 = \int_{\rm flat~~potentials} d\mu_{\rm flat}
 \sum_{\rm RR~~fluxes} \Det \cdot e^{-S_{cl} } +\cdots 
}
where $d\mu_{\rm flat}$ is a $T$-duality invariant 
measure on the flat potentials, 
$\Det$ is a product of 1-loop determinants and $S_{cl}$ is 
the classical action. Now, to investigate $T$-duality it is 
convenient to denote by $\cal F$ the collection of all fields occuring 
in \schemstruct\ which transform locally and linearly under $\CD_T$. 
These include the flat NSNS potentials
above as well as the classical fluxes $g_{(p)},p=0,\ldots,3.$
We introduce a measure  $[d\CF] $ on  $\CF$ which includes 
integration over the flat potentials and summation over the 
fluxes for $p=0,1,2,3$. This measure is $T$-duality invariant, 
and we can write 
\eqn\introii{
Z(t,g_{MN},\xi,\tau,\rho) = \int [d\CF] Z(\CF;t,g_{MN},\xi,\tau,\rho).
}
The invariance of \introii\ under the subgroup $SL(2,Z)_\tau$ 
of the T-duality group is essentially trivial. 
The relevant actions and   determinants are all 
based on $SL(2,Z)_\tau$-invariant differential operators. 
The invariance of the theory under $SL(2,Z)_\rho$ is, however, 
much more nontrivial. Therefore we simplify 
notation and  just write $Z(\CF,\rho)$ for the integrand of \introii.  
Now, checking $T$-duality invariance is reduced to 
checking the invariance of 
$Z (\CF,\rho)$. This function is constructed from 

\item{a.} The K-theoretic sum over RR fluxes of $g_{(4)}$ in the presence of $\CF$.  

\item{b.} The integration over the Fermi zeromodes in the presence of $g_{(4)}$ and 
$\CF$. 

\item{c.} The inclusion of 1-loop determinants, including determinants of the 
8D supergravity fields and the quantum corrections due to worldsheet 
instantons.

In the following subsections we sketch how each of these elements enters 
$Z (\CF,\rho)$. Briefly, the K-theoretic sum over RR fluxes $g_{(4)}$ leads 
to a theta function   $\Theta(\CF,\rho)$. This function turns out to 
transform anomalously under $T$-duality. The integration over the fermion 
zeromodes corrects this to a function $\widehat{\Theta}(\CF, \rho)$.
This function still transforms anomalously.  The 
inclusion of 1-loop effects, including the string 1-loop effects 
finally cancels the anomaly.

\subsec{The K-theoretic RR partition function }

 In order to write explicit formulae 
for the quantities in \introii\ we must turn to the K-theoretic formulation 
of RR fields. In practical terms the K-theoretic formulation alters 
the standard formulation of supergravity in two ways: First it restricts 
the allowed flux configurations through a ``Dirac quantization 
condition'' on the fluxes.   Second, it changes the supergravity 
action by the addition of important topological terms in the action. 
\foot{It also alters the overall normalization of the bosonic 
determinants by changing the nature of the gauge group for 
RR potentials, but we will not discuss this  in the 
present paper.}

In more detail, the 
 K-theoretic Dirac quantization condition states that the 
DeRham class of the total RR fieldstrength $[G/(2\pi)]$ 
is related to  a K-theory class $x\in K^0(X_{10})$  via 
\eqn\dirac{
[{G\over 2\pi}] = \ch(x) \sqrt{\hat A} 
}
The topological terms in the action can be described as follows.  
On a general 10-manifold this 
term involves the mod-two index of a Dirac operator and cannot  even be
written as a traditional local term in the  supergravity action
\refs{\wittenduality,\mw,\DMW}.
In the case of zero NS-NS fluxes,
the general expression for the phase in the supergravity theory is:
\eqn\introphi{Im(S_{10D})=-2\pi \Phi,\quad  \Phi=\Phi_1+\Phi_2}
where $e^{2\pi i \Phi_2}$ is the mod-two index and
 $\Phi_1$ is given by 
the explicit  expression
\eqn\intpf{\Phi_1=\int_{X_{10}} \Biggl\{ -{1 \over 15}
\left ( {G_2 \over 2\pi} \right)^5+
{1 \over 6}
\left( {G_2 \over 2\pi} \right)^3\Biggl[\left({G_4 \over 2\pi}\right)+
{p_1 \over 12}\Bigl(1+{G_0\over 8\pi} \Bigr)\Biggr]
}
$$-\left({G_2 \over 2\pi}\right)
\Biggl[{p_1 \over 48}\left({G_4 \over 2\pi}\right)+{{\hat A}_8\over 2}
\Bigl(1+{G_0\over 2\pi}\Bigr) +{G_0\over 4\pi  }
\left({p_1 \over 48}\right)^2\Biggr ] \Biggr\}$$
 where $G_{2j},j=0,1,2$ are RR fluxes on $X_{10},$
$p_1 = p_1(TX_{10})$ and ${\hat A}$ is expressed in terms
of the Pontryagin classes of $X_{10}$ as
\eqn\pontrintr{
{\hat A}=1-{1 \over 24}p_1+{1 \over 5760}
\left(7p_1^2-4p_2\right).}

In the case that we reduce to 8 dimensions, taking
our manifold to be of the form   $X\times T^2$ with the choice
of supersymmetric spin structure  on   $T^2$ the above 
considerations simplify and can be made much more 
concrete.

Consider first the Dirac quantization condition. 
We reduce RR fieldstrengths as:
\foot{Beware of notation! The subscript $(p)$ indicates 
form degree, while the other sub- and superscripts on $g_{(p)}$ indicate 
$\CD_T$ transformation properties. Thus, for example, $g_{(0)}^2$ is the 
second component of a doublet $g_{(0)}^{\alpha}$ of $0$-forms.} 
\eqn\zero{
\eqalign{
{{ G}_0 \over 2\pi} & = g_{(0)}^2 \cr {{ G}_2 \over 2\pi}& = g_{(0)}^1 d\s^8\wdg d\s^9
 +g_{(1)m}\wdg d\s^m + g^2_{(2)} \cr {{ G}_4 \over 2\pi} & =g_{(4)} +g_{(3)m}\wdg d\s^m+
g^1_{(2)}\wdg d\s^8\wdg d\s^9
\cr} } 
where $\sigma^m,m=8,9$ are coordinates on $T^2.$
%
%Recall once again the subscript $(p)$ on $g$'s denotes form degree
%on $X.$ The other sub(super)scripts indicate $\CD_T$ transformation
%properties. 
%
In the K-theoretic formulation  
of flux quantization the  fieldstrengths  
$g_{(4)}, g_{(3)m}, g_{(2)}^{\a}, g_{(1)m}, g_{(0)}^{\a}$ are related to
certain integral cohomology classes which we denote  as
\eqn\integrint{
a \in H^4(X,\Z),\quad f_m \in H^3(X,\Z)\otimes {\Z^2},
\quad e^{\a}=\pmatrix{e''\cr e}\in H^2(X,\Z)\otimes {\Z^2},}
$$
  \c_m\in H^1(X,\Z)\otimes {\Z^2},
\quad n^{\a}=\pmatrix{n_1 \cr n_0 \cr}\in H^0(X,\Z)\otimes{\Z}^2$$
The explicit relation between these classes and the 
$g_{(p)}$  is somewhat complicated and given in
equation (4.3) below. The K-theoretic Dirac quantization 
condition leaves all integral classes in \integrint\ 
unconstrained except for $f_m$. One finds that 
$Sq^3(f_m)=0.$
 As explained in section 3.3 and 5.2 
``turning on''   flat NSNS potentials corresponds to acting 
on the K-theory torus by an automorphism changing the 
holonomies of the flat connection on the torus.  In 
concrete terms, turning on flat potentials 
modifies the reduction formulae \zero\ according to equations 
(5.15) to (5.18) below.

Now let us consider the phase. It turns out that 
on  10-folds of the form $X \times T^2$ the phase
 $e^{2\pi i \Phi_2}$ arising from 
 the  mod 2 index may   be expressed in concrete terms as 
\eqn\itrpfii{
\exp[2\pi i \Phi_2] = 
\exp\Biggl[ i \pi \int_X \Biggl \{
g_{(3)8}\cup Sq^2(g_{(3)9})+g_{(3)8}\cup Sq^2(g_{(3)8})+
g_{(3)9}\cup Sq^2(g_{(3)9}) \Biggr\}  }
$$ + i \pi \int_X \Biggl\{ g_{(0)}^2 {\hat A}_8
+\Biggl(g_{(4)}+{g_{(0)}^2 \over 48}p_1 -{1 \over 2} \left(g_{(2)}^2 \right)^2
\Biggr)
\Biggl (\Bigl [g_{(2)}^1-g_{(0)}^1g_{(2)}^2
+g_{(1)8}g_{(1)9}\Bigr ]^2 +{p_1 \over 2}\Biggr ) $$
$$+{p_1^2\over 8} + g_{(1)8}g_{(1)9}
 \left (g_{(2)}^2 \right )^3-
\left (g_{(2)}^2 \right )^2\epsilon^{mn}g_{(1)m}g_{(3)n}\Biggr\}\Biggr] $$
This expression is cohomological although it is still
unconventional in supergravity theory since it involves the 
mod-two valued Steenrod squares, 
denoted $Sq^2(g_{(3)})$, in the first line. 

The above topological term \introphi\ is deduced from the K-theory 
theta function $\Theta_K$ defined in \refs{\wittenduality,\mw,\DMW}, 
and reviewed below. As explained above, it is convenient to fix the 
fields $\CF$. We can define a function $\Theta(\CF,\rho)$ by 
writing $\Theta_{K}$ as a sum 
\eqn\introiii{
\Theta_K = \sum 
 e^{-S_B(\CF)} \Theta(\CF,\rho)
}
The sum is over all integral classes except $a$. 
That is, we sum over $n^{\a},\c_m,e^{\a},f_m$
subject to the constraint on $Sq^3f_m$. The 
action $S_B(\CF)$ is the manifestly $T$-duality 
invariant action for the fluxes given in \rri. 
$\Theta_K$ is a function of $g_{MN}, \rho,\tau$  and 
the flat background  NSNS fluxes.  As we have mentioned, 
turning on flat 
potentials corresponds, in the  
K-theoretic interpretation, to acting by   automorphisms of the K-theory 
group $K^0(X)\otimes R$. These automorphisms act naturally on the theta function. 
We give concrete formulae for this action by showing how 
the inclusion of nonzero flat NSNS fields $B_0,B_{(2)},{\bf A}_{(1)}^{m \a}$
modifies the phase $\Phi$. The explicit formula is in  
 equations (5.20)-(5.24) below.

Since the K-theoretic constraint 
$Sq^3a =0,a \in H^4(X,\Z)$ is automatically
 satisfied on spin 8-folds $X$ 
it turns out that $\Theta(\CF)$ is, essentially, a 
Siegel-Narain theta function for the lattice $H^4(X;\Z)$. 
More precisely, there is a quadratic form on 
$H^4(X;{\bf R})$ given by $Q = \im(\rho)  HI - i \re(\rho)I $ where 
$H$ is the action of Hodge $*$ and $I$ is the integral intersection 
pairing on $H^4(X,\Z)$.  Then 
\eqn\thetasiegel{
\Theta(\CF,\rho) = e^{i 2\pi\Delta  {\widetilde \Phi(\CF)}} 
\Theta \biggl[\matrix{{\vec {\tilde \alpha}}\cr 
{\vec {\tilde \beta}}\cr}\biggr](Q) 
}
Here $\Theta \biggl[\matrix{{\vec {\tilde \alpha}}\cr 
{\vec {\tilde \beta}}\cr}\biggr](Q) $ 
is the Siegel-Narain theta function with 
characteristics. The characteristics are
written explicitly in equations (5.10), (5.20), and (5.21) below.
Finally,
the prefactor $\Delta  {\widetilde \Phi(\CF)}$ in \thetasiegel\
is defined in (5.23) and (5.24)  below.

\subsec{T-duality transformations} 
 
One of the more subtle aspects of the  K-theoretic formulation of RR fluxes, 
 is that the very formulation of the 
action depends   crucially on a choice of polarization of the 
K-theory lattice $K(X_{10})$ with respect to the
quadratic form defined by the index.
In the above discussion we have chosen 
the ``standard polarization'' for IIA theory, i.e
$\Gamma_2$ is the sublattice of $K(X_{10})$  with vanishing $G_4,G_2,G_0.$
$\Gamma_1$ is then a complementary Lagrangian sublattice such that
$K(X_{10})=\Gamma_1+\Gamma_2.$
 The standard polarization
is distinguished for any large 10-manifold in the following sense.
When the metric of $X_{10}$ is  scaled up  ${\hat g}_{\hM \hN}\to
t^2 {\hat g}_{\hM \hN}$   the action $\int_{X_{10}}\sqrt{{\hat g}}|G_{2p}|^2$
of the Type IIA RR 2p-form scales as $t^{10-4p}.$
This
allows the sensible approximation of first summing only over $G_4,$
with $G_2=G_0=0,$ then including $G_2$ with $G_0=0,$ and finally
summing over  all classical fluxes $G_4,G_2,G_0.$

In the case of $X_{10}=T^2 \times X $ with the metric \introi\ the
standard polarization is no longer distinguished.
Various equally good choices are related by the action of the T-duality
group $\CD_T  $ on $\Gamma_K:=K(X\times T^2).$
\foot{
There is also a   polarization on manifolds of the type
$S^1 \times X_9,$ (in our case $X_9=S^1\times X$ )
where the measure is  purely real and the 
imaginary part of the action
 is an integral multiple of $i \pi$ (without flat NSNS potentials).
However,  this  polarization  does not lead to a
 good long-distance approximation scheme. }
In section 4 we explain how the duality group  $\CD_T$
acts as a subgroup of symplectic transformations on the K-theory lattice
and we give an explicit embeding $\CD_T \subset Sp(2N,\Z),$ where 
$2N=rank(\Gamma_K).$
As explained in section 4.2, since
$\CD_T$ acts symplectically, the function $\Theta(\cF,\rho)$
must transform under $T$-duality as 
$\Theta(\gamma\cdot \cF,\gamma\cdot \rho) = j(\gamma, \rho) 
\Theta(\CF,\rho)$ where $j(\gamma,\rho)$ is a standard 
transformation factor for modular forms.  
Nevertheless, this transformation law leaves open 
the possibility of a T-duality anomaly through a multiplier 
system in $j(\gamma,\rho)$.  In order 
to investigate this potential anomaly more closely we 
must choose an explicit duality 
frame and perform the relevant modular transformations.

We find that, in fact, the   function $\Theta(\CF,\rho)$ 
does transform as a modular form with a nontrivial ``multiplier system'' 
under $SL(2,Z)_{\rho}$. That is, using  the standard 
generators $T,S$ of $SL(2,Z)_{\rho}$ we have: 
\eqn\introv{
\eqalign{
\Theta(T \cdot \CF, \rho+1) & = \mu(T) \Theta(\CF, \rho) \cr
\Theta(S \cdot \CF, -1/\rho) 
& =\mu(S) (-i \rho)^{\half b_4^+} (i \bar \rho)^{\half b_4^-} \Theta(\CF,\rho) \cr}
}
where $T\cdot \CF, S\cdot \CF$ denotes the linear action of $\CD_T$ on 
the fluxes. 
Here $b_4^+,b_4^-$ is the dimension of the space of self-dual and 
anti-self-dual 
harmonic forms on $X$ and  the multiplier system is 
\eqn\introvi{
\eqalign{
\mu(T) & =\exp\bigl[ {i \pi \over 4} \int_X \lambda^2 \bigr] \cr
\mu(S) & =\exp\bigl[ {i \pi \over 2} \int_X \lambda^2 \bigr] \cr}
}
where $\lambda $ is the integral characteristic  class 
of the spin bundle on $X.$ (So, $2\lambda=p_1$). 
The multiplier system  
is indeed nontrivial on certain 8-manifolds. As an
 example, on all
Calabi-Yau 4-folds  we have the relation 
\eqn\valid{{1 \over 4}\int_X \lambda^2=
62\int_X{\hat A}_8-4+{1 \over 12}\chi}
and hence $\mu$ is nontrivial if $\chi$ is not 
divisible by $12$. In particular, a 
homogeneous polynomial of degree 6 in $P^5,$   has $\chi=2610$. 
See, e.g. \Sethi. 

In more physical language, the ``multiplier system'' signals a 
potential $T$-duality anomaly. Such an anomaly would spell disaster 
for the theory since the $T$-duality group should  be 
regarded as a {\it gauge symmetry} of M-theory. Accordingly, we 
turn to the remaining functional integrals in the supergravity 
theory. We will find that the anomalies cancel, of course, 
but this cancellation 
is surprisingly intricate. 

\subsec{Inclusion of 1-loop effects}

We first  turn to the 1-loop functional determinants of the quantum 
fluctuations of the bosonic fields. We show that these are all 
manifestly $T$-duality invariant functions of $\CF$ except for the 
quantum fluctuations of $g_{(4)}$. The full bosonic 1-loop determinant 
$\Det_B$ is given in equation (6.20) below.
The 
net effect of inlcuding the bosonic determinants is thus 
to replace 
\eqn\bosonic{
e^{-S_B(\CF)} \Theta(\CF,\rho) \rightarrow Z_B(\CF,\rho) :=\Det_B 
e^{-S_B(\CF)} \Theta(\CF,\rho) 
}
Inclusion of this determinant 
alters the modular weight so that 
$Z_B(\CF,\rho)$ transforms with weight 
$({1\over 4} (\chi+\sigma), {1\over 4}(\chi-\sigma))$,
in close analogy to the theory of abelian gauge potentials 
on a 4-manifold \refs{\Wits,\Verl}. Here 
$\chi, \sigma$ are the Euler character and signature of the 
8-fold $X$. The multiplier system \introvi\ is 
left unchanged. 

Now let us consider modifications from the fermionic path integral. 
Recall that we
 may always regard a modular form as a section of a line bundle over
the modular curve $\CH/SL(2,Z)_\rho$. 
On general grounds, we expect the fermionic path integral to provide a 
trivializing line bundle. The 
gravitino and dilatino in the 8d theory transform as modular forms under 
the T-duality group $\CD_T$ with half-integral weights
 and consequently they too are subject to 
potential $T$-duality anomalies. 

The inclusion of the fermions modifies the bosonic partition 
function in two ways: through zeromodes and through determinants. 
The fermion action in the 8D supergravity has the form 
\eqn\schemferm{S_{\rm Fermi}^{(8)}=S_{\rm kinetic}+S_{\rm fermi-flux}+S_{\rm 4-fermi}
}
where kinetic terms $S_{\rm kinetic}$ as well as fermion-flux couplings $S_{\rm fermi-flux}$ 
are quadratic in fermions and $S_{\rm 4-fermi}$ denotes the four-fermion coupling.
$S_{\rm kinetic}$ is T-duality invariant but $S_{\rm fermi-flux}$ 
 and $S_{\rm 4-fermi}$ 
contain some non-invariant terms.
The non-invariant fermion zeromode couplings
 are collected together in the form
\eqn\collect{ S^{(zm)ninv }
=\int_X \Biggl\{4\pi \im \rho ~ g_{(4)}\wdg *Y_{(4)}+2\pi \im\rho ~ Y_{(4)}\wdg *Y_{(4)} \Biggr\} }
where the  harmonic 4-form $Y_{(4)}$ is  bilinear in the fermion zeromodes. 
The explicit expression for $Y_{(4)}$ can be found in equations (7.21) and 
(7.41) below.

The inclusion of the integral over the fermionic zeromodes  
of $S_{\rm kinetic}$ modifies  the partition function 
by replacing  the expression $\Theta(\CF,\rho)$ 
in \thetasiegel\ by 
\eqn\thetap{
\widehat{\Theta}(\CF,\rho) = \int d\mu_F^{(zm)}~ 
e^{i2\pi \widehat{\Delta \Phi}(\CF)}
 \Theta \biggl[\matrix{{\vec {\widehat \alpha}}\cr 
{\vec{ \widehat \beta}}\cr}\biggr](Q) 
}
Here 
$$
\Theta \biggl[\matrix{{\vec {\widehat \alpha}}\cr 
{\vec{ \widehat \beta}}\cr}\biggr](Q) 
$$
is a supertheta function for a superabelian variety based on 
the K-theory theta function. (This is explained in Appendix F.) 
In particular, the 
characteristics ${\vec {\hat \alpha}},{\vec {\hat \beta}}$ differ from
 ${\vec {\tilde\alpha}},{\vec {\tilde\beta}}$ 
by expressions bilinear in the 
fermion zeromodes. Similarly, the prefactor $\widehat {\Delta \Phi}$ differs from
$\Delta{\widetilde \Phi}$ by 
an expression quartic in the fermion zeromodes.
Finally, $d\mu_F^{(zm)}$ is a $T$-duality invariant measure 
for the  finite 
dimensional  integral over fermion  and ghost zeromodes. 
It includes the $T$-duality invariant term $e^{-S^{(zm)inv}} $ from the action.

Including the one-loop fermionic determinants of the non-zero modes 
we finally arrive at 
\eqn\introxi{
Z_{B+F}(\CF,\rho):= \Det'_B \Det'_F e^{-S_B(\CF)} 
\widehat{\Theta}(\CF,\rho)
}

The formula we derive for  \introxi\   allows a relatively straightforward check 
of the T-duality transformation laws and we find: 
\eqn\introvii{
\eqalign{
Z_{B+F}(T\cdot \CF,\rho+1) & = \mu(T) Z_{B+F}(\CF,\rho) \cr
Z_{B+F}(S\cdot \CF,-1/\rho)& = 
(-i \rho)^{{1 \over 4} \chi+
{1 \over 8}\int_X(p_2-\lambda^2)}(i \bar \rho)^{{1 \over 4} \chi-
{1 \over 8}\int_X(p_2-\lambda^2)}Z_{B+F}(\cF,\rho)} }

Perhaps surprisingly, the fermion determinants have {\it not} completely 
trivialized the RR contribution to the path integral measure. However, 
there is one final ingredient we must take into account: In the low energy 
supergravity there are quantum corrections which contribute
to  leading order in the $t\to \infty$ and 
$\xi \to -\infty$ limit. From the string worldsheet viewpoint these 
consist of a 1-loop term in the $\alpha'$ expansion together with 
worldsheet instanton corrections. From the $M$-theory viewpoint 
we must include the one-loop correction $\int C_3 X_8$ in $M$-theory
together with the effect \Piol\ of membrane instantons. The net effect is to modify the 
action by the quantum correction 
\eqn\introx{
S_{\rm quant} = \Bigl [\half \chi+{1 \over 4}\int_X(p_2-\lambda^2)\Bigr]
 log \left[\eta(\rho)\right]+
\Bigl [\half \chi-{1 \over 4}\int_X(p_2-\lambda^2)\Bigr]
 log \left[\eta(-\bar \rho)\right] }
Where   $\eta(\rho)$ is the Dedekind function.
The final combination 
\eqn\finalcombo{Z(\CF,\rho) = e^{-S_{\rm quant}} Z_{B+F}(\CF,\rho)
}
is the fully T-duality invariant low energy partition function.

\subsec{ Applictions} 

As a by-product of the above results we will make some comments 
on the  open problem of the relation of M-theory 
to massive IIA string theory.
In \Hull\ C. Hull 
made an interesting suggestion for an 11-dimensional 
interpretation of certain backgrounds in the Romans 
theory.
One version of Hull's proposal states that massive IIA 
string theory on $T^2 \times X$ is equivalent to $M$-theory 
on a certain 3-manifold, the nilmanifold.

In section 9 we review Hull's proposal. For 
reasons explained there we are motivated  
 to  introduce a modification of Hull's proposal, in 
which one does not try to set up a 1-1 correspondence 
between M-theory geometries and massive IIA geometries, 
but nevertheless, the physical partition function 
$Z (\CF,\rho)$ of the massive IIA theory 
can be  identified
with  a certain sum over M-theory geometries involving the 
nilmanifold. The detailed proposal can be found   in section 9.3.

\subsec{$U$-duality and $M$-theory} 

In the final section of the paper we comment on some of the 
issues which arise in trying to extend these considerations 
to writing the fully $U$-duality-invariant partition function. 
We summarize briefly the $M$-theory partition function 
on $X \times T^3$, we comment on the $SL(2,Z)_\rho$ duality 
invariance, and we make some preliminary remarks on how 
one can see K-theory theta functions for twisted K-theory 
from the $M$ theory formulation.

\newsec{Review of T-duality invariance in the standard formulation  
of type IIA supergravity}

We start by reviewing  bosonic part of the standard 10D IIA supergravity
action \Romans. Fermions will
be incorporated into the discussion in section 7.
\subsec{Bosonic action of the standard 10D IIA supergravity } 

The 10D NSNS fields are the dilaton $\phi,$ 2-form  potential ${\hat B}_2$
and string frame metric ${\hat g}_{\hM\hN},$    
where $\hM,\hN=0,\ldots 9.$ The
10D RR fieldstrenghts are the 4-form $G_4,$ 2-form $G_2$
and 0-form $G_0.$

We measure all dimensionful
fields in units of 11D Planck length $l_p$ and set $k_{11}=\pi,$ so

\eqn\bosact{
\eqalign{
S^{(10)}_{bos} & = {1 \over 2\pi}
\int_{X_{10}} e^{-2\phi}\Biggl( \sqrt{g_{10}}\CR({\hat g}) +
4  d\phi \wedge {\hat *} d\phi +
 \half {\hat H}_3\wedge {\hat *} {\hat H}_3 \Biggr )\cr
& +{1 \over 4\pi} \int_{X_{10}}\Biggl(
\tG_4\wdg {\hat *}\tG_4 +i{\hat B}\wdg\tG_4\wdg\tG_4 
+ \tG_2\wdg {\hat *}\tG_2+\sqrt{g_{10}}G_0^2 \Biggr) 
} }
where ${\hat *}$ stands for the 10D Hodge duality operator.
The fields in \bosact\ are defined as
$${\tilde G}_2= G_2+{\hat B}_2G_0,
\quad {\tilde G}_4= G_4+{\hat B}_2G_2+\half {\hat B}_2{\hat B}_2G_0,
\quad {\hat H}_3=d{\hat B}_2.$$

We explain the relation between our fields and those of \Romans\ 
in Appendix(B).
\subsec{Reduction of IIA supergravity on a torus} 

We now recall some basic facts about 
the reduction of the bosonic part of the 10D action 
 on $T^2.$
 Let us  consider $X_{10} = T^2 \times X$
and split coordinates  as 
$X^{\hM}=(x^M,\sigma^{m}),$
where $ M=0,\ldots,7,\quad m=8,9.$ 

The standard ansatz for the reduction of the 
10d metric  has the form:
\eqn\redmetr{ds_{10}^2=t^2 g_{MN}dx^Mdx^N+Vg_{mn}\omega^m \otimes \omega^n}
where $g_{mn}$ is defined in \mrho,
 $t^2 g_{MN}$ is 8D metric,  $det g_{MN}=1.$ $V$ is the volume of $T^2$
and 
$ \omega^m=d\s^m+\cA_{(1)}^m.$
The other bosonic fields of the $8D$  theory
are listed below. 
\item 1. 
${ g}^{\a}_{(0)},
{ g}^{\a}_{(2)}, \quad \a=1,2 \quad
{ g}_{(1)m},
{ g}_{(3)m}\quad m=8,9$ \quad
and  $g_{(4)}$
are defined from\foot{$\epsilon^{89}=1, \quad \epsilon_{89}= 1$}
\eqn\arrfieldii{
\eqalign{
{{ G}_0 \over 2\pi} & = g^{2}_{(0)} \cr
{{\tilde G}_2 \over 2\pi}& = 
\left (g^1_{(0)} + g^2_{(0)} B_0 \right) \half \epsilon_{mn}
\omega^m\omega^n +g_{(1)m}\omega^m +
g^2_{(2)} \cr
{{\tilde G}_4 \over 2\pi} & =g_{(4)} +g_{(3)m}\omega^m+
\left ( B_0g^2_{(2)}+g^1_{(2)}\right)\half \epsilon_{mn}
\omega^m\omega^n
\cr}
}
 
\item 2. The
$8D$ dilaton
$\xi$ is defined by 
\eqn\ksi{
e^{-2\xi} =e^{-2\phi} V } 
\item 3.
$B_{(2)},B_{(1)m},B_0$
are obtained from the KK reduction of the
NSNS 2-form potential in the following way 

\eqn\bfldii{{\hat B}_{2}=\half B_0\epsilon_{mn}\omega^m \omega^n+
 B_{(1)m}\omega^m+B_{(2)}+\half \cA_{(1)}^m B_{(1)m}}
Now, the real part of the $8D$ bosonic action obtained 
by the above reduction is
\eqn\standact{
Re\left(S_{\rm boson}^{(8D)}\right) = S_{NS} + \sum_{p=0}^3S_p\left ({ g}_{(p)}\right)+
S_4\left (g_{(4)}\right) 
}

where
\eqn\snsii{
\eqalign{
S_{NS} & = {1 \over 2\pi}
\int e^{-2\xi}\Biggl\{t^6\bigl( \CR(g) +4  d\xi \wedge * d\xi+
28 t^{-2}dt \wdg *dt\bigr ) +
 \half t^2H_{(3)}\wedge *H_{(3)}\cr
& + \half t^6{d\tau \wedge * d\tau \over (\im \tau)^2 } + 
\half t^6 {d\rho \wedge * d\rho \over (\im \rho)^2 } 
+\half t^4 g_{mn}{\cal G}_{\a \b}{ \bf F}^{m \a}
 \wdg * {\bf F}^{n \b} \Biggr\} \cr}
}
where ${\cal G}_{\a \b}$ is defined in \mrho\ and
 $\cA_{(1)}^m$ and $B_{(1)m}$ are combined
into 1-form as a collection of
\eqn\snsiii{
{\bf A}_{(1)}^{m \a}=
\pmatrix{\epsilon^{mn}B_{(1)n}\cr \cA_{(1)}^m \cr} }

Also, we  denote\foot{${\cal E}_{12}=1, \quad {\cal E}_{21}=-1$}
\eqn\defhiii{
H_{(3)}=dB_{(2)}-\half \epsilon_{mn}{\cal E}_{\a \b}
{\bf A}_{(1)m \a} {\bf F}_{(2)}^{n \b}
}

\eqn\rrii{
\sum_{p=0}^3 S_p\left ( g_{(p)}\right ) =\pi \int_X \Biggl \{
  t^8 {\cal G}_{\a \b} g^{\a}_{(0)}\wdg * { g}^{\b}_{(0)} +
  t^6 g^{mn} { g}_{(1)m}\wdg *{ g}_{(1)n} +}
$$ t^4{\cal G}_{\a \b}{ g}^{\a}_{(2)}\wdg *{ g}^{\b}_{(2)} + 
 t^2 g^{mn} { g}_{(3)m}\wdg *{ g}_{(3)n} \Biggr \}
$$
Finally we have
\eqn\gfouract{
S_{4}\left (g_{(4)}\right) =  \pi \int_{X} \im(\rho) g_{(4)} \wedge * g_{(4)}}

It is convenient to introduce the 
notation $S_B(\cF)=\sum_{p=0}^3 S_p\left ( g_{(p)}\right )$
for the value of the actions 
evaluated on a background flux field configuration. $S_B(\cF)$
will enter the partition sum $Z_{B+F}({\cal F},\tau,\rho)$
in equation (8.1) below.

\subsec{ T-duality action on 8D bosonic fields}

The T-duality group  of the $8D$ effective theory
obtained by reduction on $T^2$ is known to be
 $\CD_T=SL(2,{\bf Z})_{\tau}\times SL(2,{\bf Z})_{\rho}, $
where the first factor is mapping class group of $T^2$
which acts on $\tau$
\eqn\tautr{
\tau \rightarrow {{a \tau + b}\over  {c \tau + d}} }
 and the second factor acts on $\rho=B_0+iV$
\eqn\rhotr{
\rho \rightarrow {{\a \rho + \b}\over  {\c \rho + \d}} } 
Let us denote  generators of  $SL(2,{\bf Z})_{\rho}$  by  
$${ S}:\rho \rightarrow -1/\rho, \quad 
{ T}:\rho \rightarrow \rho +1 $$
and generators of  $SL(2,{\bf Z})_{\tau}$  by  
$${\tilde { S}}:\tau \rightarrow -1/\tau, \quad 
{\tilde { T}}:\tau \rightarrow \tau +1 $$

We now  recall how T-duality
acts on the remaining bosonic fields of the 8D theory \Cvet.
First, $ \xi ,t, g_{MN}$ are $T$-duality invariant. 
Next, there is the collection of fields $\CF$ mentioned 
in the introduction. These transform linearly under 
$T$-duality. They include the NS potential
$B_{(2)}$, which is   T-duality invariant, 
as well as ${\bf A}_{(1)}^{m \a}$, which 
 transform in the ${\bf  (2,2)}$. The other components 
of $\CF$ are the RR fieldstrengths ${ g}^{\a}_{(0)},{ g}^{\a}_{(2)}, \a=1,2$ 
which transform
in the ${\bf (1,2)}$ of $\CD_T$ 
and $ { g}_{(1)m},{ g}_{(3)m}, m=8,9$ which  transform
in the ${\bf (2',1)}$ of $\CD_T$. 
 
Finally, the field $ g_{(4)}$
  is singled out among all the other fields since
according to  the conventional supergravity \Cvet\  
 $SL(2,\Z)_{\rho}$ mixes  $g_{(4)}$ with its Hodge dual $*g_{(4)}$ 
and hence $g_{(4)}$ does not have a local transformation. 
More concretely, 
\eqn\doublet{
\pmatrix{ -\re\rho g_{(4)}+i\im\rho*g_{(4)}\cr g_{(4)}  }
}
transforms in the ${\bf (1,2)}$ of $\CD_T.$
Due to this non-trivial transformation  the classical bosonic 8D action
$S_{\rm boson}^{(8D)}$
is not manifestly invariant under  $SL(2,\Z)_{\rho}.$

\newsec{Review of the K-theory theta function }

In this section we review the basic flux quantization law of 
RR fields and the definition of the K-theory theta function. 
We follow closely the treatment in \refs{\wittenduality,\mw,\DMW}. 

\subsec{K-theoretic formulation of RR fluxes}

As found in \mm-\wittenduality\ RR fields in IIA 
superstring theory  are classified topologically by 
an element $x \in K^0(X_{10}).$ The relation for  ${\hat B}_2=0$ is
\eqn\topcl{\left [{G \over 2\pi}\right ]=\sqrt{{\hat A}}ch x, 
\quad G=\sum_{j=0}^{10}G_j }
where ch is 
the total Chern character and ${\hat A}$ is expressed in terms
of the Pontryagin classes as
\eqn\pontr{{\hat A}=1-{1 \over 24}p_1+{1 \over 5760}\left(7p_1^2-4p_2\right)}
In \topcl,
 the right hand side  refers to the harmonic
differential form in the specified real cohomology class.
The quantization of the RR background fluxes is understood in the sense that
they are derived  from an element of $K^0(X_{10}).$

\subsec{Definition of the K-theory theta function} 

Let us recall the general construction of a K-theory
theta function, which serves as the RR partition function in Type IIA.
One starts with the lattice $\Gamma_K=K^0(X_{10})/K^0(X_{10})_{tors}.$
This lattice is endowed with an integer-valued unimodular
antisymmetric form by the formula 
\eqn\symplstr{\omega(x,y)=I(x\otimes \bar y),}
where for any $z \in K^0(X_{10}),$ $I(z)$ is the index
of the Dirac operator with values in $z.$

Given a metric on $X_{10},$ one can define a metric on $\Gamma_K$ 
\eqn\kmetric{g(x,y)=\int_{X_{10}}
{G(x) \over 2\pi}\wdg {{\hat *}G(y) \over 2\pi}}
where ${\hat *}$ is the 10D Hodge duality operator.

Let us consider  the torus ${\bf T}=\left(\Gamma_K \otimes_{\Z}{\bf R}\right)/\Gamma_K.$
The quantities $\omega$ and $g$ can be interpreted as 
a symplectic form and a metric, respectively, on  ${\bf T}.$
To turn ${\bf T}$ into a Kahler manifold one 
defines the complex structure $J$ on  ${\bf T}$ as
\eqn\complstr{g(x,y)=\omega(Jx,y)}
Now, if it is possible to find a complex line bundle
 $\cal L$  over ${\bf T}$ with $c_1({\cal L})=\omega,$ 
then  ${\bf T}$ 
 becomes a
``principally polarized abelian variety.''
 $\cal L$  has, up to a constant multiple, a unique\foot{
The uniqueness follows from the index theorem on ${\bf T}$
using  unimodularity of $\omega$ and the fact that for
any complex line bundle $M$ over ${\bf T}$ with positive curvature
we have
$H^i\left({\bf T}; M\right )=0,\quad i>0.$ }
holomorphic section which is the contribution of the sum over  fluxes
to the RR partition function.

As  explained in detail in \wittenvbr, holomorphic
line bundles ${\cal L}$ over ${\bf T}$ with  constant
curvature $\omega$ are in one-one correspondence with U(1)-valued
functions $\Omega$ on $\Gamma_K$ such that
\eqn\omegak{  \Omega(x+y)= \Omega(x) \Omega(y)(-1)^{\omega(x,y)}.}
For 
weakly coupled Type II superstrings
one can take $\Omega$ to be valued in $\Z_2.$ 
Motivated   by T-duality, and the requirements of 
anomaly cancellation on D-branes \mw,  Witten proposed that the 
natural  $\Z_2-$ valued function $\Omega$ for the RR partition 
function is given by a mod two index  \wittenduality. 
For any $x \in K^0(X_{10}),$ $x \otimes {\bar x}\in KO(X_{10})$ 
lies in the   real K-theory group on $X_{10}$, and for any
 $v \in KO(X_{10}),$ there is a well-defined
 mod 2 index $q(v)$ \asv. We take 
\eqn\omegakii{\Omega(x)=(-1)^{j(x)}}
where
$j(x)=q(x\otimes \bar x)$. 

As explained in \refs{\wittenduality,\mw,\DMW}\ there is an 
anomaly in the theory unless $\Omega(x)$ is  identically 1
on the torsion subgroup  of $K(X_{10}).$ In the absence of 
this anomaly it descends to  a function on $\Gamma_K=
K^0(X_{10})/K^0(X_{10})_{tors}$
and can be used to define a line bundle ${\cal L}$ and hence
the RR partition function.

To define the theta function one  must choose a 
decomposition of $\Gamma_K$ as a sum $ \Gamma_1 \oplus \Gamma_2,$
where $\Gamma_1$ and $\Gamma_2$ are ``maximal Lagrangian''
sublattices.
$\omega$ establishes a duality between $\Gamma_1$ and $\Gamma_2,$
and therefore there exists $\theta_K \in \Gamma_1/2\Gamma_1$ such that
\eqn\omegakiii{\Omega(y)=(-1)^{\omega(\theta_K,y)},\quad \forall y\in \Gamma_2}
Following \DMW\ we choose 
the standard polarization:the sublattice $\Gamma_2^{std}$
 is defined as the set of
$x $ with  vanishing $G_0,G_2,G_4.$ This choice implies that  $G_0,G_2,G_4$
are considered as independent variables. This  is a distinguished
choice for  every large 10-manifold in the sense that
it allows for a good large volume semiclassical approximation scheme 
on any 10-manifold ( see sec.5). 

It was demonstrated in \DMW\ that  $ \Gamma_1^{std}$
in the standard polarization consists of K-theory
classes of the form $x=n_0\one+x(c_1,c_2).$
 $\one$ is a trivial complex line bundle 
and $x(c_1,c_2)$ is defined for $c_1 \in H^2(X_{10},\Z)$
and $c_2 \in H^4(X_{10},\Z)$  with $Sq^3c_2=0,$ as 
\eqn\defx{ ch(x(c_1,c_2))=c_1+(-c_2+\half c_1^2 )+\ldots .}
 The higher Chern classes indicated by $\ldots$  are such that $x(c_1,c_2)$
 is in a maximal Lagrangian sublattice $\Gamma_1^{std}$ complementary to
$\Gamma_2^{std}.$
Then, $\theta_K $ for
 the standard polarization can be chosen to satisfy
\eqn\sat{ch_0(\theta_K)=0,\quad ch_1(\theta_K)=0,
\quad ch_2(\theta_K)=-\l+2{\hat a}_0,\quad I(\theta_K)=0}
where $\quad \lambda=\half p_1$ and $ {\hat a}_0$ is a fixed element
of $H^4(X_{10},\Z)$ such that 
\eqn\sati{
\forall {\hat c}\in L'
\quad f({\hat c})=
\int_{X_{10}} {\hat c} \cup 
Sq^2{\hat a}_0} 
where $L'=\Bigl\{{\hat c}\in 
H_{tors}^4(X_{10},\Z)/2H_{tors}^4(X_{10},\Z),\quad Sq^3({\hat c})=0\Bigr\}$ and
$f({\hat a})$ stands for the mod 2 index of the Dirac operator 
coupled to an $E_8$ bundle 
on the 11D manifold $X_{10}\times S^1$  with the characteristic class 
${\hat a} \in H^4(X_{10},\Z)$ and supersymmetric spin structure on the $S^1$. 
(We will show in section 5.1 
below that for $X_{10}= X \times T^2$ in fact $\hat a_0 =0$.)

The K-theory theta function in the standard polarization is 
\eqn\intro{
\Theta_K=
e^{iu}\sum_{x \in \Gamma_1}
e^{i\pi \tau_K(x+\half \theta_K) } \Omega(x) }
where $u=-{ \pi \over 4}\int_{X_{10}} 
ch_2(\theta_K)ch_3(\theta_K) $ and
the explicit form of the period matrix $\tau_K$  
is given by
\eqn\tauk{
Re\tau_K(x+\half \theta_K)={1\over (2\pi)^2}\int_{X_{10} }\left 
(G_0G_{10}-G_2G_8+G_4G_6\right)}
\eqn\taukim{
Im\tau_K(x+\half \theta_K)=\sum_{p=0}^{2}
{1\over (2\pi)^2}\int_{X_{10}}  G_{2p}\wdg {\hat *}G_{2p} }
 The RR fields which enter \tauk,\taukim\ are: 
\eqn\fldstrngth{
\eqalign{
{1\over 2\pi} G_0(x+\half \theta_K) & = n_0 \cr
{1\over 2\pi}G_2(x+ \half \theta_K) & =  {\hat e} \cr
{1\over 2\pi}G_4(x+\half \theta_K) & = {\hat a} + 
 \half {\hat e}^2    - \half (1+n_0/12) \lambda\cr
}
}
where we denote ${\hat e}=c_1(x), \quad {\hat a}=-c_2(x)+{\hat a_0}. $

From \intro\ and \tauk,\taukim\  the following
topological term was found in \DMW\ to be the
K-theoretic corrections 
to the 10D IIA supergravity action.

\eqn\defPhi{e^{2\pi i \Phi(n_0,{\hat e},{\hat a})}=
exp\Bigl[-2 \pi i n_0 \int_{X_{10}}{\hat e}
\left( \sqrt{\hA} \right )_8 \Bigr ]\left(\Omega(\one)\right )^{n_0}
e^{2\pi i \Phi({\hat e},{\hat a})}}
\eqn\defPhii{
e^{2\pi i \Phi({\hat e},{\hat a})}=
(-1)^{f({\hat a}_0)}(-1)^{f({\hat a})}exp \Biggl [ 2\pi i \int_{X_{10}}
\Bigl ( {{\hat e}^5\over 60}+ {{\hat e}^3{\hat a}\over 6}-
 {{11{\hat e}^3\l}\over 144}
- {{{\hat e}{\hat a}\l}\over 24}+ {{{\hat e}\l^2}\over 48}-
 \half {\hat e}\hA_8 \Bigr ) \Biggr ]}

\subsec{Turning on the NSNS 2-form flux with $[{\hat H}_3]=0$}

In the presence of an $H$-flux we expect $K$-theory to be replaced 
by twisted $K$-theory $K_H$ classifying bundles of algebras with 
nontrivial Dixmier-Douady class. The Morita equivalence class of 
the relevant algebras only depends on the cohomology class of $H$, 
but this does not mean that the choice of ``connection'' that is, 
the choice of $B$ field is irrelevant to formulating 
the $K$-theory theta function. Indeed, when $[\hat H]=0$, the 
choice of trivialization 
$\hat B$ in $\hat H = d \hat B$ 
changes the action in supergravity and ``turning on'' this 
field in supergravity corresponds to acting with an automorphism 
on the K-theory torus. In this section we describe  this change explicitly. 
See \Mathai\MathaiB\ for recent mathematical 
results relevant to this issue.

Let us turn on ${\hat B}_2\in H^2(X_{10},R).$
We normalize ${\hat B}_2$ so that it is defined mod $H^2(X_{10},\Z)$ under global 
tensorfield gauge transformation.
By Morita equivalence, the RR fields are still classified topologically by
 $x \in K^0(X_{10}).$ The standard coupling to the D-branes 
implies that the cohomology class of the RR field is
\eqn\defgtilde{{\tG(x)\over 2\pi}=e^{{\hat B}_2}ch(x)\sqrt{\hat A}
}
Let us define
\eqn\bartg{{{\overline {\tG(x)}}\over 2\pi}:=
e^{-{\hat B}_2}ch(\bar x)\sqrt{\hat A}}
The bilinear form on $\Gamma_K=K^0(X_{10})/K^0(X_{10})_{tors}$
 is still given by the index:
\eqn\modsympl{\omega(x,y)={1 \over (2\pi)^2}
\int_{X_{10}}\tG(x)\wdg {\overline {\tG(y)}}=I(x\otimes {\bar y})}
while the  metric on $\Gamma_K$ is modified to be
\eqn\modmetr{{\tilde g}(x,y)={1 \over (2\pi)^2}
\int_{X_{10}}\tG(x)\wdg {\hat *}\tG(y)}
and the $\Z_2$ valued function $\Omega(x)$ is unchanged.
If we continue to use the standard polarization then $\theta_K \in 
\Gamma_1/2\Gamma_1$ is unchanged as well.

The net effect to modify \intro\ is that the period matrix
$\tau_K$ should be substituted for $\widetilde {\tau_K}={\tau_K}( G \to \tG ).$
\eqn\introii{
\Theta_K\left({\hat B}_2\right )=
e^{iu}\sum_{x \in \Gamma_1}
e^{i\pi \widetilde {\tau_K}(x+\half \theta_K) } \Omega(x) }
Note, that the constant phase $e^{iu}$ in front of the sum 
remains the same as in  \intro\

The imaginary part of the 10D Type IIA supergravity action
now becomes $ Im(S_{10D})=-2\pi {\widetilde \Phi},$ where
\eqn\modphase{{\widetilde \Phi}=\Phi+{1 \over 8\pi^2}\Bigl[{\hat B}_2G_4^2+
{\hat B}_2^2G_2G_4+{1 \over 3}{\hat B}_2^3\left (G_2^2+G_0G_4\right )
+{1 \over 4}{\hat B}_2^4G_0G_2+{1 \over 20}{\hat B}_2^5G_0^2 \Bigr ],}
 $\Phi$ is defined in \defPhi,\defPhii\ and $G_{2p}(x+\half \theta_K),
\quad p=0,1,2$ are given in \fldstrngth.

From \modphase\ we find that corrections to $\Phi$ depending on ${\hat B}_2$ 
coincide with the imaginary part of the standard  supergravity action (see, for
example \BRG.)

Note, that $\tG$ defined in \bartg\ is a gauge invariant field if
the  global tensorfield gauge transformation 
\eqn\glgauge{{\hat B}_2\to{\hat B}_2+ f_2,\quad f_2\in H^2(X_{10},\Z)} also  
acts on $K^0(X_{10})$  as: 
\eqn\aut{x \to  L(-f_2)\otimes x, \quad x \in K^0(X_{10})}
where the line bundle $ L(-f_2)$ has $c_1\left( L (-f_2)\right)=-f_2.$

Thus, according to \aut\ a tensorfield gauge transformation acts as
 an automorphism of $\Gamma_K,$ 
 preserving the symplectic form $\omega.$ \aut\ 
acts on theta function \introii\ by multiplication by 
a constant phase:
\eqn\aitii{\Theta_K\left({\hat B}_2+f_2\right )=e^{i {\pi \over 4}
\int_{X_{10}}f_2 (\l-2{\hat a}_0)^2}\Theta_K\left({\hat B}_2\right )}

\newsec{ Action of T-duality in K-theory} 

In this section we consider $X_{10}=T^2 \times X $ and describe the action
 of T-duality on the K-theory variables. 

As we have mentioned,  the standard polarization
is distinguished for any large 10-manifold in the following sense.
When the metric of $X_{10}$ is  scaled up  ${\hat g}_{\hM \hN}\to
t^2  {\hat g}_{\hM \hN}$   the action $\int_{X_{10}}\sqrt{{\hat g}}|G_{2p}|^2$
of the Type IIA RR 2p-form scales as $t^{10-4p}.$
This
allows the successive approximation of keeping only $G_4$
whose periods have the smallest action, then including $G_2$
and finally keeping all $G_4,G_2,G_0.$

In the case of $X_{10}=T^2 \times X $ with the metric \introi, the
standard polarization is no longer distinguished.
Various equally good choices are related by the action of the T-duality
group $\CD_T  $ on $\Gamma_K=K^0(T^2 \times X )/K_{tors}^0(T^2 \times X ).$

We argue below that $\CD_T  $ can be considered as a subgroup of $Sp(2N,\Z),$
where  $N$ denotes the complex dimension of the K-theory torus
${\bf T}=K^0(T^2 \times X )\otimes_{\Z}{\bf R}/ \Gamma_K $ 
and $Sp(2N,\Z)$ stands for the group of   symplectic transformations
of the lattice $\Gamma_K.$

\subsec{Background RR fluxes in terms of integral classes on $X.$}

To describe the action of $\CD_T$ on K-theory variables, we
will  write RR fields in terms of integral classes on $X.$ 
Let us  start from the standard polarization \foot{ $\Gamma_1^{std}$ and
$\Gamma_2^{std}$ are defined on page 19.} 
and 
write a general element of $\Gamma_1^{std}$ as
\eqn\newx{
x=n_0 {\bf 1} + \Bigl (  L(n_1e_0+e+ \c_m d\s^m)-{\bf 1} \Bigr )+
x(e_0e'+a+h_md\s^m)+\Delta}
where $e_0 = d\sigma^8 \wedge d\sigma^9$, so that $\int_{T^2} e_0 = 1$.
 $ L({\hat e})$ is a line bundle 
with $c_1( L)= {\hat e}\in H^2(X_{10};\Z)$,
\one\ is a trivial line bundle, and
 for any ${\hat a}\in H^4(X_{10};Z)$, $x({\hat a})$
 is a $K$-theory lift (if it exists). 
In \newx\ $\Delta$ puts $x$ into the Lagrangian lattice $\Gamma_1^{std}$
and we also  introduce the notations:
\eqn\newxii{ 
a\in H^4(X;\Z),\quad  e,e'\in H^2(X;\Z),
\quad h_m \in H^3(X,{\bf Z}),\quad \c_m  \in H^1(X;\Z)\quad m=8,9}

%
%Note that if ${\hat a}_0\in H^4(X\times T^2,\Z),$ defined in \sati,
% is nonzero, it 
% has the property 
% $Sq^3{\hat a}_0\ne 0,$ so that it must have the form
%${\hat a}_0=a_{m}d\s^m, a_m\in H^3(X,\Z).$
%It is convenient to redefine $h_m$ by  including $a_m,$
%which will be assumed from now on.
%

The RR fields entering \tauk,\taukim\ are  given by
\eqn\newG{\eqalign{
{1\over 2\pi}G_0(x+\half \tht_K)& = n_0,\cr
{1\over 2\pi}G_2(x+\half \tht_K)& = n_1e_0+e+\c_m d\s^m,\cr
{1\over 2\pi}G_4(x+\half \tht_K)& = a+\half e^2 + e_0{e}''+f_md\s^m
-\half(1+n_0/12)\l\cr
}}
where
\eqn\redef{
{e}''=n_1e+e'- \c_1 \c_2, \quad f_m=h_m+a_m+e\c_m}
Note that \tauk\ is in fact only a function of these variables, 
by the Lagrangian property. 

From the 10D constraint $Sq^3{\hat a}=Sq^3{\hat a}_0,$
valid in the case $[{\hat H}_3]=0,$ we find
the constraints on the integral cohomology classes:
$Sq^3f_m=Sq^3a_m, \quad m=8,9.$ We will show that
actually $Sq^3f_m=0,\quad m=8,9$ ( see comment below 5.8).

\subsec{The embedding $\CD_T \subset Sp(2N,\Z)$}
From the  transformation  rules of the RR fields under the T-duality
group \Myers\ we 
find that $f_m$ and $\c_m$ transform in the ${\bf (2',1)}$ of $\CD_T$
and we can form a representation ${\bf (1,2)}$ out of $n_0,n_1$ and $e,e''$ 
in the following way:
\eqn\formrep{n^{\a}=\pmatrix{n_1 \cr n_0},\quad
e^{\a}=\pmatrix{e'' \cr e} }

We would like to reformulate the transformation rules
for RR fields in terms of the action on $\Gamma_K.$
\foot{Some discussion of $T$-duality in the 
$K$-theoretic context can be found in \hori.}
The action of  $SL(2,\Z)_{\tau}$  on $\Gamma_K$ is via 
standard pullback under topologically nontrivial 
diffeomorphisms. The action of $SL(2,\Z)_\rho$ is more 
novel.

We will explain the action of the two generators 
$S,T$ of $SL(2,\Z)_\rho$ separately. 
To begin, the action of  $T$ on $\Gamma_K$ is a particular case of
the global gauge transformation  \glgauge,\aut\
with $f_2=e_0$ and for this reason $T \in Sp(2N,\Z).$
The action of  $T$ preserves the standard polarization since
it maps $\Gamma_2^{std} \to \Gamma_2^{std}$:
\eqn\since{G_{2p}\left(y\otimes  L(-e_0)\right)=0,\quad
 \forall y \in \Gamma_2^{std} \quad p=0,1,2} 
The action of the generator $S$ on $\Gamma_K$ is more interesting. 
By the Kunneth theorem we can decompose
\eqn\addition{K^0(X\times T^2)=K^0(X)\otimes K^0(T^2)\oplus
K^1(X)\otimes K^1(T^2)}
Both $K^0(T^2)=\Z \oplus \Z $ and $K^1(T^2)=\Z \oplus \Z$ 
have natural symplectic 
bases on which $S$ acts as the standard symplectic operator $i \sigma_2$. 
For $K^0(T^2)$ we choose basis $\one$ and $L(e_0)-\one$, 
and for $K^1(T^2)$ we denote the basis as $\zeta^m,m=8,9$. 
We now have a   Lagrangian decomposition of
$\Gamma_K=\Gamma_1\oplus \Gamma_2:$
\eqn\additioni{\Gamma_1=K^0(X)\otimes \one \oplus K^1(X)\otimes \zeta^8,\quad
\Gamma_2=K^0(X)\otimes (L(e_0)-\one) \oplus K^1(X)\otimes \zeta^9} 
on which the $T$-duality generator $S$ acts simply. However, 
the decomposition   \additioni\
is not compatible with the standard polarization, and hence the action
of $S$ in the standard polarization appears complicated. 
We now give an explicit description of the action of $S$ in 
the standard polarization. 

Let us write
a generic element $y\in \Gamma_2^{std}$ as 
\eqn\elementy{
y=x({\tilde a})\otimes \Bigl (  L(e_0)-\one \Bigr )+ z_1+z_2+
z_3\otimes \Bigl (  L(e_0)-\one \Bigr ), \quad {\tilde a} \in 
H^4(X,\Z)}
In \elementy\ $z_1,z_2,z_3$ are such that 
\eqn\condz{{G \over 2\pi}(z_1)=j_m d\s^m, \quad {G \over 2\pi}(z_2)=k,\quad
{G \over 2\pi}(z_3)=k'}
where
$ j_m \in H^5(X,{\bf R})\oplus H^7(X,{\bf R}),\quad
 k,k' \in H^6(X,{\bf R})\oplus H^8(X,{\bf R})$
According to the transformation rules of RR fields \Myers\ 
$S$ acts on $y$ as
\eqn\actsony{S: y \to y', \quad y'=x({\tilde a})+z_1+z_3-
z_2\otimes \Bigl (  L(e_0)-\one \Bigr )}
From \actsony\ we  find that the image $\Gamma_2':=S(\Gamma_2^{std})$ 
 differs from $\Gamma_2^{std}$. 
\foot{In following \Myers\ we have actually combined the transformation 
$S$ with the transformation $\tilde S$ from $SL(2,\Z)_\tau$. This is 
a more convenient basis for checking the invariance of the theory.}

Since we have an  embedding $\CD_T \subset Sp(2N,\Z),$ we can deduce
 the existence of well-defined transformation laws under $\CD_T$
of the function $\Theta(\cF,\rho),$
related by \introiii\ to the K-theory
theta function $\Theta_K.$ 
This follows from  
the fact that $\Theta_K$ is an holomorphic section of the 
the line bundle ${\cal L}$ over the K-theory torus with 
$c_1({\cal L})=\omega$. Since ${\cal L}$ is not affected by
 symplectic transformations, and has a one-dimensional 
space of holomorphic sections, it follows that 
under T-duality transformatons $\Theta_K$ 
can at most  be  multiplied by a constant. Nevertheless, this leaves
open the possibility of a T-duality anomaly, as indeed takes place.

To conclude this section we show how the multiplier system 
of \introv\introvi\ is related to
the standard $8^{th}$ roots of unity appearing in theta function
transformation laws.
Let us recall the general transformation rule under $Sp(2N,\Z)$ for 
the theta function $ \theta[m ]\left( \tau \right)$
of a principally polarized lattice $\Lambda=\Lambda_1+\Lambda_2$  
of  rank $2N.$ 
Here  $m=\pmatrix{m' \cr m''}\in R^{2N}$ are the characterstics and the period
matrix 
 $\tau\in M_N(\bf C),\quad \tau^T=\tau$ is a quadratic form
on $\Lambda_1.$

It was found in \Ibusa\ that under symplectic transformations
\eqn\ibusa{
\s\cdot \tau ={A\tau + B \over  C\tau + D}, \quad \s\in Sp(2N,\Z)}
the general  
$\theta[m](\tau)$  transforms   as 
\eqn\modtmn{
\vartheta[\sigma \cdot m ]\left ( \sigma \cdot \tau \right )=
 {\kappa}(\s)e^{2\pi i \phi(m ,\s)} det(C \tau+D)^{1/2}
 \vartheta[m ]( \tau)}
where
$$\sigma \cdot m =m\s^{-1}+\half \pmatrix{\left (C^TD \right )_d\cr
\left (A^TB \right )_d\cr}$$
$$ \phi(m ,\s)=-\half \left (m^{'T} DB^T m'-2m^{'T} BC^T m''+
m^{''T}CA^Tm''\right) +$$
$$+\half \left (m^{'T} D-m^{''T}C\right )
\left(A^TB\right)_d$$
where $(A)_d$ denotes a vector constructed out of diagonal
elements of matrix $A.$

The factor ${\kappa}(\s)$ in \modtmn\ 
has  quite nontrivial properties
\Ibusa. 
In particular ${\kappa}^2(\s)$ is a character of
 $\Gamma(1,2)\subset Sp(2N,\Z),$ where
\eqn\ibusaii{
\s \in \Gamma(1,2)\quad  iff \quad  \left(A^TB\right)_d \in 2\Z,\quad
\left(C^TD\right)_d \in 2\Z} 

One can easily check that $SL(2,Z)_{\rho} \subset  \Gamma(1,2)$ by 
writing out explicit representations $\s(S)$ and $\s(T)$ 
 in $Sp(2N,\Z)$ . We give
 $\s(S)$ and $\s(T)$  in Appendix(A). 
 
Using the explicit expressions for $\s(S)$ and $\s(T)$  
as well as the definition of  $\tau_K$ \tauk,\taukim\
we find that in \modtmn\
\eqn\Ibusaiii{
 det(C(S) \tau_K+D(S))^{1/2}=e^{i{\pi\over 4} b_4} 
(-i \rho)^{\half b_4^+}(i \bar \rho)^{\half b_4^-},\quad
\phi(m,\s(S))=0}
\eqn\phasefact{ det(C(T) \tau_K+D(T))^{1/2}=1,
\quad \phi(m,\s(T))=0}
Now comparing  \modtmn\
and the explicit formulae (5.31) for the 
transformation laws of $\Theta(\cF,\rho)$
derived in the next section we find the relation
between  $\kappa(\sigma)$ 
and the multiplier system $\mu(S),\mu(T)$
\eqn\relation{\kappa(S) e^{i{\pi \over 4} b_4}=\mu(S),\quad
\kappa(T)=\mu(T)}

\newsec{$\Theta(\cF,\rho)$  as a modular form} 
In this section we derive an explicit expression for
$\Theta(\cF,\rho)$ using its relation \introiii\ to the  K-theory
theta function $\Theta_K$ and we check that $\Theta(\cF,\rho)$ 
transforms under the T-duality group $\CD_T$ as a modular form.
  
\subsec{Zero NSNS fields}
We first 
assume that all NSNS background fields are zero.
In this case 
$\Theta(\cF,\rho),$ defined in \introiii\
 is given by an expression of the form
\eqn\thetain{\Theta(\cF,\rho)=
\sum_{a \in H^4(X,\Z)}e^{i2\pi \Phi(a,\cF)}
e^{-\pi \int_{X} Im(\rho) g_{(4)} \wdg *g_{(4)} }}
where the imaginary part of the 8D effective action $2\pi \Phi(a,\cF)$
is derived as follows. We  substitute 
\eqn\subst{
{\hat a}= a+e_0e'+h_md\s^m,\quad  {\hat e}=
e+n_1e_0+\c_m d\s^m}
 into the definition \defPhi\ of  $e^{i2\pi \Phi(n_0,{\hat e},{\hat a})}.$

We need to evaluate $f(a+e_0e'+h_md\s^m).$
We   use the bilinear identity from \DMW\
\eqn\bilidii{f(u+v)=f(u)+f(v)+\int_{X_{10}}u Sq^2v,\quad \forall u,v 
\in H^4(X_{10};\Z)}
to find 
\eqn\fii{
f(a+e_0e'+h_md\s^m)=f(a+e_0e')+f(h_md\s^m).
}
Let us consider $f(h_md\s^m)$ first. Again using the 
bilinear identity we obtain: 
\eqn\fnew{
f(h_md\s^m)= f(h_8d\s^8)+f(h_9d\s^9)+
\int_{X}h_8Sq^2(h_9)} 
From  \bilidii\ it follows that $f(h d\s^m),m=8,9$
are linear functions of $h\in H^3(X,\Z)$. Moreover, from the 
diffeomorphism invariance of the mod two index 
we see that $f(h d\s^8 ) = 
f(h d\s^8 +  \ell h d\s^9 )$, for any integer $\ell$
and, using the bilinear identity once more we find that 
 $f(h d\s^m)=r(h), m=8,9$
where 
\eqn\nontriv{r(h)=\int_X h Sq^2 h, \quad h\in H^3(X,\Z)}
is a spin-cobordism invariant $\Z_2$-valued function. 
In fact, 
$r(h)$  is a nontrivial invariant 
 since for $X=SU(3)$ and $h= x_3$ 
the generator of $H^3(SU(3),\Z)$ we have $r(h)=1$. 
In conclusion: 
\eqn\fnewp{
f(h_md\s^m)= 
\int_{X} \biggl[ h_8 Sq^2 h_8 + h_9 Sq^2 h_9 + h_8Sq^2(h_9)\biggr] } 

Now we  consider $f(a+e_0e')$: 
\eqn\mdotwo{
\eqalign{
f(e_0 e' + a) & = f(a) + f(e_0 e') + \int_{X_{10}} e_0 e' Sq^2 a \cr
& = \int_{X} (a)^2  - \half (e')^2\lambda + (e')^2 a=
 \int_{X} a\lambda    + (e')^2 (a- \half \lambda) \cr}
}
This uses the bilinear identity \bilidii, 
the reduction of the mod two index along $T^2$, and the formula eq.(8.40) for 
$f(u\cup v)$ from \DMW. 

We can now evaluate $\hat a_0$ defined in \sati. The kernel of 
$Sq^3$ is given by those elements $a + e_0 e' + h_m d\sigma^m$ 
such that $h_8 \cup h_8 =h_9 \cup h_9 =0$. If we add the condition 
that the element is a torsion class then $f(a+e_0e')=0$ and
we need only evaluate \fnewp.
Now, since 
$Sq^3(h_m) = h_m \cup h_m =0 $ it follows that $Sq^2(h_m)$ 
has an integral lift. Using again the condition that $h_m$ is 
torsion we find that the right hand side of \fnewp\ is zero. 
It follows that $\hat a_0 =0$.

We can now evaluate the phase. 
Using \redef\ 
we reexpress \fnew\ as
\eqn\fnewi{f(h_md\s^m)=
\int_{X}\Bigl (
f_8Sq^2(f_9)+f_8Sq^2(f_8)+f_9Sq^2(f_9)
+e^2(\c_9f_8-\c_8f_9)+e^3\c_8\c_9 \Bigr)} 
Taking into account \fnewi\ and \mdotwo\
 we find the total phase  $\Phi(a,\cF)$ in \thetain\ is given by: 
 \eqn\totphi{\Phi(a,\cF)=\Delta \Phi+\int_X  (a+\a)\b,}
where the characteristics are defined as:
\eqn\defchr{
\eqalign{
\a & = \half \left(e \right)^2 + \half\left ( 1-n_0/12\right)
\lambda+\half\left(e''+
e\right)\epsilon^{mn}\c_m\c_n\cr
\b & = \half \left(e'' \right)^2 + 
\half \left(1-n_1/12 \right)\lambda+
\half \left(e''- e \right)\epsilon^{mn}\c_m\c_n\cr}
}
and we recall that $e''=n_1e+e'-\half \epsilon^{mn}\c_m\c_n.$
Note that for convenience 
we have made a shift of the summation variable in \thetain\
 $a\to a+\l+\half \left (e + e'' \right ) \epsilon^{mn}\c_m\c_n. $

The prefactor $\Delta \Phi$  is given by
\eqn\delphi{
exp[2 \pi i \Delta \Phi]  =
exp\Biggl[\pi i \int_X \Bigl(f_8Sq^2(f_9)+
f_8Sq^2(f_8)+f_9Sq^2(f_9) \Bigr )\Biggr ]}
$$exp\Biggl[ 2 \pi i \int_X \Bigl (
-{1\over 4} \left(e'' e \right)^2  -{1\over 24}  e'' e \lambda + 
{1\over 6} e^3 e''  
 - {1\over 4}   e^2\lambda +
{1\over 48} n_0 \lambda \left( e''\right)^2 
+{1\over 4} (1+n_0 /12)\lambda^2 +$$
$$ +\half(n_0  - n_1 )
 \hat A_8  - \half n_0 n_1   \left[ \hat A_8 + 
\left({\lambda\over 24}\right)^2\right] +{\l\over 24}\epsilon^{mn}\c_mf_n+$$
$$+{1 \over 48} \Bigl [n_0(e''-e)\l 
-12 e^2e'' -
4e\l-4e^3 \Bigr ]\epsilon^{mn}\c_m\c_n \Bigr )\Biggr ]$$

In deriving $\Delta \Phi$ we have used
$$
\bigl(\sqrt{\hat A}\bigr)_8 = 
\half \left[ \hat A_8 - \left({\lambda\over 24}\right)^2\right] 
$$
Also, in bringing $\Delta \Phi$ to the form \delphi\
we have used the congruences 
\eqn\indxii{\int_X
{1\over 6} \left[\left(e''\right)^3 e + e''e^3\right] +
 {1\over 4} \left(e''\right)^2 e^2 - 
{1\over 12} \lambda e'' e \in \Z
}
\eqn\indxiii{\int_X \left(e'' e\right)^2\in 2\Z,\quad
\int_X e'' e \lambda \in 2 \Z .
}
which follow 
from the index theorem on $X$ :
\eqn\indxi{
\int_X 
{1\over 24} e^4  - {1\over 24} \lambda e^2 \in  \Z,\qquad  \forall e \in H^2(X,\Z).
}

\subsec{Including flat NSNS potentials } 

Let us now take  into account globally defined  NSNS fields:
$${\hat B}_{2}=\half B_0\epsilon_{mn}\omega^m \omega^n+
 B_{(1)m}\omega^m+B_{(2)}+\half \cA_{(1)}^m B_{(1)m}, \quad \cA^m_{(1)}$$ 
and recall that $\cA^m_{(1)}$ and $B_{(1)m}$ are combined
into the ${\bf (2,2)}$ of $\CD_T$ as in \snsiii. 
 
We define a gauge invariant fieldstrength $\tG=e^{{\hat B}_2} G$ as in \defgtilde\
where $G$ are given in \newG\
and we
expand $\tG\left(x+ \half \theta_K\right)$ as
\eqn\arrfield{
\eqalign{
{{\tilde G}_0 \over 2\pi}\left(x+ \half \theta_K\right) & = g_{(0)}^2 \cr
{{\tilde G}_2 \over 2\pi}\left(x+ \half \theta_K\right)& = 
\left (g_{(0)}^1+  g_{(0)}^2B_0 \right) \half \epsilon_{mn}
\omega^m\omega^n +g_{(1)m}\omega^m +
g_{(2)}^2 \cr
{{\tilde G}_4 \over 2\pi}\left(x+ \half \theta_K\right) & =
g_{(4)} +g_{(3)m}\omega^m+
\left ( B_0g_{(2)}^2+g_{(2)}^1\right)\half \epsilon_{mn}
\omega^m\omega^n
\cr}}
The first effect of including flat NSNS fields is to
modify the fields which enter $S_B(\cF).$
These fields  
$g_{(0)}^{\a},g_{(1)m},g_{(2)}^{\a},g_{(3)m}$  are now  linear
combinations 
 of the integral classes $\c_m,f_m,e^{\a},n^{\a}$ 
defined in \newxii,\redef\ with coefficients
constructed from ${\bf A}_{(1)}^{m \a}$ and $B_{(2)}$:

\eqn\harmfl{g_{(0)}^{\a}=\pmatrix{n_1 \cr n_0\cr},\quad
g_{(1)m}= \c_{m}+\xi_{(1)m},\quad 
g_{(2)}^{\a}=e^{\a}+{\bf A}_{(1)}^{m \a}\left (\c_{m}+
\half \xi_{(1)m} \right)
+B_{(2)}g_{(0)}^{\a}}
\eqn\harmflii{g_{(3)m}=f_{m}+B_{(2)}g_{(1)m}
+\lambda_{(3)m}+\half k_{(3)m}+{1\over 6}\epsilon_{mn}
{\cal E}_{\a \b} {\bf A}_{(1)}^{p\a}\xi_{(1)p} {\bf A}_{(1)}^{n \b}}
where we denote
\eqn\harmfliiii{
\xi_{(1)m}=\epsilon_{mn}{\cal E}_{\a \b} g_{(0)}^{\a}{\bf A}_{(1)}^{n \b},\quad
\lambda_{(3)m}=\epsilon_{mn}{\cal E}_{\a \b} e^{\a}{\bf A}_{(1)}^{n \b},\quad
k_{(3)m}=\epsilon_{mn}{\cal E}_{\a \b}
{\bf A}_{(1)}^{p \a}\c_p{\bf A}_{(1)}^{n \b}}

The other effect of
 including flat NSNS fields
is to  shift the characteristics and the prefactor of $ \Theta(\CF,\rho).$ 
Now $ \Theta(\CF,\rho)$  has the form:
\eqn\newtht{
\Theta(\CF,\rho)=
e^{2\pi i  \Delta {\widetilde \Phi}}\sum_{a\in H^4(X,{\bf Z})}
exp\Biggl [\int_{X}\Bigl ( -\pi Im(\rho) g_{(4)} \wdg *g_{(4)} +
i\pi  Re(\rho) g_{(4)} \wdg g_{(4)}
+2\pi i g_{(4)} {\tilde \b} \Bigr ) \Biggr ]}
where
$[g_{(4)}]=a+{\tilde \a}, \quad a \in H^4(X,\Z)$, and the shifted characteristics
${\tilde \a},{\tilde \b}$ are 
\eqn\tila{ {\tilde \a}=\a+\vph^2,\quad {\tilde \b}= \b+\vph^1}
where $\a,\b$ are defined in  terms of integral classes
 $n_0,n_1,\c_m,e^{\a}$
in \defchr, while $\vph^{\a}$ transform in the 
${\bf  (1,2)}$ of  $\CD_T.$
Explicitly,
\eqn\tilai{
\vph^{\a}={\bf A}_{(1)}^{m \a}\left (f_{m} +
\half \lambda_{(3)m}+{1\over 6}k_{(3)m} \right )  + 
B_{(2)}\left [e^{\a}+{\bf A}_{(1)}^{m \a}
\left (\c_m +\half \xi_{(1)m} \right )\right ]+}
$$+\half B_{(2)}B_{(2)}g_{(0)}^{\a} 
-\zeta_{(4)} g_{(0)}^{\a}
$$
where $\xi_{(1)m},\lambda_{(3)m},k_{(3)m}$ are given in \harmfliiii\ 
and we also denote 
\eqn\tilaiii{
\zeta_{(4)}={1 \over 64}{\cal E}_{\b_1 \b_2}
{\cal E}_{\b_3 \b_4}
{\bf A}_{(1)}^{n_1 \b_1}\epsilon_{n_1n_2}{\bf A}_{(1)}^{n_2 \b_3}
{\bf A}_{(1)}^{m_1 \b_2}\epsilon_{m_1m_2}{\bf A}_{(1)}^{m_2 \b_4}}
The shifted prefactor 
$ \Delta  {\widetilde \Phi}$ in \newtht\ is given by
\eqn\deltaphi{
\Delta  {\widetilde \Phi}= \Delta  \Phi -\int_X\Biggl[
{\b}\wdg \vph^2 +\half \vph^1\wdg \vph^2\Biggr]+
\left (\Delta  \Phi\right )_{inv}
}
where $\Delta  \Phi$ is defined in terms of 
integral classes
$n_0,n_1\c_{m},e^{\a},f_{m}$
 in \delphi\ and  $\left (\Delta  \Phi\right )_{inv}$
is the part of the phase which is manifestly invariant
under the T-duality group $\CD_T$. Explicitly, 
\eqn\invphase{
\left (\Delta  \Phi\right )_{inv}=\int_X B_{(2)}^3\Bigl [
{1 \over 12}{\cal E}_{\a \b}g_{(0)}^{\a}e^{\b} -{1 \over 6}
\epsilon^{mn}\c_{m}\c_{n}
-{1 \over 4}
\epsilon^{mn}\xi_{(1)m}\c_{n}
-{1 \over 8}
\epsilon^{mn}\xi_{(1)m}\xi_{(1)n} \Bigr]+}
$$\int_X B_{(2)}^2\Bigl[-{1 \over 4}
\epsilon^{mn}\xi_{(1)m}f_{n}-{1 \over 2}
\epsilon^{mn}\lambda_{(3)m}\c_{n}
-{3 \over 8}
\epsilon^{mn}\lambda_{(3)m}\xi_{(1)n}-{1\over 24}
\epsilon^{mn}k_{(3)m}\xi_{(1)n}\Bigr ]+$$
$$\int_X B_{(2)}\Bigl[-{1 \over 2}
\epsilon^{mn}f_{m}f_{n}-{1 \over 2}
\epsilon^{mn}\lambda_{(3)m}f_{n}
-{1 \over 4}
\epsilon^{mn}\lambda_{(3)m}\lambda_{(3)n}-{1\over 6}
\epsilon^{mn}\lambda_{(3)m}k_{(3)n}+$$
$$+{1 \over 12} \xi_{(1)m} q^m_{(5)}+
{1 \over 2} \zeta_{(4)}{\cal E}_{\a \b}e^{\a}g_{(0)}^{\b}+
 \zeta_{(4)}\epsilon^{mn}\c_{m}\c_{n}\Bigr ]+
\int_X \Bigl [{1 \over 12}
\lambda_{(3)m}q_{(5)}^m
+\zeta_{(4)}
\epsilon^{mn}\c_{m}f_{n}\Bigr ]$$
where 
$q_{(5)}^m={\cal E}_{\a \b}
{\bf A}_{(1)}^{p \a}f_{p}{\bf A}_{(1)}^{m \b}$

\subsec{Derivation of T-duality transformations.} 

Let us study transformations of $\Theta(\cF,\rho)$ defined in \newtht\
under $\CD_T.$
First, we note that $\Theta(\cF,\rho)$  is  invariant
under $SL(2,Z)_{\tau}.$
Next, we consider the action of the generator $S.$
 For any function $h(\cF)$ of fluxes $\cF,$ we denote
$$
S\Bigl[h(\cF)\Bigr ] :=h( S \cdot \cF )$$ 
and 
$$\delta_S[h] := S[h]-h$$
where $S \cdot \cF$ denotes the linear action on fluxes. 
 To check the transformation under $S$ we 
need to do a Poisson resummation on the self-dual lattice
$H^4(X,\Z)$. 
The basic transformation law is: 
\eqn\modtmnii{
\vartheta\biggl[\matrix{\theta\cr \phi\cr}\biggr](0 \vert -1/\tau) =
 (-i \tau)^{1/2} 
e^{2\pi i \theta\phi} 
\vartheta\biggl[\matrix{-\phi \cr \theta \cr}\biggr](0 \vert \tau)
}
and its generalization to self-dual lattices \modtmn.

 After the Poisson resummation and a shift 
of   summation variable
$a \rightarrow a+e^2 + \lambda$ we find that
$\Theta(\cF,\rho)$ 
 transforms  under $ S$ as
\eqn\res{
\Theta(S \cdot \cF, -1/\rho)
=e^{ 2\pi i \left \{\int_X 
 S\bigl[{\tilde \a} \bigr] 
{ S}\bigl[{\tilde \b} \bigr]
 +\delta_S\left[\Delta {\widetilde \Phi} \right]\right \}}  
(-i \rho)^{\half b_4^+}(i \bar \rho)^{\half b_4^-}
\Theta(\cF,\rho) }
Now using the definitions of ${\tilde \a},{\tilde \b} $ \tila,\tilai\
and $\Delta {\widetilde \Phi}$ \deltaphi\
as well as the transformation rules for $\cF,$ 
we find after some tedious algebra
\eqn\resi{
\delta_S\left[\Delta {\widetilde \Phi} \right]=- \int_X 
S\bigl[{\tilde \a} \bigr] { S}\bigl[ {\tilde \b} \bigr] + 
\int_X{\l^2 \over 4}+\Z }

We conclude 
that the generator $S$ acts  as 
\eqn\finresnew{
\Theta(S \cdot \cF, -1/\rho)
= e^{i\pi \int_X \l^2 /2}(-i \rho)^{\half b_4^+}(i \bar \rho)^{\half b_4^-}
\Theta(\cF,\rho)}

To check how $\Theta(\cF,\rho)$ transforms under the generator $T$
we use its  relation \introiii\ to the K-theory theta function $\Theta_K$
as well as the transformation of  $\Theta_K$ under 
global gauge transformation ${\hat B}_2 \to {\hat B}_2+f_2$ \aitii\ 
where the action of the generator $T$ corresponds to 
$f_2=e_0.$
In this way we find from \aitii\ that
\eqn\finresnew{
\Theta(T \cdot \cF, \rho+1)
= e^{i\pi \int_X \l^2 /4}\Theta(\cF,\rho)}

\subsec{Summary of T-duality transformation laws}
Below we  summarize the transformation laws of the function $\Theta(\cF,\rho)$
under the generators of T-duality group $\CD_T.$

 $\Theta(\CF,\rho)$ is invariant under $SL(2,Z)_{\tau}$:
 \eqn\summarv{
\eqalign{
\Theta({\tilde T} \cdot \CF, \rho) & = \Theta(\CF, \rho) \cr
\Theta({\tilde S} \cdot \CF, \rho) 
& = \Theta(\CF,\rho) \cr}
}

 $\Theta(\CF,\rho)$ 
transforms as a modular form with a nontrivial ``multiplier system'' 
under $SL(2,Z)_{\rho}$. That is, using  the standard 
generators $T,S$ of $SL(2,Z)_{\rho}$ we have: 
\eqn\summarvii{
\eqalign{
\Theta(T \cdot \CF, \rho+1) & = \mu(T) \Theta(\CF, \rho) \cr
\Theta(S \cdot \CF, -1/\rho) 
& =\mu(S) (-i \rho)^{\half b_4^+} (i \bar \rho)^{\half b_4^-} \Theta(\CF,\rho) \cr}
}
where $T\cdot \CF, S\cdot \CF$ denotes the linear action of $\CD_T$ on 
the fluxes. 
Here $b_4^+,b_4^-$ is the dimension of the space of self-dual and 
anti-self-dual 
harmonic forms on $X$ and  the multiplier system is 

\eqn\introvi{
\eqalign{
\mu(T) & =\exp\bigl[ {i \pi \over 4} \int_X \lambda^2 \bigr] \cr
\mu(S) & =\exp\bigl[ {i \pi \over 2} \int_X \lambda^2 \bigr] \cr}
}
where $p_1 = p_1(TX)$. These define the ``T-duality anomaly of RR fields.''
\newsec{The bosonic determinants} 
In this section we compute bosonic quantum determinants
around the background specified in section 2. 

Let us factorize bosonic quantum determinants as:
 $Det_B={\cal D}_{RR}{\cal D}_{NS},$
where ${\cal D}_{RR}({\cal D}_{NS})$ denotes the 
contribution from RR (NSNS) fields.

\subsec{ Quantum determinants ${\cal D}_{RR}$ for RR fields}

Quantum determinants ${\cal D}_{RR}$ for RR fields  have the form
\eqn\dform{{\cal D}_{RR}=\prod_{p=1}^4Z_{RR,p}}
where  $Z_{RR,p}$ is  the quantum determinant for $g_{(p)}.$
First, we present the contribution
$Z_{RR,4}$ arising from the  fluctuation $dC_{(3)}$ of $g_{(4)}.$
From \gfouract\ we find the kinetic term for $C_{(3)}$ 
\eqn\Ciii{
S_{3,cl}=\pi Im(\rho) \bigl( dC_{(3)}, dC_{(3)}\bigr )}
where 
$(~,~)$ denotes the standard inner product on the
 space of p-forms on $X$, 
constructed with the background metric $g_{MN}.$

We use  the   standard  procedure  \refs{\Siegel,\Geg}
for path-integration over  p-forms, which
can be summarized as follows.
Starting from the classical action for  the p-form
$ S_{p,cl}=\a \bigl( dC_{(p)}, dC_{(p)}\bigr )$
one constructs the quantum action as\foot{
Factors ${\a}^{1 \over m+1}$  should be understood
as a mnemonic rule to keep track of the dependence
on ${\a}$ which follows from the analysis of various
cancellations between ghosts and gauge-fixing fields}:  
 \eqn\genform{S_{p,qu}= \a \bigl( C_{(p)},\Delta_p C_{(p)}\bigr )+
\sum_{m=1}^p {\a}^{1 \over m+1}\sum_{k=1}^{m+1}\left (u_{(p-m)}^k,\Delta_{p-m}
u_{(p-m)}^k \right )}
where $u_{(p-m)}^k,\quad k=1,\ldots m+1,m=1,\ldots p $ 
 are ghosts  of alternating statistics.
For example, $u_{(p-1)}^k, \quad k=1,2$ are fermions,
$u_{(p-2)}^k, \quad k=1,2,3$ are bosons, etc.
In \genform\
 $ \Delta_p $ 
is  the Laplacian acting on p-forms and constructed with $g_{MN}$\foot{
$\Delta=dd^{\dg}+d^{\dg}d$}.

To compute  $Z_{RR,4}$ we apply \genform\ for
 $p=3, \quad \a=\pi \im(\rho) $ and use the measure
$[DC_p]$  normalized as $\int [DC_p]e^{-(C_p,C_p)}=1$: 
\eqn\answ{
Z_{RR,4}=\left (\a\right)^{-\half(B'_3-B'_2 +B'_1-B'_0)}
\Bigl [{ det' \Delta_3 \over V_3 } \Bigr ]^{-\half}
\Bigl[ {det' \Delta_2 \over V_2} \Bigr ]
\Bigl [ {det' \Delta_1 \over V_1} \Bigr ]^{-3/2}
\Bigl[ {det' \Delta_0 \over V_0} \Bigr ]^2
}
where
 $det'\Delta_p $ 
is the determinant of nonzero modes
of the Laplacian acting on p-forms. $B'_p=B_p-b_p,$ where $B_p$
 denotes the (infinite )
number of eigen-p-forms and $b_p$ and $V_p$  are   
the dimension and   the determinant of the metric  
 of the harmonic torus  $T_{harm}^p={\cal H}^p/ {\cal H}_{\bf Z}^{p}.$ 
The appearance of $V_p$ in \answ\ 
 is due to the appropriate treatment of zeromodes and is explained
in Appendix(E).

The determinants $det'\Delta_p$ together with 
the infinite powers depending on $B_p$, here and below, require 
regularization and renormalization, of course. These can be 
handled using, for example, the techniques of \birreldavies. In particular the 
expression
\eqn\regul{q(Im\rho):=( \im \rho)^{-\half(B_3-B_2 +B_1-B_0)}}
is
a local counterterm of the form   $e^{-\pi \im \rho \int_X(u \lambda^2+vp_2)},$
where the numbers $u,v$ depend on the regularization.
From now on we will assume that  $\pi Im\rho \int_X(u \lambda^2+vp_2)$ 
is included into the
1-loop action:
\eqn\assume{S_{1-loop}=\pi \im \rho \int_X(u \lambda^2+vp_2)+
{i \pi\over 24} \re \rho \int_X \bigl(p_2-\lambda^2\bigr)}
In section 8 we will show that T-duality invariance
determines $u$ and $v$ uniquely.

Next, we consider the contributions to ${\cal D}_{RR}$ from
 $dC_{(2)m},\quad dC_{(1)}^{\a},\quad d{\tilde C}_{(0)m}$
which are the fluctuations for 
$g_{(3)m}, \quad g_{(2)}^{\a}, \quad g_{(1)m}$ respectively.
Let us also make
field redefinition of the quantum fields ${\tilde C}_{(0)m},m=8,9$
to   fields $C_{(0)m},m=8,9$ 
which have well defined transformation properties
under the full U-duality group\foot{For some  discussion
of U-duality see sec.10}
\eqn\connto{
C_{(0)8}=\sqrt{\tau_2}e^{\xi}{\tilde C}_{(0)8}, \quad
C_{(0)9}={1 \over \sqrt{\tau_2}}e^{\xi} {\tilde C}_{(0)9}}
From \rrii\ we find classical action quadratic in the above
   fluctuations:

$$
S_{0,cl}=\pi  {\tilde t}^6 g'^{mn} 
\left ({ C}_{(0)m},d^{\dg}d{ C}_{(0)n}\right ),\quad
S_{1,cl}=\pi t^4{\cal G}_{\a \b}\left ({ C}^{\a}_{(1)},
 d^{\dg}d{ C}^{\b}_{(1)}\right ) $$
$$S_{2,cl}= \pi t^2 g^{mn} \left ({ C}_{(2)m},d^{\dg}d{ C}_{(2)n}\right )
$$
where $\tilde t=te^{-\xi/3}$ is U-duality invariant,
and $g'^{88}={1 \over \tau_2} g^{88},\quad
g'^{99}=\tau_2 g^{99},\quad g'^{89}=g^{89}.$
Now, using \genform\ with 
$a=\pi  {\tilde t}^6g'^{mn},\pi t^4{\cal G}_{\a \b},\pi t^2g^{mn}$ and $p=0,1,2$
correspondingly  we 
find:
\eqn\answi{
Z_{RR,1}=\Bigl (\pi {\tilde t}^6 \Bigr)^{ -B'_0}
\Bigl [{ det' \Delta_0 \over V_0 } \Bigr ]^{-1}
}
\eqn\answii{
Z_{RR,2}=\Bigl (\pi t^4 \Bigr)^{B'_0-B'_1}
\Bigl [{ det' \Delta_1 \over V_1 } \Bigr ]^{-1}
\Bigl[ {det' \Delta_0 \over V_0} \Bigr ]^2
}
\eqn\answiii{
Z_{RR,3}=\Bigl (\pi t^2 \Bigr)^{-B'_2+B'_1-B'_0}
\Bigl [{ det' \Delta_2 \over V_2 } \Bigr ]^{-1}
\Bigl[ {det' \Delta_1 \over V_1} \Bigr ]^2
\Bigl [ {det' \Delta_0 \over V_0} \Bigr ]^{-3}
}
In computing \answi-\answiii\ we also used that 
$det_{m,n}g^{mn}=1,\quad det_{m,n}g'^{mn}=1$ and
$det_{\a ,\b}{\cal G}_{\a \b}=1.$

Collecting together \answ\ and \answi-\answiii\
we find that ${\cal D}_{RR}$ has the form:
\eqn\answiiii{
{\cal D}_{RR}=r_{RR}(t,\rho) 
\Bigl [{ det' \Delta_3 \over V_3 } \Bigr ]^{-\half}
\Bigl [ {det' \Delta_1 \over V_1} \Bigr ]^{-\half}
}
where 
$$r_{RR}(t,\rho)=(e^\xi)^{2B_0'} \Bigl ( Im\rho \Bigr)^{\half(b_3-b_2+b_1-b_0)}
t^{-2B'_2-2B'_1-4B'_0}(\pi)^{-\half(B'_0+B'_1+B'_2+B'_3)}$$
and we recall that $q(Im\rho)$ was included into
$S_{1-loop}.$

We have computed the quantum determinants ${\cal D}_{RR}$ 
treating RR fluctuations as differential forms.
It would be more natural if these determinants
had a K-theoretic formulation.
This might be an interesting application to physics of
differential  K-theory.

\subsec{Quantum determinants for NSNS fields}

Let us first consider  fluctuations $d{\bf a}_{(1)}^{m \a}$ and
$db_{(2)}$ 
of the NSNS field ${\bf F}_{(2)}^{m \a}$ and $H_{(3)}.$
From \snsii\ we find the quadratic action  for fluctuations:
 \eqn\aii{S_{cl}={1 \over 4\pi}e^{-2\xi}\Biggl\{
 t^4 g_{mn}{\cal G}_{\a \b}\left(
{\bf a}_{(1)}^{m \a},d^{\dg} d{\bf a}_{(1)}^{n \b} \right)+
t^2 \left(b_{(2)},d^{\dg}d b_{(2)} \right) \Biggr \}}

Now, again using \genform\ 
 we find
\eqn\qdet{Z_{NS,2}= \Bigl ({ t^4 \over 4\pi}e^{-2\xi} \Bigr)^{2(B'_0-B'_1)}
\Bigl [{ det' \Delta_1 \over V_1 } \Bigr ]^{-2}
\Bigl[ {det' \Delta_0 \over V_0} \Bigr ]^4}
and
\eqn\qdetii{
Z_{NS,3}=\Bigl ({ t^2\over 4\pi}e^{-2\xi} \Bigr)^{\half(B'_1-B'_2-B'_0)}
\Bigl [{ det' \Delta_2 \over V_2 } \Bigr ]^{-\half}
\Bigl[ {det' \Delta_1 \over V_1} \Bigr ]
\Bigl [ {det' \Delta_0 \over V_0} \Bigr ]^{-3/2}
}

Let us now consider fluctuations of scalars: $\d \xi, \d \tau,\d \rho.$
From \snsii\ we write the  action quadratic in these fluctuations:  
\eqn\scal{
S_{scal}= \b \int_{X}
\Biggl \{ 8\p^M\d \xi \p_M \d \xi + 
{1 \over (\tau_2)^2} \p^M\d \tau \p_M { \d {\bar \tau}}+
{1 \over (\rho_2)^2} \p^M \d \rho \p_M { \d {\bar \rho}} \Bigr \}
}
where $\b={1\over 4\pi} e^{-2\xi}t^6.$
Now using  the scalar
measures  defined as
\eqn\also{
\int [D\d\rho][D\d\bar \rho]
e^{-  \int_X{\d \rho\wdg *{\d \bar  \rho} \over (\im \rho)^2}}=1,
\quad
\int [D \d\tau][D\d\bar \tau]
e^{- \int_X {\d \tau \wdg *{\d \bar \tau} \over (\im \tau)^2 }}=1}
\eqn\alsoi{
\int [D\d\xi]e^{- 8 \int_X \d \xi\wdg*\d \xi}=1}
we
find the quantum determinants for the NSNS scalars $Z_{NS,0}$:
\eqn\qtau{
Z_{NS,0} = \beta^{-{5\over 2}B'_0}
\Bigl [ {det' \Delta_0 \over V_0} \Bigr ]^{-{5\over 2}}}

Finally, we consider the fluctuation $h_{MN}$ of the metric $t^2g_{MN}.$
Recall that we work in the limit $e^{-\xi}\to \infty$
so that in computing the quantum determinant for the metric
we drop couplings to RR background fluxes.

From \snsii\ we find the quadratic  action:
\eqn\metrfl{S_{metr}=\b  
\int_{X}\Biggl \{ 
(D_N h_{MP})P^{MPQS}\left (D^N h_{QS}\right )
+h^{MP}\cR_{MNPQ}h^{NQ}}
$$-\left ( D^M h_{MN}- \half D_Nh \right )^2 \Biggr \}
$$
where $h=g^{MN}h_{MN}$ and
$$ P^{MPQS}=\half g^{MQ}g^{PS}-{1\over 4}g^{MP}g^{QS}$$

In \metrfl\ $\cR_{MNPQ}$  is the Riemann  tensor
of the 
Ricci-flat\foot{If the background metric is not Ricci-flat
there are terms involving the Ricci-tensor
in \metrfl\ as well as in (6.22) below.}
background  metric $g_{MN}.$  The covariant derivative $D_M$
is performed with the background metric, and indices are raised and lowered
with this metric.

 Following standard procedure \refs{\Deser,\Hooft} we 
first insert the gauge fixing  condition into the path-integral
$\d \left (\kappa_N- (D^M h_{MN}-\half D_Nh) \right )$
Then, we insert the unit
\eqn\metrunit{1=\sqrt{ det\bigl (\b  \one_1\bigr)}\int D\kappa_{(1)} 
e^{- \b \left( \kappa_{(1)},\kappa_{(1)}\right ) }  }
and integrate over $\kappa_{(1)}$ in the path-integral.
 This procedure brings  
the kinetic term for the fluctuation $h_{MN}$ to the form
\eqn\newform{
\b
\int_{X}h_{MP}P^{MPNR}{\cal K}_{NR}^{QS} h_{QS},\quad
{\cal K}_{NR}^{QS}=-\d_N^Q \d_R^S D_LD^L +2\cR_{N~~R}^{~~Q~~S} }
Gauge fixing also introduces fermionic ghosts $k_{(1)},l_{(1)}$ 
with the action
\eqn\metrgh{S_{gh}=\b^{1/2} \left (l_{(1)},\Delta_1  k_{(1)}\right )}
Using the measure 
$\int [Dh_{MN}]e^{-\int_X h_{MN}P^{MNPQ}h_{PQ}}=1$
we obtain
the  result for the quantum determinant $Z_{metr}$ of the 
metric:
\eqn\finform{
Z_{metr}=\left( \beta \right)^{-\half 
\left( N'_{{\cal K}}-B_1'\right )}
\Bigl[det'{\cal K}\Bigr]^{- \half}{det'\Delta_1 \over V_1}}
where $det'{\cal K}$ is a regularized determinant of nonzero modes
of the operator ${\cal K}$  defined in
\newform\ and  $N'_{{\cal K}}=N_{{\cal K}}-n_{{\cal K}},$
where $N_{{\cal K}}$  
stands for the dimension (infinite ) of the space of the second rank
symmetric tensors and $ n_{{\cal K}}$ is the number of
zeromodes of the operator  ${\cal K}.$
 We will explain how
we regularize $det'{\cal K}$ shortly.

Combining all NSNS determinants together we find: 

\eqn\nsnsdet{{\cal D}_{NS}=r_{NS}(t,\xi)
\Bigl[det'{\cal K}\Bigr]^{- \half}
\Bigl[ {det'\Delta_2 \over V_2}\Bigr]^{-\half} }
where
\eqn\nsnsetii{r_{NS}(t,\xi)=\left(4\pi\right)^{{1\over 2}N'_{{\cal K}}+
B'_0
+B'_1+{1\over 2}B'_2}
 \left (e^{\xi}\right )^{N'_{{\cal K}}+B'_2
+2B'_1+2B'_0} t^{-3N'_{{\cal K}}-B'_2
-4B'_1-8B'_0}}

Finally, from \answiiii\ and \nsnsdet\ we find 
the full expression for bosonic determinants
\eqn\findet{Det_B=Q( t,g_{MN})
\Bigl ( Im\rho \Bigr)^{\half(b_3-b_2+b_1-b_0)}}
where 
$Q$ is a function only of the T-duality invariant
variables $g_{MN},$ $t$ and $\xi.$
Explicitly,
\eqn\normq{Q( t,g_{MN})=r_{tot}
\Bigl[det'{\cal K}\Bigr]^{- \half}
\Bigl [{ det' \Delta_3 \over V_3 } \Bigr ]^{-\half}
\Bigl[ {det'\Delta_2 \over V_2}\Bigr]^{-\half} 
\Bigl [ {det' \Delta_1 \over V_1} \Bigr ]^{-\half}}
where we regularized $det'{\cal K}$ in a way that eliminates
dependence on ifinite numbers $B_p$ and $N_{\cal K}$
so that
\eqn\normqii{r_{tot}=
( \tilde t)^{3(n_{{\cal K}}+b_2+2b_1+4b_0)}}
where we recall $\tilde t=te^{-\xi/3}.$ 

Now, let us check the transformation laws of $Det_B$ under $\CD_T.$
From \findet\ it is obvious that
 $Det_B$ is manifestly invariant under all generators of $\CD_T$
except generator $S.$

Using,
\eqn\Im{
Im(-1/\rho)={{Im(\rho)}\over{\rho {\bar \rho}} }}
we find that under $S$,  $Det_B$ transforms as
 \eqn\quant{
{Det_B}(-1/\rho)=s_B {Det_B}(\rho), \quad
s_B=\bigl(\rho{\bar \rho}
\bigr)^{\half (b_0-b_1+b_2-b_3)}}

\newsec{Inclusion of the fermion determinants}

In this section we include the effects of the fermionic 
path integral. We recall the fermion content in the 10-dimensional 
and 8-dimensional supergravity theories and derive their 
actions. In the presence of nontrivial fluxes these fermionic 
path integrals are nonvanishing, even for the supersymmetric spin 
structure on $T^2$.

\subsec{Fermions in 8D theory and their T-duality transformations.}

Let us begin by listing the fermionic content in the 8-dimensional 
supergravity theory (this content will be derived from the 10-dimensional 
theory below.) 

The fermions   in the 8D theory  include two gravitinos
$\psi^{A},\quad \eta^{A},\quad A=0,\ldots,7$ and spinors 
$ \Sigma,\quad
\L_R,\quad l,\quad \mu, \quad {\tilde l},\quad {\tilde \mu}.$
\foot{These fields are MW in Lorentzian signature. 
We supress   16 component spinor indices
below} The relation of these fields to the 10D fields is explained
in (7.13),(7.14) below.
There are also bosonic spinor ghosts $b_1,c_1,\Upsilon_2 $ and $b_2,c_2,\Upsilon_1 $
 which
accompany  $\psi^{A}$ and $ \eta^{A}$ respectively.

The fermions and ghosts transform under T-duality generators as follows. 
The generators $T,{\tilde  T},{\tilde S}$ act trivially on fermions and ghosts
while the under the  generator
 $ S$ they  transform as 
\eqn\trferm{\psi^{A}\rightarrow e^{i\a \bGamma}\psi^{A},\quad
\eta^{A}\rightarrow \eta^{A},
 \quad \L_R\rightarrow e^{-i\a \bGamma}\L_R, \quad
\Sigma \rightarrow \Sigma }
\eqn\trfermii{l \rightarrow e^{2 i\a \bGamma} l, \quad
{\tilde l} \rightarrow e^{-2 i\a \bGamma} {\tilde l}, \quad
 \mu \rightarrow e^{i\a \bGamma}\mu,\quad
{\tilde \mu} \rightarrow e^{i\a \bGamma}{\tilde \mu} } 
and ghosts transform as
\eqn\upstr{ \Upsilon_1 \to \Upsilon_1,\quad
 \Upsilon_2 \to e^{-i \a \bGamma}\Upsilon_2}
\eqn\upstr{ \{c_1,b_1\} \to e^{i \a \bGamma} \{c_1,b_1\}\quad
\{c_2,b_2\} \to  \{c_2,b_2\} } 
where  $\a $ is defined by
\eqn\defnu{\a=\nu+\half\pi,\quad
i{\bar \rho}=e^{i\nu}\vert \rho \vert}
and  $\bGamma$ is the $8D$ chirality matrix.

The above transformation rules for space-time fermions follow from
the transformation  rules for 
the appropriate vertex operators on the world-sheet (as discussed for example
in \GPR). 
The only
generator of $\CD_T$ acting non-trivially on fermions is $S$.    
The components ${\cal V}_{NS}^a,a=8,9$ of the
right-moving NS vertex 
are rotated by $2 \a,$   while the components
${\cal V}_{NS}^{A}$ are invariant.
This follows since $ S$ does not act on the left-moving components of
vertex operators. In this way we find
 the
transformation rules for  $\eta^A,b_2,c_2,\Sigma,\Upsilon_1, l,{\tilde l},$
which originate from $R \otimes NS$ sector.
To account for the transformation rules for 
 $\psi^A,b_1,c_1,\Lambda, \Upsilon_2,\mu,{\tilde \mu}$
we recall that these fields originate from $NS \otimes R$ sector
and that 
the right-moving R vertex ${\cal V}_{R}$ transforms
under  $ S$ 
as 
\eqn\vertex{
S: {\cal V}_{R}\rightarrow e^{i\a  \bGamma }{\cal V}_R. }

\subsec{10D fermion action}

We start from the part of the $10D$ IIA supergravity
action  quadratic in fermions\Romans. We work in the string frame.
\foot{ We explain the relation between our conventions
and those of \Romans\ in Appendix(B).} 
$$S_{ferm}^{(10)}=\int \sqrt{-g_{10}} e^{-2\phi} 
\Biggl [ \half
  {\bar {\hat \psi}}_{\hA}{\hat \Gamma}^{\hA \hN \hB}
D_{\hN} {\hat \psi}_{\hB}+
\half {\bar {\hat \Lambda}}{\hat \Gamma}^{ \hN }
D_{\hN} {\hat \Lambda}-{1 \over \sqrt{2}}(\p_{\hN}\phi){\bar {\hat \Lambda}}
{\hat \Gamma}^{\hA}{\hat \Gamma}^{\hN}{\hat \psi}_{\hA}\Biggr ] $$
$$+{1\over 16 } \int \sqrt{-g_{10}}e^{-\phi} \tG_{\hA \hC}
\Biggl [{\bar {\hat \psi}}^{\hE}{\hat \Gamma}_{[\hE}{\hat \Gamma}^{\hA  \hC}
{\hat \Gamma}_{\hF ]}{\hat \Gamma}^{11}{\hat \psi}^{\hF}+{3 \over \sqrt{2}}
{\bar {\hat \Lambda}}{\hat \Gamma}^{\hE}{\hat \Gamma}^{\hA  \hC}
{\hat \Gamma}^{11}{\hat \psi}_{\hE}+{5 \over 4}{\bar {\hat \Lambda}}
{\hat \Gamma}^{\hA  \hC}
{\hat \Gamma}^{11}{\hat \Lambda} \Biggr ] $$
\eqn\tendferm{+ \int \sqrt{-g_{10}} e^{-\phi} G_0
\Biggl [
{1 \over 8}{\bar {\hat \psi}}_{\hA}{\hat \Gamma}^{\hA  \hB}
{\hat \psi}_{\hB}+
{5 \over 8\sqrt{2}} {\bar {\hat \Lambda}}{\hat \Gamma}^{\hA}
{\hat \psi}_{\hA}-{21 \over 32}{\bar {\hat \Lambda}}{\hat \Lambda}
 \Biggr ]+ }
$$+{ 1\over 192  }\int \sqrt{-g_{10}} 
 e^{-\phi}\tG_{\hA \hB \hC \hD }\Biggl [
{\bar {\hat \psi}}^{\hE}{\hat \Gamma}_{[\hE}{\hat \Gamma}^{\hA \hB \hC \hD}
{\hat \Gamma}_{\hF ]}{\hat \psi}^{\hF}+ {1 \over \sqrt{2}}  
{\bar {\hat \Lambda}}{\hat \Gamma}^{\hE}{\hat \Gamma}^{\hA \hB \hC \hD}
{\hat \psi}_{\hE}+{3 \over 4}{\bar {\hat \Lambda}}
{\hat \Gamma}^{\hA  \hB \hC \hD }{\hat \Lambda} \Biggr ] $$
$$+{1 \over 48}\int \sqrt{-g_{10}} 
 e^{-2\phi}H_{\hA \hB \hC  }\Biggl [
{\bar {\hat \psi}}^{\hE}{\hat \Gamma}_{[\hE}{\hat \Gamma}^{\hA \hB \hC }
{\hat \Gamma}_{\hF ]}{\hat \Gamma}^{11}{\hat \psi}^{\hF}+  \sqrt{2}  
{\bar {\hat \Lambda}}{\hat \Gamma}^{\hE}{\hat \Gamma}^{\hA \hB \hC }
{\hat \Gamma}^{11}{\hat \psi}_{\hE}\Biggr ]
$$
where ${\hat \Lambda}$ and ${\hat \psi}^{\hA}$ are the 
Majorana  dilatino
and gravitino and covariant derivatives act on them as
$$D_{\hN}{\hat \psi}^{\hA}=\p_{\hN}{\hat \psi}^{\hA}+
\omega^{~~\hA}_{\hN~~\hB}{\hat \psi}^{\hB}+{1 \over 4}
\omega_{\hN  \hB \hC}\Gamma^{\hB \hC}{\hat \psi}^{\hA}$$
$$D_{\hN}{\hat \Lambda}=\p_{\hN}{\hat \Lambda}
+{1 \over 4}
\omega_{\hN  \hB \hC}\Gamma^{\hB \hC}{\hat \Lambda}$$

There are also  terms quartic in fermions in the action.
It turns out that it is important to take them into account to check
the T-duality invariance of partition sum. 
We recall the 4-fermionic terms  in Appendix(C). 

\subsec{Reduction on $T^2$.}

To carry out the reduction of the fermionic action to 8D 
we choose the gauge for the 10D veilbein as
\eqn\veil{
{\hat E}_{\hM}^{\hA}=\pmatrix{
tE_M^A & \cA_M^m V^{\half}e_m^a  \cr
0 &V^{\half} e_m^a\cr},  
}
(recall $a=8,9$ and $A=0,...,7$) 
and use the following basis  
of 10D $32\times 32 $ matrices ${\hat \Gamma}^{\hA},$ 
\eqn\gam{{\hat \Gamma}^{A}=\sigma_2\otimes\Gamma^{A}
\quad A=0,\ldots 7,\quad
{\hat \Gamma}^{8}=
\sigma_1\otimes \one_{16},\quad
{\hat \Gamma}^{9}=
\sigma_2\otimes \bGamma, \quad \bGamma=\Gamma^0\ldots \Gamma^7}
Here $\Gamma^{A}$ are symmetric 8D Dirac 
matrices, which in Euclidean signature
can be all chosen to be real, and $\sigma_{1,2,3}$ are Pauli matrices.
In this basis the 10D chirality ${\hat \Gamma}^{11}$ and charge 
conjugation matrices $C^{(10)}$  
have the form
\eqn\chcon{{\hat \Gamma}^{11}=  \sigma_3 \otimes \one_{16},\quad
\quad C^{(10)}=i\sigma_2 \otimes \one_{16} . }

The 8D fermions listed in section 7.1
 are related to 10D fields ${\hat \psi}^{\hA}$ and ${\hat \Lambda}$
in the following way
\foot{${\hat \Lambda}$ and ${\hat \Gamma}_{a}{\hat \psi}^a$
are mixed to give the 8D ``dilatino'', 
the superpartner of $e^{-2\xi}=e^{-2\phi}V$. }: 
\eqn\fermred{\pmatrix{ \psi^{A} \cr  \eta^{A} \cr}={\hat \psi}^{A}
+{1 \over 6}{\hat \Gamma}^{A}{\hat \Gamma}_{a}{\hat \psi}^a,\quad
\pmatrix{\Sigma \cr \L_R \cr }={3 \over 4}{\hat \Lambda}
+{\sqrt{2} \over 4}{\hat \Gamma}_{a}{\hat \psi}^a,
}
\eqn\fermredii{
\pmatrix{ l \cr \mu\cr}={\hat \Gamma}_a{\hat \psi}^{ a}-
{\sqrt{2} \over 2}{\hat \Lambda},\quad
\pmatrix{ {\tilde \mu} \cr {\tilde l} \cr }={\hat \psi}_{8}-
{\hat \Gamma}^{89}{\hat \psi}_{9}}

\subsec{8D fermion action}
Now we present the 8D action
 $S_{quad}^{(8)}=S_{kin}+S_{fermi-flux}$
quadratic in fermionic fluctuations
\foot{In Minkowski signature $\bpsi_A=\psi_A^{\dg}\Gamma^0.$
In Euclidean  signature $\bpsi_A$ and $\psi_A$ are treated
as independent fields.} 
over  the 8D background specified in section 2.2. The kinetic term is
standard

\eqn\ferm{   
S_{kin}=
 \int_{X}  e^{-2\xi} 
 t^7 \Biggl \{{ 1 \over 2} \bpsi_{A}
\Gamma^{AMB}D_{M} \psi_B +{ 1 \over 2} \bareta_{A}\Gamma^{AMB}D_{M}\eta_B
+{2  \over 3} \bSigma \Gamma^{M}D_M \Sigma +
{2  \over 3}\bLambda \Gamma^{M}D_M {\L_R}}
$$+{1\over 4}\barl \Gamma^{M}D_M l
+{1\over 4}\barmu \Gamma^{M}D_M \mu
 +{1\over 4}\bartl \Gamma^{M}D_M {\tilde l}+
{1\over 4}\bartmu \Gamma^{M}D_M {\tilde \mu} \biggr \}$$
The coupling of fluxes to fermion bilinears is:
\eqn\fermii{S_{fermi-flux}={\pi\over 4} \int_{X} e^{-\xi} \Biggl \{
t^8  \Bigl [ { n_0\rho+n_1 \over \sqrt{\im\rho} }X_{(0)}
-{ n_0\bar \rho+n_1 \over \sqrt{\im\rho} }{\tilde X}_{(0)} \Bigr ]
+ t^7 g_{(1) m}\wdg *X_{(1)}^m+ }
$$+ t^6 \Bigl [ {g_{(2)}^2 \rho+g_{(2)}^1 \over  \sqrt{\im\rho} }\wdg *X_{(2)}-{ g_{(2)}^2\bar \rho+ g_{(2)}^1\over  \sqrt{\im\rho} }
\wdg *\tX_{(2)} \Bigr]+
t^5 g_{(3) m}\wdg *X_{(3)}^m $$
$$+t^4 \sqrt{\im\rho} g_{(4)}\wdg * \Bigl[X_{(4)}+{\tilde X}_{(4)} \Bigr] 
\Biggr \}$$
where  the harmonic fluxes $g_{(p)},p=0,\ldots,4$  were defined in 
\arrfield. These harmonic fields couple to differential
p-forms $X_{(p)},\tX_{(p)}$ constructed out of fermi bilinears. 
We now give explicit formulae for $X_{(p)}$: 

\eqn\defx{
X_{(0)}=-\bPsi_A^{(-)}\Gamma^{AB}{\bf W}_B^{(-)}
-\bW_B^{(-)}\Gamma^{AB}{\bf \Psi}_A^{(-)}+
i\sqrt{2}\bLambda^{(+)}\Gamma^{A}{\bf W}_A^{(+)}}
$$-i\sqrt{2}\bW_A^{(-)}\Gamma^{A}{\L_R}^{(+)}+
i\sqrt{2}\bSigma^{(+)}\Gamma^{A}{\bf \Psi}_A^{(-)}
-i\sqrt{2} \bPsi_A^{(-)}\Gamma^{A}\Sigma^{(+)}$$
$$+{i \over 2} \barl^{(-)}\Gamma^{A}{\bf \Psi}_A^{(+)}
-{i \over 2} \bPsi_A^{(+)}\Gamma^{A}l^{(-)}+
{i \over 2}\barmu^{(-)}\Gamma^{A}{\bf W}_A^{(+)}-{i \over 2}
\bW_A^{(+)}\Gamma^{A} \mu^{(-)}$$
$$+4\bSigma^{(+)}\L_R^{(+)}
-4\bLambda^{(+)}\Sigma^{(+)}
-\half\bartl^{(+)}{\tilde \mu}^{(+)}+ 
\half{\bartmu}^{(+)}{\tilde l}^{(+)}$$

\eqn\defxii{
{\bigl(X_{(2)}\bigr)}_{MN}= \bPsi_A^{(-)}
\Gamma^{[A}\Gamma_{MN}\Gamma^{B]}{\bf W}_B^{(-)}+
{\bW}_A^{(-)}
\Gamma^{[A}\Gamma_{MN}\Gamma^{B]}{\bf \Psi}_B^{(-)}+}
$$i\sqrt{2}\bLambda^{(+)}\Gamma_{MN}\Gamma^{A}{\bf W}_A^{(-)}-
i\sqrt{2}{\bW}_A^{(-)}\Gamma^{A}\Gamma_{MN}{\L_R}^{(+)}
+i\sqrt{2}\bSigma^{(+)}\Gamma_{MN}\Gamma^{A}{\bf \Psi}_A^{(-)}$$
$$+i\sqrt{2}{\bf \bPsi}_A^{(-)}\Gamma^{A}\Gamma_{MN}{\Sigma}^{(+)}
+ {i \over 2} \barl^{(-)}\Gamma^{A}\Gamma_{MN}{\bf \Psi}_A^{(+)}
+{i \over 2} \bPsi_A^{(+)}\Gamma_{MN}\Gamma^{A}l^{(-)}$$
$$-{i \over 2} \barmu^{(-)}\Gamma^{A}\Gamma_{MN}{\bf W}_A^{(+)}
-{i \over 2}{\bW}_A^{(+)}\Gamma_{MN}\Gamma^{A}{\mu}^{(-)}
+4\bSigma^{(+)}\Gamma_{MN}\L_R^{(+)}$$
$$+4\bLambda^{(+)}\Gamma_{MN}\Sigma^{(+)}
-\half{\bartl }^{(+)}\Gamma_{MN}{\tilde \mu}^{(+)}-
\half{\bartmu}^{(+)}\Gamma_{MN}{\tilde l}^{(+)} $$

where $\psi_{A}^{(\pm)}=\half \left(\one_{16}\pm \bGamma \right)\psi_{A},$etc.
and
we use the combinations of $8D$ fields
$${\bf \Psi}_A=\psi_A+i {\sqrt{2} \over 3}\Gamma_A \L_R,
\quad {\bf W}_A=\eta_A-i {\sqrt{2} \over 3}\Gamma_A \Sigma$$
to make the expressions for $X_{(0)},X_{(2)}$ 
  have nicer coefficients.

The forms $\tX_{(0)},\tX_{(2)}$ can be obtained from
$X_{(0)},X_{(2)}$ by exchange of 8D chiralities $(-)\leftrightarrow (+). $

Under the T-duality generator $ S$ the above  forms transform as
\eqn\formtransf{\Bigl \{ X_{(0)},X_{(2) }\Bigr \}\rightarrow
  e^{-i\a}\Bigl \{ X_{(0)},X_{(2)} \Bigr \}
,\quad
\Bigl \{ \tX_{(0)},\tX_{(2)} \Bigr \}\rightarrow
  e^{i\a}\Bigl \{ \tX_{(0)},\tX_{(2)} \Bigr \} }
so that the combinations $ {1 \over \sqrt{\im\rho}}( n_0\rho+n_1)X_{(p)}, 
\quad {1 \over \sqrt{\im\rho}}( n_0\bar \rho+n_1)\tX_{(p)}$ for
 $ p=0,2 $
 which appear in the action \ferm\ are invariant under $ S.$

Also we have defined the  1-form
\eqn\defxi{
{\bigl(X_{(1)}^m\bigr)}_{M}=
e^m_{+}\Bigl [\bPsi_A^{(-)}\Gamma^{[A}\Gamma_{M}\Gamma^{B]}
{\bf W}_B^{(+)}-\bW_A^{(+)}\Gamma^{[A}\Gamma_{M}\Gamma^{B]}
{\bf \Psi}_B^{(-)}}
$$
-i\sqrt{2}\bLambda^{(+)}\Gamma_{M}\Gamma^{A}{\bf W}_A^{(+)}
+i\sqrt{2}\bW_A^{(+)}\Gamma^{A}\Gamma_{M}{\L_R}^{(+)}
-i\sqrt{2}\bSigma^{(-)}\Gamma_{M}\Gamma^{A}{\bf \Psi}_A^{(-)}$$  
$$+i\sqrt{2} \bPsi_A^{(-)}\Gamma^{A}\Gamma_{M}{\Sigma}^{(-)}
-{i \over 2} \bartl^{(+)}\Gamma^{A}\Gamma_{M}{\bf \Psi}_A^{(+)}+
{i \over 2} \bPsi_A^{(+)}\Gamma_{M}\Gamma^{A}{\tilde l}^{(+)}$$
$$-{i \over 2} \bartmu^{(-)}\Gamma^{A}\Gamma_{M}{\bf W}_A^{(-)}
+{i \over 2} \bW_A^{(-)}\Gamma_{M}\Gamma^{A}{\tilde \mu}^{(-)}
-4\bSigma^{(-)}\Gamma_{M}\L_R^{(+)}$$
$$+4\bLambda^{(+)}\Gamma_{M}\Sigma^{(-)}
-\half \barmu^{(+)}\Gamma_{M}l^{(-)} 
+\half \barl^{(-)}\Gamma_{M}\mu^{(+)} \Bigr ]+e^m_{-}
\Bigl[(+)\leftrightarrow (-)\Bigr] $$
and the 3-form
\eqn\defxiii{
{\bigl(X_{(3)}^m\bigr)}_{MNP}=
e^m_{+}\Bigl [- \bPsi_A^{(-)}\Gamma^{[A}\Gamma_{MNP}\Gamma^{B]}
{\bf W}_B^{(+)}-{\bW}_A^{(+)}
\Gamma^{[A}\Gamma_{MNP}\Gamma^{B]}{\bf \Psi}_B^{(-)}}
$$
+i\sqrt{2}\bLambda^{(+)}\Gamma_{MNP}\Gamma^{A}{\bf W}_A^{(+)}
+i\sqrt{2}\bW_A^{(+)}\Gamma^{A}\Gamma_{MNP}{\L_R}^{(+)}
-i\sqrt{2}\bSigma^{(-)}\Gamma_{MNP}\Gamma^{A}{\bf \Psi}_A^{(-)}$$  
$$-i\sqrt{2} \bPsi_A^{(-)}\Gamma^{A}\Gamma_{MNP}{\Sigma}^{(-)}
-{i \over 2} \bartl^{(+)}\Gamma^{A}\Gamma_{MNP}{\bf \Psi}_A^{(+)}-{i \over 2}
 \bPsi_A^{(+)}\Gamma_{MNP}\Gamma^{A}{\tilde l}^{(+)}$$
$$+{i \over 2} \bartmu^{(-)}\Gamma^{A}\Gamma_{MNP}{\bf W}_A^{(-)}
+{i \over 2} \bW_A^{(-)}\Gamma_{MNP}\Gamma^{A}{\tilde \mu}^{(-)}
-4\bSigma^{(-)}\Gamma_{MNP}\L_R^{(+)}$$
$$-4\bLambda^{(+)}\Gamma_{MNP}\Sigma^{(-)}
+\half \barmu^{(+)}\Gamma_{MNP}l^{(-)} 
+\half \barl^{(-)}\Gamma_{MNP}\mu^{(+)} \Bigr ]+e^m_{-}
\Bigl[(+)\leftrightarrow (-)\Bigr] $$
where we denote $e^m_{\pm}=e^m_8 \mp ie^m_9.$

The forms $X_{(1)}^m$ and $X_{(3)}^m$ transform in the  ${\bf 2}$ of 
$SL(2,\Z)_{\tau}.$
Also from \defxi,\defxiii\ we  find 
that $X_{(1)}^m$ and $X_{(3)}^m$ are invariant under  $SL(2,\Z)_{\rho}$
if we accompany the action of the generator  $ S$ by
 the U(1) rotation of $e^m_a$
\eqn\transfbein{ e^m_{\pm}\rightarrow e^{\pm i\a } e^m_{\pm}}
Since there are no local Lorentz anomalies, we can make this 
transformation. 

The most important objects in \fermii\ are the self-dual\foot{
In our conventions $\Gamma_{A_1A_2A_3A_4}=-{1 \over 4!}
\epsilon_{A_1A_2A_3A_4B_1B_2B_3B_4}\Gamma^{B_1B_2B_3B_4}\bGamma$} 
form $X_{(4)}$ and the anti-self-dual form ${\tilde X}_{(4)}$
 which couple to the flux $g_{(4)}.$
$X_{(4)}$ is defined by  
\eqn\defxiiii{
{\bigl(X_{(4)}\bigr)}_{MNPQ}=
 -i\bPsi_A^{(+)}\Gamma^{[A}\Gamma_{MNPQ}\Gamma^{B]}{\bf W}_B^{(+)}
+i\bW_A^{(+)}\Gamma^{[A}\Gamma_{MNPQ}\Gamma^{B]}{\bf \Psi}_B^{(+)}}
$$-\sqrt{2}\bLambda^{(-)}\Gamma_{MNPQ}\Gamma^{A}{\bf W}_A^{(+)}
+\sqrt{2}\bW_A^{(+)}\Gamma^{A}\Gamma_{MNPQ}{\L_R}^{(-)}$$
$$
-\sqrt{2}\bSigma^{(-)}\Gamma_{MNPQ}\Gamma^{A}{\bf \Psi}_A^{(+)}
+\sqrt{2} \bPsi_A^{(+)}\Gamma^{A}\Gamma_{MNPQ}\Sigma^{(-)}
-\half \barl^{(+)}\Gamma^{A}\Gamma_{MNPQ}{\bf \Psi}_A^{(-)}$$
$$+\half \bPsi_A^{(-)}\Gamma_{MNPQ}\Gamma^{A}l^{(+)}
-\half \barmu^{(+)}\Gamma^{A}\Gamma_{MNPQ}{\bf W}_A^{(-)}
+\half \bW_A^{(-)}\Gamma_{MNPQ}\Gamma^{A}\mu^{(+)}$$
$$
+4i\bSigma^{(-)}\Gamma_{MNPQ}\L_R^{(-)}-
4i\bLambda^{(-)}\Gamma_{MNPQ}\Sigma^{(-)}
-{i \over 2}{\bartl}^{(-)}\Gamma_{MNPQ}{\tilde \mu}^{(-)}+{i \over 2}
 {\bartmu}^{(-)}\Gamma_{MNPQ}{\tilde l}^{(-)}   $$
and $\tX_{(4)}$ can be obtained from $X_{(4)}$ by the exchange of
8D chiralities $(+)\leftrightarrow (-). $

Under the T-duality generator $ S$ these forms transform as
\eqn\formtransf{ X_{(4)} \rightarrow
  e^{i\a} X_{(4)},\quad
 \tX_4\rightarrow e^{-i\a} \tX_{(4)} }

We have also checked using  Appendix(C)
that the 4-fermion terms in the 8D action can be written as
\eqn\fourferm{  
S_{4-ferm}^{(8D)}=S_{4-ferm}'+S_{4-ferm}'',\quad
S_{4-ferm}'={\pi \over 128 }\int_{X}  e^{-2\xi}t^8
 \Bigl[ X_{(4)}\wdg * X_{(4)}+{\tilde X}_{(4)}\wdg * {\tilde X}_{(4)}\Bigr ] }
While $S_{4-ferm}''$ is manifestly invariant under T-duality, we will see
that the
 non-invariant term $ S_{4-ferm}'$
 is required for T-duality invariance of the total partition
sum $Z(\CF,\rho)$ of \finalcombo. 

\subsec{$T$-duality invariance of the ghost interactions}

The classical 8D action obtained from the reduction
of 10D IIA supergravity on $T^2$  is invariant under local
supersymmetry (all 32 components survive the reduction ).
To construct the quantum action we have to impose
a gauge fixing condition on the gravitino ${\hat \psi}_{(8D)}:=
\pmatrix{ \psi_A \cr \eta_A \cr}$
and include ghosts. Since the susy transformation laws involve fluxes,
there is a potential T-duality anomaly from the ghost sector.
In fact no such anomaly will occur as we now demonstrate.
 There are two generic properties of supergravity
theories:
\item 1.) In addition to a pair of Faddeev-Popov ghosts 
associated to the local susy gauge transformation
${\hat \psi}_{(8D)} \to {\hat \psi}_{(8D)}^A+
\d_{\hat \epsilon}{\hat \psi}_{(8D)}^A$
a ``third ghost,'' the Nielsen-Kallosh ghost,  appears \Nielsen. 

\item 2.) Terms quartic in Faddeev-Popov ghosts 
are required \Kallosh.

Let us recall first how the ``third ghost'' appears.
Following the standard procedure
we fix the local susy gauge 
by inserting $\d \left (f-{\hat \Gamma}_A{\hat \psi}_{(8D)}^A \right )$
into the path integral.
Then we also insert the unit\foot{We use the measure $\int [df]e^{i\int_X 
{\bar f}f}=1$. }
\eqn\fermifp{
 \one={1 \over \sqrt{det \left(\half e^{-2\xi}t^7{\hat D}\right )} }\int [df]
e^{{i\over 2} \int_{X}e^{-2\xi}t^7
{\bar f}{\hat D}f},\quad
{\hat D}={i\hat \Gamma}^ND_N }
and integrate over $[df].$ (If ${\hat D}$ has zeromodes this
expression is formally $0/0,$ but (7.27) below still makes sense.)

As a result we first find that the gravitino kinetic term
gets modified to 
\eqn\corkin{
 -{i\over 2} \int_{X}  e^{-2\xi} 
 t^7 \Biggl \{ \bpsi^{A}{\cal M}_{AB}\psi^B 
+ \bareta^{A}{\cal M}_{AB}\eta^B \Biggr\}}
where the operator ${\cal M}_{AB}$  acts on 
sections of the bundle \foot{$Spin(X)$ and $TX$
are spinor and tangent bundles on $X$ }  $Spin(X) \otimes TX $ as
\eqn\opmab{{\cal M}_{AB}=\d_{AB}i\Gamma^{M}D_{M}-2i\Gamma_{A}D_{B}}
where $D_A=E^M_AD_M.$
The determinant in \fermifp\ is expressed as the partition function
for  the ``third ghost''$\hat \Upsilon$
with action
\eqn\ups{
S_{\hat \Upsilon}= -{i\over 2}\int_{X}e^{-2\xi}t^7
{\bar {\hat \Upsilon}}{\hat D}\hat \Upsilon }
 
$\hat \Upsilon$ is a bosonic 32 component spinor,
which we decompose  into 16 component spinors as
$${\hat \Upsilon}=\pmatrix{\Upsilon_1 \cr \Upsilon_2}$$

Now we come to the most interesting part of quantum
action 
 which involves  Faddeev-Popov ghosts ${\hat b},{\hat c}.$
  
\eqn\fadd{S_{bc}=S_{bc}^{(2)}+S_{bc}^{(4)}}
where $S_{bc}^{(2)}\left (S_{bc}^{(4)} \right)$ denotes
the parts of the action quadratic (quartic) in FP ghosts.
Let us discuss the quadratic  part first.
According to the standard FP procedure we have
\eqn\faddtwo{
S_{bc}^{(2)}=\int_{X_8}t^7e^{-2\xi}
{\bar  {\hat b}}{\hat \Gamma}_A\d_{\hat c}{\hat \psi}_{(8D)}^A }

 We  decompose bosonic 32 component spinors ${\hat b},{\hat c}$ as
$${\hat c}=\pmatrix{c_1 \cr c_2},\quad
{\hat b}=\pmatrix{b_1 \cr b_2}.$$

We can write the action as a sum of two pieces 
$$S_{bc}^{(2)}=S_{bc}^{(2)0}+S_{bc}^{(2)2}$$   
Here $S_{bc}^{(2)0}$ does not contain fermionic matter fields
while $S_{bc}^{(2)2}$ is quadratic in fermions.
We now present  $S_{bc}^{(2)0}$ and  put $S_{bc}^{(2)2}$
in Appendix(D).

\eqn\sfp{S_{bc}^{(2)0}=\int_{X_8}t^7e^{-\xi}
{\bar  {\hat b}}(-i \hat D){\hat c}-\pi e^{-\xi}
\Biggl \{
{2\over 3}t^8 \Bigl [ { n_0\rho+n_1 \over \sqrt{\im\rho} }X^{gh}_{(0)}
-{ n_0\bar \rho+n_1 \over  \sqrt{\im\rho} }\tX^{gh}_{(0)} \Bigr ]}
$$+\half t^7 g_{(1) m}\wdg *X^{gh~m}_{(1)} 
+{1 \over 3} t^6 \im\Bigl [ {g_{(2)}^2 \rho+g_{(2)}^1 \over \sqrt{\im\rho} }
\wdg *X^{gh}_{(2)}
 -{ g_{(2)}^2\bar \rho+ g_{(2)}^1\over \sqrt{\im\rho} }
\wdg *\tX^{gh}_{(2)} \Bigr ]$$
$$+
{1 \over 8} t^5 g_{(3)m}\wdg *X^{gh ~m}_{(3)} \Biggr \} $$
where we  define forms bilinear in FP ghosts as
\eqn\defxgh{X^{gh}_{(0)}=
\half\Biggl\{{\overline {b_2}}^{(-)}c_1^{(-)}-
{\overline {c_1}}^{(-)}b_2^{(-)} -
{\overline {b_1}}^{(-)}c_2^{(-)}+ {\overline {c_2}}^{(-)}b_1^{(-)}\Biggr \}}
\eqn\defxghii{{\bigl(X^{gh}_{(2)}\bigr)}_{MN}=\half\Biggl\{
{\overline {b_2}}^{(-)}\Gamma_{MN}c_1^{(-)}+
{\overline {c_1}}^{(-)}\Gamma_{MN}b_2^{(-)}+
{\overline {b_1}}^{(-)}\Gamma_{MN}c_2^{(-)}+
{\overline {c_2}}^{(-)}\Gamma_{MN}b_1^{(-)}\Biggr \}}

\eqn\defxghi{{\bigl(X^{gh~m}_{(1)}\bigr)}_{M}=\half
e^m_{+}\Bigl [{\overline {b_2}}^{(+)}\Gamma_{M}c_1^{(-)}-
{\overline {c_1}}^{(-)}\Gamma_{M}b_2^{(+)} -
{\overline {b_1}}^{(-)}\Gamma_{M}c_2^{(+)}+
{\overline {c_2}}^{(+)}\Gamma_{M}b_1^{(-)} \Bigr ]}
$$+\half e^m_{-}
\Bigl[(+)\leftrightarrow (-)\Bigr] $$
\eqn\defxghiii{{\bigl(X^{gh ~m}_{(3)}\bigr)}_{MNP}=\half
e^m_{+}\Bigl [{\overline {b_2}}^{(+)}\Gamma_{MNP}c_1^{(-)}+
{\overline {c_1}}^{(-)}\Gamma_{MNP}b_2^{(+)}+
{\overline {b_1}}^{(-)}\Gamma_{MNP}c_2^{(+)}}
$$+
{\overline {c_2}}^{(+)}\Gamma_{MNP}b_1^{(-)} \Bigr ] +\half e^m_{-}
\Bigl[(+)\leftrightarrow (-)\Bigr] $$ 
The forms $\tX^{gh}_{(0)},\tX^{gh}_{(2)}$ can be obtained from
$X^{gh}_{(0)},X^{gh}_{(2)}$ 
by exchange of 8D chiralities $(-)\leftrightarrow (+). $
Note, that ${\hat b},{\hat c}$ do not couple to the flux  $g_{(4)}$. 

Let us now present the  part of the quantum 8D
action which is quartic in ghosts (as obtained by following
 the procedure of \Kallosh): 
\eqn\quartic{S_{bc}^{(4)}=e^{-2 \xi}t^8\Bigl\{
{1 \over 84}\left( \bar {\hat b} 
{\hat \Gamma}^{ABC}{\hat c} \right )\left( \bar {\hat b} 
{\hat \Gamma}_{ABC}{\hat c} \right )+{1 \over 3}
\left( \bar {\hat b} 
{\hat \Gamma}^{A}{\hat c} \right )\left( \bar {\hat b} 
{\hat \Gamma}_{A}{\hat c} \right )
\Bigr \} }
The presence of this quartic  action
 is due to the fact that
gauge symmetry algebra is open in supergravity:
$\left [\d_{\hat \epsilon_1},\d_{\hat \epsilon_1}\right ]{\hat \psi}_{(8D)}^A$
contains a term proportional to the equation of motion of
 ${\hat \psi}_{(8D)}^A.$

The T-duality invariance of $S_{bc}^{(4)},S_{bc}^{(2)0}$  and $S_{\hat \Upsilon}$
is manifest and we have also checked that $S_{bc}^{(2)2}$ is T-duality
invariant, so we conclude that
the part of the 8D quantum action which contains ghosts
is T-duality invariant.

\subsec{Computation of the determinants} 

We can now compute the fermionic quantum determinants including ghosts.  
Let us expand the fields $ \L_R,\Sigma,l,{\tilde l},\mu,{\tilde \mu}, b_1,b_2,c_1,c_2,
\Upsilon_1,\Upsilon_2$ and $\psi_A,\eta_A$ in the full orthonormal  basis 
of the operators ${\breve D}=i\Gamma^ND_N$ and ${\cal M}$ respectively,
where
the operator ${\cal M}$ was defined in
\opmab.
Note that  since we are assuming that background fluxes
are harmonic,  fermionic non-zero modes do not  couple to them.
Moreover,we can rescale  non-zero modes by  a factor of $e^{-\x}t^{7/2}$
so that kinetic terms appear without any dependence on $\xi$ and $t,$
but four-fermionic terms are supressed as $e^{2 \xi}t^{-6}$ 
with respect to the kinetic terms. Since kinetic
terms are manifestly T-duality invariant the integration over
nonzero modes will just give a factor ${\Det}'_F $ depending only
on
the Ricci flat metric $g_{MN}$ and the constants  $t$ and $\xi,$
all of which are T-duality invariant.
${\Det}'_F $ has the form
\eqn\fermdetf{
{\Det}'_F=r_F(\xi,t)det'{\cal M}}
where $ det'{\cal M} $ is 
determinant of the operator ${\cal M}$  defined in \opmab\ 
regularized in a way that 
\eqn\xidepferm{
r_F(\xi,t)=const \left ( e^{-2 \xi}t^7 \right )^{-n_{\cal M} }
}
where $n_{\cal M} $ denotes the number of zero modes of ${\cal M}.$

Note, that determinants of nonzero modes of
the fermions $\Sigma, \L_R,l,\mu,{\tilde l},{\tilde \mu}$ and
bosons $\Upsilon_1,\Upsilon_2, b_1,b_2,c_1,c_2$
 cancel each other and do not contribute to ${\Det}'_F. $

The situation is quite different for zero-modes: the kinetic terms are
zero but there is nonzero coupling to harmonic fluxes, so that
if we  rescale fermion zeromodes by $e^{-\half \xi}t^{2}$ 
we make both the 
fermion coupling to $g_{(4)}$ and the fermion 
quartic terms independent of $\xi$ and $t.$
We will also rescale ghost zeromodes by $e^{-\half \xi}t^{2}$   
and  include
the factor $\left (e^{-\xi}t^4\right )^{n_{\cal M}}$
  which comes from the rescaling of fermion and ghost 
zeromodes into the definition of $ {\Det}'_F,$ i.e. we
define new $r_F$:
\eqn\fermdetfi{
r_F^{new}(\xi,t):=r_F(\xi,t)\left (e^{-\xi}t^4\right )^{n_{\cal M}}=
const  ( t)^{-3n_{\cal M}}(e^{\xi })^{n_{\cal M}}}

From \normqii\ and \fermdetfi\ we find
that the full quantum determinants depend on $t$ and $\xi$ 
in the following way
\eqn\bytheway{
({\tilde t}^{-3})^{n_{\cal M}-n_{\cal K}-b_2-2b_1-4b_0}}
where we recall that $\tilde t=te^{-\xi/3}$ is the U-duality invariant
combination.\foot{
For any Ricci-flat spin 8-manifold the numbers
$n_{\cal M}$ and $n_{\cal K}$ can be expressed in terms
of topological invariants.}  Note that the dependence  on ${\tilde t}$
in \bytheway\  comes entirely from the
volume of the space of zero modes.
The volume of bosonic zero modes is  blowing up 
in the limit ${\tilde t}\to \infty,$
but the volume of fermion zero modes is  shrinking.
Since \bytheway\ is an overall factor in the partition sum, 
 it is a question of a net balance
between fermion and boson zero modes
whether the partition sum  blows up or vanishes
in the limit ${\tilde t}\to \infty.$

\subsec{Integration over the space of fermion zeromodes} 

We can split the action of the rescaled
fermion and  ghost  zeromodes as
$$ S^{(zm) }=S^{(zm)inv }+S^{(zm)ninv }.$$
Here the part $ S^{(zm)inv }$ is invariant under T-duality
 and includes all the ghost zeromode interactions,
the coupling of the fermion zeromodes to all RR fluxes except for $g_{(4)}$ and
the invariant part of the 4-fermion zeromode couplings, denoted
$S^{(zm)''}_{4-ferm}.$ 

$S^{(zm)ninv }$
 transforms non-trivially under the generator ${ S}$
of T-duality and can be recast in the following way:
\eqn\recast{ S^{(zm)ninv }
=\int_X \Biggl\{4\pi \im\rho g_{(4)}\wdg *Y_{(4)}+
2\pi \im\rho Y_{(4)}\wdg *Y_{(4)}\Biggr\}}
where we define the harmonic 4-form $Y_{(4)}$ as
\eqn\fermchar{
 Y_{(4)}={1 \over  16 } 
{1 \over \sqrt{\im\rho}}\Bigl [ X^{(zm)}_{(4)} 
+{\tilde X}^{(zm)}_{(4)} \Bigr ] .}
This object transforms under  $ S$ as
\eqn\fermcharii{
  S\cdot Y_{(4)}=-\re\rho Y_{(4)} +i \im\rho *Y_{(4)} .}

We now expand  the  harmonic 4-forms
in the basis $\omega_i$ of $H^4(X,\Z)$ 
$$g_{(4)}=(n^i+{\tilde \a}^i )\omega_i,\quad Y_{(4)}=y^i\omega_i,\quad
{\tilde \b}={\tilde \b}^i\omega_i$$
where the chracteristics ${\tilde \a}, {\tilde \b}$ are given in \tila.
Next, we define  
\eqn\thetapii{
\widehat{\Theta}(\CF,\rho) = \int d\mu_F^{(zm)}{\hat h}
e^{i 2\pi \widehat{\Delta \Phi}} \Theta \biggl[\matrix{\widehat{\alpha}\cr 
\widehat{\beta}\cr}\biggr](Q) 
}
where the shifted characterstics are defined as
${\widehat \alpha}^i={\tilde \a}^i +y^i,\quad {\widehat \b}^i={\tilde \b}^i + S\cdot y^i,$
and 
$d\mu^{(zm)}_F$ denotes the measure  of the  rescaled
fermion and  ghost   zeromodes. Recall that  
$Q(\rho)=[H\im\rho-i\re\rho]I.$
In \thetapii\ 
$ {\hat h}=e^{-S^{(zm)inv}} $ is a T-duality 
invariant expression which
depends on $\tau,\rho,t,g_{MN}$ as well as
 fermion and  ghost  zeromodes. 
The dependence on
$\tau,\rho,t,g_{MN}$ comes entirely from
the coupling of the rescaled zeromodes (of fermions and ghosts) to
the fluxes $g_{(p)},p=0,1,2,3.$ Finally, we have also defined  
\eqn\anotherdfn{
\widehat{\Delta \Phi}( \CF,\rho, \vec y) :=\Delta {\widetilde \Phi}
-\half \vec y I  S\cdot \vec y-\vec y I 
\vec \b 
}
where $\Delta {\widetilde \Phi}$ was defined in \deltaphi.

$\widehat{\Theta}(\CF,\rho)$ is invariant 
under $SL(2,\Z)_{\tau}$ and transforms 
under $SL(2,\Z)_{\rho}$    as

\eqn\trwides{
\widehat{\Theta}(S\cdot \CF,-1/\rho)=
s_F\mu(S)(-i \rho)^{\half b_4^+}(i \bar \rho)^{\half b_4^-}
\widehat{\Theta}( \CF,\rho)}
\eqn\trwidet{
\widehat{\Theta}(T\cdot \CF,\rho+1)=
\mu(T)\widehat{\Theta}( \CF,\rho)}
We  do Poisson ressumation to find  \trwides\ and
the extra phase $s_F$ is due to the transformation\foot{
Here we use the fact that the 10D fermions are
Majorana fermions in Minkowski signature.}
 of $d\mu_F^{zm}$
\eqn\fermiphase{
s_F=\Bigl(e^{i\a}\Bigr )^{I\bigl({\cal M}\bigr)}=
(i)^{I\bigl({\cal M}\bigr)}
(-i \rho)^{-\half  I\bigl(\cal M\bigr)}
(i{\bar \rho})^{\half I\bigl({\cal M}\bigr)} }
where $I\bigl({\cal M}\bigr)$  is the index
of the operator ${\cal M}$ defined in \opmab. 
As in the standard computation of the chiral anomaly \Fuj, only the zeromodes
 contribute to the  transformation
of fermionic measure. Indeed,   the contribution of 
 the bosonic
ghosts $c_1,b_1,\Upsilon_2$ to the transformation
of the measure cancels that of  the contribution of the fermions 
$\mu,{\tilde \mu},\L_R,l,{\tilde l}.$

\newsec{T-duality invariance}

\subsec{ Transformation laws for $Z_{B+F}(\cF,\tau,\rho)$ }

Now we study the transformation laws for 
\eqn\pretot{
Z_{B+F}(\cF,\tau,\rho)=Det_BDet'_Fe^{-S_B(\cF)}{\widehat \Theta}(\cF,\rho)}
where ${\widehat \Theta}(\cF,\rho)$
 is defined in \thetapii, while
 $Det_B$ and $Det'_F$ are defined in \findet\ and \fermdetf,\fermdetfi\
respectively.
We also recall that $S_B(\cF)$ is the real part of the classical action
evaluated on the background field configuration.  
 
First, we note that $Z_{B+F}(\cF,\tau,\rho)$
 is invariant under $SL(2,Z)_{\tau}.$  
Second, we learn how 
 $Z_{B+F}(\cF,\tau,\rho)$ transforms  under $SL(2,Z)_{\rho}$
by using the  transformation rules of $Det_B$ \quant\ 
and  $\widehat{\Theta}( \CF,\rho)$ \trwides,\trwidet.\
We find: 
\eqn\trzbf{
Z_{B+F}( S\cdot \cF,\tau,-1/\rho)=
s_Bs_F\mu(S)(-i \rho)^{\half b_4^+}(i \bar \rho)^{\half b_4^-}
Z_{B+F}(\cF,\tau,\rho)}
\eqn\trzbfii{
Z_{B+F}( T\cdot \cF,\tau,\rho+1)=\mu(T)Z_{B+F}(\cF,\tau,\rho)}
where  $s_B$ is taken from  the transformation of 
$D_B$.

Now, using the definition of $\chi$ and $\s$ 
\eqn\ident{
\half (b_0-b_1+b_2-b_3+b_4^{\pm})={1\over 4}
(\chi\pm \s),\quad \s=b_4^{+}-b_4^{-}}
as well as the index theorem:
$$I\bigl({\cal M}\bigr)+\int_{X}\lambda^2=\int_{X}248\hA_8$$
we obtain  the final result for the transformation under the generator $S$ 
\eqn\finresi{
Z_{B+F}(S \cdot \cF,\tau,-1/\rho)=
(-i \rho)^{{1 \over 4} \chi+
{1 \over 8}\int_X(p_2-\lambda^2)}(i \bar \rho)^{{1 \over 4} \chi-
{1 \over 8}\int_X(p_2-\lambda^2)}Z_{B+F}(\cF,\tau,\rho)}
From \trzbfii\ and \finresi\ we find that there is a T-duality
anomaly.

Let us note in passing  that 
the transformations \trzbfii,\finresi\ are
consistent for any  8-dimensional  spin manifold.
This can be seen by computing \foot{
The branches for the $8-th$ roots of unity
are chosen in such a way that $S^2=(-)^{F_R},$ where $F_R$
is a space-time fermion number in right-moving sector of type IIA
string}
\eqn\cosistent{Z_{B+F}\bigl ( (ST)^6 \cdot \cF,\tau,\rho)=
e^{i {\pi \over 4}\int_X (7\lambda^2-p_2)}
Z_{B+F}\bigl ( \cF,\tau,\rho)}
$$Z_{B+F}\bigl ( S^4 \cdot \cF,\tau,\rho)=
Z_{B+F}\bigl ( \cF,\tau,\rho)$$
and then noting that the index  
 theorem for 8-dimensional spin manifolds implies
\eqn\trivial{\int_X(7\lambda^2-p_2) \in 1440\Z.}

Incidentally,  when $X$  
 admits a nowhere-vanishing  Majorana  spinor of $\pm$ chirality 
the Euler characteristic is given by \Warn:
\eqn\spec{\chi=\pm {1 \over 2}\int_X (p_2-\lambda^2)}
and the transformation rule \finresi\ simplifies to: 
\eqn\finresii{
Z_{B+F}(S \cdot \cF,\tau,-1/\rho)=
(-i \rho)^{\half \chi}Z_{B+F}(\cF,\tau,\rho)}
\eqn\finresii{
Z_{B+F}(S \cdot \cF,\tau,-1/\rho)=
(i \bar \rho)^{\half \chi}Z_{B+F}(\cF,\tau,\rho)}
for   positive and negative chirality, respectively.

\subsec{Including quantum corrections}
Now we recall that there is a 1-loop correction to the effective $8D$
action:

\eqn\Bchi{
 S_{1-loop}=\pi \im \rho \int_X(u \lambda^2+vp_2)+
{i \pi\over 24}\re \rho \int_X \bigl(p_2-\lambda^2\bigr)  }
where we recall that $\pi \im \rho \int_X(u \lambda^2+vp_2)$ comes from
the regularization of $q(\im \rho)$ in\regul\
 and the numbers $u$ and $v$
depend on the regularization.

We now demonstrate that to construct a T-duality invariant 
partition function  
this term should be replaced with 
\eqn\newS{
 S_{quant}=\Bigl [\half \chi+{1 \over 4}\int_X(p_2-\lambda^2)\Bigr]
 log \left[\eta(\rho)\right]+
\Bigl [\half \chi-{1 \over 4}\int_X(p_2-\lambda^2)\Bigr]
 log \left[\eta(-\bar \rho)\right] }
where $\eta(\rho)$ is Dedekind function.
Taking the limit $Im\rho \to \infty$ one can uniquely
determine $u=-{1\over 24}$ and $v={1\over 24}$ in \Bchi.

$\eta$ has the following transformation laws:
\eqn\trJ{
\eta\left(-{1/\rho}\right)=(-i \rho )^{\half}\eta(\rho),
 \quad \eta(\rho+1)=e^{{\pi i \over 12}} \eta(\rho) }
so that
$e^{- S_{quant}}$ transforms as
\eqn\trS{
e^{- S_{quant}}\left(-{1/\rho} \right)=(-i \rho)^{-{1 \over 4} \chi-
{1 \over 8}\int_X(p_2-\lambda^2)}(i \bar \rho)^{-{1 \over 4} \chi+
{1 \over 8}\int_X(p_2-\lambda^2)}e^{- S_{quant}}\left(\rho \right)}
\eqn\trSii{
e^{- S_{quant}}\left(\rho+1 \right)=
e^{-i{\pi \over 24}\int_X(p_2 -\lambda^2) }e^{- S_{quant}}\left(\rho \right)}

Finally, we find that the total partition function 
\eqn\total{ Z(\cF,\rho):=
e^{-S_{quant}}Z_{B+F}(\cF, \rho)}
is invariant:
\eqn\finn{
 Z\bigl ( T\cdot \cF, \rho+1\bigr)=
 Z\bigl(\cF, \rho\bigr),}
\eqn\finnii{
  Z\bigl ( S\cdot \cF, -1/\rho \bigr)= 
Z\bigl (\cF,\tau,\rho \bigr).}
This is our main result.

As a consistency check  consider( for simplicity)
the case when $X$ admits a nowehere-vanishing  spinor
of positive chirality and take 
 the limit $Im\rho=V \rightarrow \infty $
\eqn\expJ{
S_{quant}\rightarrow \Biggl ({i \pi\over 12} \rho + 
\sum_{n\ge 1}\sum_{m\ge 1}{1\over m}e^{2\pi i nm\rho}\Biggr )\chi . }
We recognize the multiple cover formula for world-sheet
instantons on $T^2$ from \Piol.

\newsec{Application: Hull's proposal for interpreting the Romans 
mass in 
the framework of $M$-theory}

As a by-product of the above results we will make some comments 
on an interesting open problem concerning the relation of M-theory 
to IIA string theory. 

It is well known that IIA supergravity admits a massive 
deformation, leading to the Romans theory. The proper 
interpretation of this massive deformation in 11-dimensional 
terms is an intriguing open problem. In \Hull\ C. Hull 
suggested an 11-dimensional 
interpretation of certain backgrounds in the Romans 
theory. His interpretation involved T-duality in an 
essential way, and in the light of the above discussion 
we will make some comments on his proposal. 
(
For a quite different proposal for interpreting
 this massive deformation see \evslin. )

\subsec{Review of the relation of M-theory to IIA supergravity}

Naive Kaluza-Klein reduction says that for an appropriate 
transformation of fields $\bigl \{ g_{M-theory}, C_{M-theory} \bigr\} 
\rightarrow
\bigl \{ g_{IIA}, H_{IIA}, \phi_{IIA}, C_{IIA}\bigr\}$ we have 

\eqn\actionsagree{
S_{M-theory} = S_{IIA} 
}

One of the main points of \DMW\ was that, 
in  the presence of topologically nontrivial fluxes
equation \actionsagree\ is not true!
Indeed, given our current understanding of these fields,
there is not even a 1-1 correspondence between classical M-theory 
field configurations and classical IIA field configurations.
Rather, certain {\it sums} of IIA-theoretic field configurations 
were asserted to be equal to certain sums of M-theoretic field 
configurations. In this sense, the equivalence of type IIA 
string theory to  $M$-theory on a circle fibration is a 
quantum equivalence. 

To be more precise, in \DMW\ it was shown 
that for product manifolds $Y= X_{10}\times S^1$, 
the sum over $K$-theory lifts $x(\hat a)$ 
 of a class $\hat a\in H^4(X_{10};\Z)$ 
is proportional 
 to the sum over  torsion shifts of the M-theory 4-form
of $Y$. We have: 
\eqn\dmwsum{
{N (-)^{{\rm Arf}(q)+f({\hat a}_0)} \over \sqrt{N_2} N_K}
\sum_{x(\hat a)} e^{-S_{IIA}} = 
exp\left ( -\vert\vert G_{M-theory}(\hat a)\vert\vert^2 \right )
\sum_{\hat c \in H^4_{tors}(X_{10},\Z)}(-1)^{f(\hat a+\hat c)}
}
The above formula is the main technical result of \DMW.
We  recall that  
$[G_{M-theory}(\hat a)] =
2 \pi\bigl( \hat a-\half \lambda \bigr )$ and the equivalence
class of $\hat a$ is defined to contain 
$M$-theory field configurations with fixed kinetic energy
$$\vert\vert G_{M-theory}(\hat a) \vert\vert^2={1 \over 4\pi} \int_{X_{10}} 
G_{M-theory}(\hat a)\wdg {\hat *} G_{M-theory}(\hat a),$$
from which follows that these fields are characterized by 
${\hat a}'=\hat a+\hat c,\quad \hat c \in H^4_{tors}(X_{10},\Z).$
Also, in \dmwsum\  $N_K$ and $N$ is the order of $K^0_{tors}(X_{10})$ and
$H^4_{tors}(X_{10};\Z)$ respectively,
$N_2$ stands for the number of elements
in the quotient $L''=L/L',$ where
$L=H^4_{tors}(X_{10};\Z)/2H^4_{tors}(X_{10};\Z)$ and 
$L'=\Bigl \{ {\hat c}\in L,\quad  Sq^3{\hat c}=0\Bigr\}.$
Finally,
 ${\rm Arf}(q)$ is the Arf
invariant of the quadratic form $q({\hat c})=f({\hat c})+
\int_{X_{10}}{\hat c}\cup Sq^2 {\hat a}_0$
on $L''.$
The identity \dmwsum\ extends to the case where  $Y$ is a nontrivial
circle bundle over $X_{10}$ \DMW. 
  
As we have mentioned, we
 interpret the fact that we must sum over field configurations 
in \dmwsum\ as a statement that IIA-theory on $X_{10}$
 and M-theory on $Y= X_{10} \times S^1$ are really only 
quantum-equivalent. This point might seem somewhat tenuous, 
relying, as it does, on the fact that the torsion groups in cohomology 
and K-theory are generally different. Nevertheless, as we will now show, 
a precise version of Hull's proposal again requires equating 
sums over IIA and M-theory field configurations. In this case, however, 
the sums are over non-torsion cohomology classes, and in this 
sense the claim that IIA-theory and M-theory are only 
quantum equivalent becomes   somewhat more dramatic.

\subsec{Review of Hull's proposal}

One version of Hull's proposal states that massive IIA 
string theory on $T^2 \times X$ is equivalent to $M$-theory 
on a certain 3-manifold which is a nontrivial circle bundle 
over a torus. The proposal is based on T-duality invariance, which allows 
one to transform away $G_0$ at the expense of introducing 
$G_2$ along the torus, combined with the interpretation of 
$G_2$ flux as the first Chern class of a nontrivial M-theory 
circle bundle \DMW. 
We now describe this in  more detail.

Hull's proposal is based on the result \BRG\
that dimensional reduction of massive IIA supergravity 
with mass $m$ on 
a circle of radius $R$, (denoted  $S^1_{R}$),  gives the same theory as 
Scherk-Schwarz 
reduction of IIB supergravity on $S^1_{1/R}$. 
The IIB fields  are twisted by 
\eqn\twits{
g(\theta) = \pmatrix{1 & m \theta \cr 0 & 1\cr}
}
where the coordinate  on $S^1_{1/R}$ is $z={2 \pi \over R}\theta, \quad
\theta \in [0,1]$ 
and  the monodromy is
\eqn\hulls{
g(1)g(0)^{-1} = \pmatrix{1 & m \cr 0 & 1\cr} \in SL(2,\Z)
}
Schematically: 
\eqn\shrkschw{
{IIA_m \over S^1_R \times X_9} = 
\left ({IIB \over S^1_{1/R} \times X_9}\right )_{g(\theta)}
}
where $X_9$ is an arbitrary 9-manifold.
Note, in particular, that the twist acts on the IIB 
axiodil $\tau_{B}=C_0+ie^{-\phi_B}$ as 
\eqn\monodromy{
\tau_{B}(\theta) = \tau_{B}(0) + m \theta 
}
which implies that the IIB RR field $G_1$ has a nonzero period. 

Let  us also recall 
the duality between IIB on a circle and M-theory on $T^2$: 
\eqn\duality{
{IIB \over   S^1_{R'}\times S^1_{1/R} \times X} =
{ M \over T^2\left(\tau_M ,A_M \right)\times S^1_{1/R}\times X }
}
where the $T^2(\tau_M,A_M)$ on the M-theory side
 has complex structure $\tau_M=\tau_B(0)$ 
and area $A_M=e^{\phi_B \over 3}(R')^{-{4\over 3}}.$

Now, invoking
the adiabatic argument we have: 
\eqn\adiab{
\left({IIB \over S^1_{1/R} \times S^1_{R'} \times X}\right)_{g(\theta)}
 = {M \over B(m;R',R) \times X}
}
where $B(m;R',R)$ is a 3-manifold with metric:
\eqn\nilgeom{
ds^2 = \left({2 \pi \over R}\right)^2 (d\theta)^2 +  
A_M\biggl[{1 \over \im \tau_M} 
\bigl(dx+ (\re \tau_M+ m\theta) dy\bigr)^2 + \im \tau_M dy^2 \biggr] 
}
where   $x,y$ are periodic $x \sim x+1$ and $y \sim y+1.$
\foot{It is not entirely obvious that the invocation is justified, 
since for a large $M$-theory torus 
the twist is carried out over a small radius on 
the IIB side.} 

Combining \shrkschw\ with \adiab\ we get the basic statement of 
Hull's proposal: 
\eqn\hullspr{
{IIA_m \over S^1_R \times S^1_{R'} \times X} ={ M \over B(m;R',R) \times X}
}
We can now see the connection between Hull's proposal and T-duality. A 
duality transformation exchanges $G_0$ for a flux of $G_2$ through 
the torus. Then we can interpret the nontrivial flux $G_2$ as the 
first chern class of a line bundle in the $M$-theory setting.

\subsec{A modified proposal} 

In view of what we have discussed in the present paper, the equivalence of 
classical actions - when proper account is taken of the 
various phases of the supergravity action - cannot be true.
This is reflected, for example, in the asymmetry of the 
phase  \delphi\ in exchanging $n_0$ for $n_1$.  
However, we follow the lead of \dmwsum\ and therefore  
 modify Hull's proposal by identifying 
 sums over certain geometries 
on the IIA and M-theory side.
\foot{In making these statements we are including the K-theoretic phase 
as part of the ``classical'' action. Since the phase is formally at 
1-loop order it is possible that one could associate it with a 1-loop 
effect in such a way that classical equivalence does hold.} 

A modified proposal identifies   $Z(\CF,\rho,\tau)$
defined in \pretot,\total\ 
with a sum over M-theory geometries as follows.
Recall first that 
in the 8D theory there is a doublet of zeroforms $g_{(0)}^{\a}$,
arising from $G_0$ and $G_2$.
Next, let us factor $g_{(0)} = \ell \pmatrix {p \cr q}$ 
where $p,q$ are relatively 
prime integers and $\ell$ is an integer. Then we take a matrix 
 ${\cal N}\in SL(2,\Z)_{\rho}$ 

\eqn\sltz{
{\cal N}= \pmatrix{r& -s \cr -q& p\cr}\qquad rp - sq = 1 
}
such that 
\eqn\noenzero{
{\cal N}g_{(0)} = \pmatrix{\ell\cr 0 \cr}
}
This is the T-duality transformation that eliminates Romans flux.

Now, thanks to the  invariance of $Z(\cF,\tau,\rho)$
under T-duality transformations (see \finn\finnii\ above)
we find: 
\eqn\rmvno{
Z\left(\cF,   \rho \right) =
Z\left({\cal N}\cdot \cF,
   {p \rho + s \over q \rho + r} \right) 
}
By the results of \DMW\ the right hand side
of \rmvno, having 
$G_0=0$, {\it does} have an interpretation as 
a sum over M-theory geometries. The M-theory 
geometry is indeed a circle bundle over  
$T^2 \times X$ defined by $c_1 = \ell e_0 +pe - q e''+\c_md\s^m $
(as in Hull's proposal), but in addition it is 
necessary to sum over $E_8$ bundles on the 
11-manifold $B\times X$. While it is   essential to sum over $g_{(4)}$, 
all other fluxes $\CF$ may be
treated as classical - that is, they may be fixed and it is not 
necessary to sum over them. 

Both sides of  \rmvno\ should be regarded as 
wavefunctions in the quantization of self-dual fields. 
For this reason we propose that   there is only an intrinsically 
{\it quantum mechanical} equivalence between IIA theory and M-theory 
in the presence of $G_0$. 

\newsec{Comments on the U-duality invariant partition function}

The present paper has been based on weakly coupled  string theory. 
However, our motivation was understanding the relationship between 
K-theory and U-duality. In generalizing our considerations to 
the full U-duality group  $\CD=SL(3,\Z)\times SL(2,\Z)_{\rho}$
of toroidally compactified IIA theory it is necessary to 
go beyond the weak coupling expansion. Thus, it is 
appropriate to start with the $M$-theory formulation.
In the present section we make a few remarks on the $U$-duality 
of the $M$-theory partition function and its relation to the 
$K$-theory partition functions of type IIA strings. 
In particular, we will address the following points:

a.) The invariance 
of the $M$-theory partition function under the nongeometrical 
 $SL(2,Z)_\rho$ is not obvious and appears to require surprising 
properties of $\eta$ invariants. In section 10.2 we state this 
open problem in precise terms. 

b.) We will sketch how one can recover 
``twisted $K$-theory theta functions,'' at weak coupling
cusps when the $H$-flux is nonzero in section 10.3.

We believe that one can clarify the 
 relation between K-theory and U-duality by studying the 
behavior of the M-theory partition function at different 
cusps of the M-theory moduli space. At a given cusp
the  summation over 
fluxes is supported on  fluxes which can be related to 
$K$-theory. (See, for example, \dmwsum.) A
 U-duality invariant formulation of the theory 
must map the equations defining the  support at one cusp 
to those at any other cusp. This should define the $U$-duality 
invariant extension  of the $K$-theory constraints. 

\subsec{The $M$-theory partition function }

Let us consider the contribution to the M-theory
partition function from a background $Y$
 which is a $T^3$
 fibration over $X.$
\eqn\relation{
ds_{11}^2=V^{-{1\over 3}}{\tilde t}^2g_{MN}dx^Mdx^N+
V^{{2\over 3}}{\tilde g}_{\bf mn}\tht^{\bf m}\tht^{\bf n}}
where  $\tht^{\bf m}=dx^{\bf m} +\cA^{\bf m}_{(1)}$ and $x^{\bf m}\in [0,1].$  
${\tilde t}^2g_{MN}$ is an 8D Einstein   metric with  $det g_{MN}=1.$
${\tilde g}_{\bf mn}$ and $V$ are the shape and the volume
of the $T^3$ fiber. We denote  world indices on  $T^3$ by 
${\bf m} =(m,11),m=8,9$ 
and $M=0, \ldots,7$ as before. 

Topologically, one can specify the $T^3$ fibration over $X$ 
by a triplet of line bundles $L^{\bf m}$ 
 which transform in the representation
${\bf 3}$ of $SL(3,\Z)$ and 
have  first Chern classes
$c_1(L^{\bf m})={\cF}_{(2)}^{\bf m},$ where 
${\cF}_{(2)}^{\bf m}=d\tht^{\bf m}.$
 Such a specification
is valid         
up to possible monodromies. These 
are characterized by a homomorphism $ \pi_1(X) \to SL(3,\Z)$. 
  
On a manifold $Y$ of the type \relation\
we reduce the M-theory 4-form  $G_{M-theory}$ as
\eqn\cfldiii{
{ G_{M-theory}\over 2\pi}=G_{(4)} +G_{(3){\bf m}}\tht^{\bf m} +
\half \left (F_{(2){\bf mn}}+ 
\varepsilon_{\bf mnk}B_0\cF^{\bf k}_{(2)}\right)\tht^{\bf m} \tht^{\bf n} }
We also   include the   flat potential 
\eqn\czero{ c_{(0)}={1 \over 6}B_0 
\varepsilon_{\bf mnk}\theta^{\bf m}\theta^{\bf n}\theta^{\bf k}}
in the Kaluza-Klein reduction.\foot{$\varepsilon^{11,8,9}=\varepsilon_{11,8,9}=1$.}
(We will list the full set of flat potentials in this background 
below.)

From the Bianchi identity $dG_{M-theory}=0$  we have 
\eqn\bi{
dG_{(4)}=\cF_{(2)}^{\bf m} G_{(3){\bf m}}, \quad
dG_{(3){\bf m}}=\cF_{(2)}^{\bf n}F_{(2){\bf mn}}\quad dF_{(2){\bf mn}}=0
\quad d\cF_{(2)}^{\bf m}=0 }
which implies that fluxes $G_{(4)}$ and $G_{(3){\bf m}}$ are in 
general  not closed forms.\foot{In IIA at weak coupling we assumed 
$G_{(3)11}=0$ and $\cF_{(2)}^{ n}=0, n=8,9$, so that all background
fluxes are closed forms.}

Let us recall how the various fields transform
under  $\CD=SL(3,\Z)\times SL(2,{\bf Z})_{\rho}$ \Cvet.

$\bullet \quad {\tilde t}, g_{MN}$ are U-duality invariant.

$\bullet \quad SL(2,{\bf Z})_{\rho}$  acts on $\rho=B_0+iV\in \CH$
by fractional linear transformations.

$\bullet \quad SL(3,{\bf Z})$  acts  on the scalars ${\tilde g}_{\bf mn}$
parametrizing $SL(3,R)/SO(3)$ via the  mapping class group of $T^3.$

$\bullet \quad {\bf F}_{(2)}^{m \a}=
\pmatrix{F_{(2)}^{\bf m} \cr \cF_{(2)}^{\bf m}}$ 
transform in the ${\bf  (3,2)}$ of $\CD,$ 
where $ F_{(2)}^{\bf m}:=\half\varepsilon^{\bf mnk}F_{(2){\bf nk}}.$

$\bullet \quad  G_{(3){\bf m}}$ transform
in the ${\bf (3',1)}$ of $\CD$ 

$ \bullet \quad G_{(4)}$
  is singled out among all the other fields since
according to conventional supergravity \Cvet\  
 $SL(2,\Z)_{\rho}$ mixes  $G_{(4)}$ with its Hodge dual $*G_{(4)}.$ 
More concretely, 
\eqn\doubletag{
\pmatrix{ -\re\rho~ G_{(4)}+i\im\rho ~ *G_{(4)}\cr G_{(4)}  }
}
 transforms in the ${\bf (1,2)}$ of $\CD.$
Due to this non-trivial transformation  the classical bosonic 8D action
is not manifestly invariant under  $SL(2,\Z)_{\rho}.$
In detail, the action has real part: 
\eqn\realsi{
Re(S_{8D})= \pi\int_X \Biggl \{
\im\rho G_{(4)}\wdg *G_{(4)}+
 {\tilde t}^2 {\tilde g}^{\bf mn}G_{(3){\bf m}}\wdg*G_{(3){\bf n}}+ 
{\tilde t}^4 {\tilde g}_{\bf mn} {\cal G}_{\a \b}
{\bf F}_{(2)}^{{\bf m} \a}\wdg *{\bf F}_{(2)}^{{\bf n}\b} \Biggr \} }
$$+ {1 \over 2\pi}\int_{X}{\tilde t}^6  \Biggl \{ \CR+ 
28{\tilde t}^{-2} \p_M{\tilde t}\p_M{\tilde t} +
{1 \over 2\rho_2^2 }\p_M \rho  \p^M {\bar \rho}
+{1 \over 3}{\tilde g}^{\bf mn}{\tilde g}^{\bf kl}\p_M {\tilde g}_{\bf mk}\p^M 
{\tilde g}_{\bf nl} \Biggr \} $$
where ${\cal G}_{\a \b}$ is defined in \mrho, ${\tilde g}^{\bf kl}$ is
inverse of ${\tilde g}_{\bf mk}$ and
 $\CR$ is the Ricci-scalar of the  metric $g_{MN}.$

The imaginary part of the 8D bosonic action   follows from
the reduction of the M-theory phase $\Omega_M(C)$. This phase is  
subtle to define in topologically nontrivial 
field configurations of the $G$-field.   It may be formulated in two ways.
The first formulation was given in \fourflux. 
It uses Stong's result that  the spin-cobordism group
$\Omega_{11}(K(\Z,4))=0$ \stong. That is,   given 
a spin 11-manifold $Y$ and a 4-form flux ${G\over 2\pi}$
one can always   find a bounding spin 12-manifold $Z$ 
and an extension $\tilde G$ of the   the flux to $Z$. 
In these terms  the M-theory phase $\Omega_M(C)$ is given 
as:
\eqn\twelve{\Omega_M(C)=\epsilon \exp\biggl[{2\pi i \over 6}\int_Z {\tilde G^3} - 
{2\pi i \over 48} \int_Z \tilde G( p_2 - \lambda^2) \biggr] }
Here $\epsilon$ is the sign of the Rarita-Schwinger determinant. The 
phase does not depend on the choice of bounding manifold $Z$, but 
does depend on the ``trivializing'' $C$-field at the boundary $Y$. 

A second formulation  \refs{\DMW,\DM,\bs}
proceeds from the observation of \fourflux\ that 
the integrand of \twelve\ may be identified as the index density 
for a Dirac operator coupled to an $E_8$ vector bundle.  
The M-theory 4-form can be formulated in the following terms \refs{\DMW,\DM,\bs}. 
We set:
\eqn\newsb{{G_{M-theory}\over 2\pi}={\bar G}+dc }
where ${\bar G}={1 \over 60}Tr_{248}{F^2\over 8\pi^2}+
{1 \over 32\pi^2}TrR^2,$ $F$ is the curvature of a connection $A$
on an $E_8$ bundle $V$ on $Y$ and $ R$ is the curvature of the metric connection
on $TY.$ $G_{M-theory}$ is a real differential form, and 
$c\in \Omega^3(Y,R)/\Omega_{Z}^3(Y),$ where $\Omega_{Z}^3(Y)$
are 3-forms  with integral periods. The pair $(A,c)$ is subject to an 
equivalence relation. In these terms the M-theory phase is expressed as: 
\eqn\mthphase{\Omega_M(C)=exp\Biggl[2\pi i \Bigl({\eta\left(D_{V}\right)+
h\left(D_{V}\right) \over 4}+{\eta\left(D_{RS}\right)+
h\left(D_{RS}\right) \over 8}\Bigl) \Biggr]\omega(c)}
where $D_V$ is the Dirac operator coupled to the connection $A$,
 $D_{RS}$ is the Rarita-Schwinger operator,
$h(D)$ is the number of zeromodes of the operator $D$ on $Y$, and
$\eta(D)$ is the $\eta$ invariant of Atiyah-Patodi-Singer. 
The phase $\omega(c)$ is given by 
\eqn\phasec{
\omega(c)=exp\Bigl[\pi i \int_Y \Bigl(c({\bar G}^2 +X_8)+cdc{\bar G}
+{1 \over 3}c(dc)^2\Bigl)\Biggr]}

\subsec{The semiclassical expansion}

For  large $\tilde t$  there is a well-defined semiclassical expansion of the
 M-theory  partition function, which follows from the appearance of kinetic terms 
in the action \realsi\ 
scaling as ${\tilde t}^{2k}$ for  $k=0,1,2,3$. 
 In the  leading approximation we can fix  all the fields
except $G_{(4)}$, but this last  field must be treated quantum mechanically.
Note that this semiclassical expansion can differ from that described 
in the previous sections
because  we do not necessarily require weak string coupling.
In the  second approximation we  treat $G_{(4)}$ and $G_{(3){\bf n}}$ 
as quantum fields, and so on.

In the leading approximation in addition to the  sum over 
fluxes  $G_{(4)}$ we must integrate over the  flat potentials. 
 These include flat connection $\cA_{(1)}^{\bf m}$
of the $T^3$ fibration and  potentials coming from
the  KK reduction
of $c$ 
\eqn\switvh{ c = C_{(3)}'+C'_{(2)\bf m}\theta^{\bf m}+
\half C_{(1)\bf mn}\theta^{\bf m}\theta^{\bf n}+c_{(0)}}
where
$C'_{(2)\bf m}=C_{(2)\bf m}-\half C_{(1)\bf pm}\cA_{(1)}^{\bf p}$
and 
$C'_{(3)}=C_{(3)}-C'_{(2)\bf m}\cA_{(1)}^{\bf m},$
and $c_{(0)}$ is  defined in \czero. 
$C_{(3)}$ is invariant under U-duality,
$C_{(2)\bf m}$
 transforms in the   ${\bf (3,1)}$ of $\CD.$
We can   combine the flat potentials
$C_{(1)\bf mn}$ and $\cA_{(1)}^{\bf m}$ in a $U$-duality multiplet  
of $\CD$ transforming as ${\bf (3,2)}$ by writing
\eqn\combinenew{ 
{\bf A}_{(1)}^{{\bf m} \a}=
\pmatrix{\half\varepsilon^{\bf mnk}C_{(1){\bf nk}} \cr \cA_{(1)}^{\bf m}}. }

The duality invariance in the leading 
approximation is straightforward to check. We  
 keep only $G_{(4)}$. 
The flux is quantized by $[G_{(4)} ]=a-\half \lambda,$ 
where  
$a \in H^4(X,\Z)$ is the characteristic class of the 
$E_8$ bundle and   $ \lambda$ is the characteristic class of 
the spin bundle. We sum over   $a\in H^4(X,\Z).$
The   8D action, including the imaginary part is $SL(3,\Z)$ 
invariant. The imaginary part of the 8D effective
action in this case  takes a simple form 
which can be  found  from \mthphase :
\eqn\simplcase{Im(S_{8D})=-\pi \int_X \Bigl(  a\cup\lambda+
B_0\left(a-\half\lambda\right)^2 \Bigr )}
The invariance under $SL(2,\Z)_{\rho}$ then 
follows    in the same way as in our discussion in 
the weak string 
coupling regime. 

Let us now try to go beyond  the first approximation. In the second
approximation $[G_{(4)}]=a-\half \lambda +[{\cal A}_{(1)}^{\bf m}
G_{(3)\bf m}].$
We allow nonzero fluxes $G_{(3)\bf m}$, but still set to 
zero the fieldstrengths $F_{(2)}$ and $\CF_{(2)}$. We thus 
have a family of tori with flat connections.  
 Already in the second approximation, when we switch on
nonzero fluxes $G_{(3)\bf m}$ there does not appear to be 
 a simple expression for
the M-theory phase. 

Nevertheless, one can get some information about
the M-theory phase from the
 requirement of U-duality invariance. 
We know that $SL(3,\Z)$ invariance
is again manifest from the definition of $\Omega_M(C)$ and
 $\re(S_{8D}).$ But the expected $SL(2,\Z)_{\rho}$ invariance 
is nontrivial.  We would simply like to state this precisely. 
To do that we
write M-theory partition function in the second approximation
as
\eqn\thetamth{Z_{M-theory}({\tilde g}_{\bf mn},\rho) :=
\int d\mu_{flat}
\sum_{G_{(3){\bf m}}}
Z_{M-theory}({\tilde g}_{\bf mn},G_{(3){\bf m}},\rho)}
where $Z_{M-theory}({\tilde g}_{\bf mn},G_{(3){\bf m}},\rho)$ 
is the partition function with fixed, but nonzero, 
flux, $G_{(3){\bf m}}$,
$d\mu_{flat}$ stands for the integration over 
\eqn\flatpotmth{
{{\cal H}^3(X)\over {\cal H}^3_{\Z}(X)}\times
 \left({{\cal H}^2(X)\over {\cal H}^2_{\Z}(X)}\right )^3 \times
\left({{\cal H}^1(X)\over {\cal H}^1_{\Z}(X)}\right )^6, 
}
where ${\cal H}^p(X)$ is a space of harmonic p-forms on $X$
 and ${\cal H}^p_{\Z}(X)$
is the lattice of integrally normalized harmonic p-forms on $X.$
The first factor is for $C_{(3)},$ the second factor
for $C_{(2)\bf m}$ and the third factor is
 for the fields
${\bf A}^{{\bf m} \a}_{(1)}$ transforming in the ${\bf (3,2)}$ of $\CD.$ 
The integration measure $d\mu_{flat}$ is U-duality
invariant. 

The summand in \thetamth\ with fixed  $G_{(3){\bf m}}$ is given by
\eqn\thetamthi{Z_{M-theory}({\tilde g}_{\bf mn},G_{(3){\bf m}},\rho)=
\sum_{a \in H^4(X,\Z)}Det(G_{(4)},G_{(3){\bf m}})e^{-S_{quant }}e^{-S_{cl }}}
where
$$e^{-S_{cl }}=\Omega_M\left (G_{(4)},G_{(3){\bf m}},B_0 \right )
e^{-\pi \int_{X}\Biggl( Im(\rho) G_{(4)} \wdg *G_{(4)}+
{\tilde t}^2{\tilde g}^{\bf mn}
G_{(3){\bf m}}\wdg * G_{(3){\bf n} }\Biggr )}$$
and $Det(G_{(4)},G_{(3){\bf m}})$ denotes 1-loop determinants. 
These depend implicitly on the scalars $\rho,{\tilde g}_{\bf mn},{\tilde t}$ 
as well as on
the  metric $g_{MN}.$ We include 1-loop corrections in $S_{quant}$ (see below).

The M-theory phase
$\Omega_M$ in \thetamthi\
 depends on the fieldstrenths  $G_{(4)},G_{(3){\bf m}}$ 
and the   flat potentials, but 
 it is metric-independent, and hence should be a  
topological invariant. 
The dependence  of $\Omega_M$ on flat potentials is explicit 
from  \phasec\ for $c$ as in \switvh.
For example  dependence of $\Omega_M$ on $B_0$ has the form 
\eqn\simplyfollows{e^{i\pi\int_X B_0 G_{(4)}G_{(4)}}} 
It is conveneient to include 1-loop
corrections $\int_X B_0 X_8$  together with  effect of membrane instantons
in $S_{quant }.$
The nontrivial question is dependence
on $G_{(4)}$ and $G_{(3)\bf m}$ which also comes from $\eta(D_V)+h(D_V)$.

The independence of  
$\Omega_M$   on the metric on $Y=X\times T^3$
(in the second approximation) 
follows from the standard variation formula for
$\eta$-invariant. To show this let us fix the connection
on the $E_8$ bundle $V$ with curvature $F$
 and consider the family of veilbeins $e(s)$
on $Y=X\times T^3$ parametrized by $ s \in [0,1]$ such that
the metric on $T^3$ remains flat and independent of
the coordinates on $X.$ The corresponding family of
Rieman tensors ${\cal R}(s)$ gives an A-roof genus $\hat A(s)$ 
which is a pullback from 
 $X \times [0,1].$ 
Now we can write the standard formula for the
change in $\eta$-invariant under the variation of veilbein \APS:
\eqn\standardeta{\eta(e(1))-\eta(e(0))=j+
\int_{Y\times [0,1]} ch(V)\hat A(s)} 
where integer $j$ is a topological invariant of  $Y\times [0,1]$
and $ch(V):={1\over 30}[Tr_{248}e^{iF \over 2\pi}]$.
In the second approximation we only switch on
$\bar G=G_{(4)}+G_{(3)\bf m}dx^m$ so that neither
$ch_2(V)=-2(\bar G+\half \lambda) $ nor 
$ch_4(V)={1 \over 5}(\bar G+\half \lambda)^2$ 
have a piece $\sim dx^8dx^9dx^{11}$ and integral in  \standardeta\
vanishes. 

Now we come to the main point.
The requirement of the invariance
under the standard generators $S,T$ of $SL(2,\Z)_{\rho}$
\eqn\nontrconst{Z_{M-theory}({\tilde g}_{\bf mn},
-1/\rho)=
Z_{M-theory}({\tilde g}_{\bf mn},
\rho)}
\eqn\nontrconsti{Z_{M-theory}
({\tilde g}_{\bf mn},
\rho+1)=
Z_{M-theory}({\tilde g}_{\bf mn},
\rho)}
gives a nontrivial statement about the properties of
the function $\Omega_M(G_{(4)},G_{(3){\bf m}},B_0).$

The sum over fluxes $G_{(3){\bf m }} 
\in H^3(X,\Z)$ in \thetamth\ might be entirely supported
by classes which 
satisfy  a system of $SL(3,\Z)$ invariant constraints. These constraints
can in principle be determined by summing over torsion classes once the phase
 $\Omega_M$ is known in sufficently explicit terms. 
In the simple case when  $G_{(3)\bf m}$ are all  2-torsion classes,
one can act by the generators of $SL(3,\Z)$ 
on  the constraint
\eqn\guess{
Sq^3(G_{(3)9})+Sq^3(G_{(3)11})+G_{(3)9}\cup G_{(3)11}=0}
which follows from \DMW. If we assume that this constraint
is part of $SL(3,\Z)$ invariant system of constraints
then we find
\eqn\guessi{
 G_{(3)\bf m}\cup G_{(3)\bf n}=0, \quad {\bf m,n}=8,9,11}

\subsec{Comment on the connection with twisted K-theory}

In this section we discuss the behavior of the partition function 
near a weak-coupling cusp. 
There is a twisted version of K-theory which is 
thought to be related to the classification of 
D-brane charges in the presence of nonzero NSNS 
$H$-flux
\refs{ \wittenk,\Mathaii, \wittenstrings, \MMS}. 
 It is natural to ask if the contributions 
to the $M$-theory 
partition function $Z_{M-theory}({\tilde g}_{\bf mn},\rho)$
 from fluxes with nonzero $H_{(3)}:=G_{(3)11}\in H^3(X,\Z)$
are related, in the weak string-coupling cusp, 
to some kind of twisted K-theory theta function.

The weak-coupling cusp may be described by relating the 
fields in \relation\ to the fields in IIA theory. First,
the scale $\tilde t$ is related 
to the expansion parameter used in our previous sections by 
 ${\tilde t}^2=e^{-{2 \over 3}\xi}t^2.$ Next, we 
 parametrize the shape of $T^3$ as 
${\tilde g}_{\bf mn}=e^{\bf a}_{~\bf m}e^{\bf b}_{~\bf n}\d_{\bf ab}$
where
\eqn\prevsec{ e^{\bf a}_{~\bf m} = 
\pmatrix{{e^{-\xi/3}\over \sqrt{\tau_2}} &0 &0\cr
0& {e^{-\xi/3}\sqrt{\tau_2}} & 0 \cr
0 & 0 &  e^{2\xi/3}\cr }
\pmatrix{1 & \tau_1 & C_{(0)8} \cr 0 & 1  & C_{(0)9} \cr 0 & 0 & 1\cr}}
We denote frame indices by 
${\bf a}=(a,11),a=8,9$. The weak coupling cusp may be written as 
\eqn\weakcusp{R \times R^2  \times SL(2,R)/SO(2)} 
where the first factor is for the dilaton $\xi$, the second for  
$C_{(0)8},C_{(0)9},$\foot{ These are related to
the RR potentials ${\tilde C}_{(0)m}$
transforming in the ${\bf 2'}$ of $SL(2,\Z)_{\tau}$
as $C_{(0)8}=e^{\xi}\sqrt{\tau_2}{\tilde C}_{(0)8},\quad
C_{(0)9}=e^{\xi}{1\over \sqrt{\tau_2}}{\tilde C}_{(0)9}$}
  and the third for the modular 
parameter $\tau$ of the IIA torus.

As far as we know, nobody
 has precisely defined  what should be meant by the 
`` $K_H$ theta function.'' Since the 
Chern character has recently been formulated in 
\Mathai\MathaiB, this should be possible. Nevertheless, 
even without  a precise definition we do expect it to
be a sum over a ``Lagrangian'' sublattice of $K_H(X \times T^2)$.
At the level of DeRham cohomology, this should be a ``maximal Lagrangian'' 
sublattice of $\ker d_3/\im d_3$ where $d_3: H^*(X_{10},\Z) \to H^*(X_{10},\Z)$ 
is the differential $d_3(\omega) = \omega \wedge [H_{(3)}]$. 
Using  the filtration implied by the semiclassical expansion, and 
working to the approximation of  $e^{-{\tilde t}^2}$ this means that we 
should first define a sublattice of the cohomology lattice by 
the set of integral cohomology classes $(a, G_{(3)8}, G_{(3)9})$  such
that $(G_{(4)}, G_{(3)8}, G_{(3)9})$ are in the
kernel of $d_3$:
\eqn\kernel{
  H_{(3)}\wedge G_{(4)} =0,\quad  H_{(3)}\wedge G_{(3)m}=0,
\quad m=8,9}
Then the theta function should be a sum over the quotient lattice 
obtained by modding out by the    image of $d_3$     
\eqn\image{
G_{(3)8} \sim G_{(3)8} - p H_{(3)} ,\quad
 G_{(3)9} \sim G_{(3)9} - s H_{(3)} ,\quad
G_{(4)} \sim G_{(4)} - \omega_{(1)} H_{(3)}. }
Here $p,s \in \Z$ and $\omega_{(1)} \in H^1(X,\Z).$
Thus, our exercise is to describe how 
 a sum over this  quotient lattice   emerges from \thetamth. 

Let us consider the couplings of flat potentials 
 $C_{(1)89}$ and $C_{(2)m}$
to the fluxes 
which follow from  \phasec:
\eqn\simplyfollii{e^{2\pi i \int_X C_{(1)89} H_{(3)}G_{(4)}}
e^{2\pi i \int_X \epsilon^{mn}C_{(2)m}G_{(3)n}H_{(3)}}}
 Integrating over $C_{(1)89}$ and $C_{(2)m}$
 gives $H_{(3)}\wdg G_{(4)}=0$
and $\epsilon^{mn}H_{(3)}\wdg G_{(3)n}=0 $ respectively.

Next, we note that, due to the $SL(3,\Z)$ invariance of the M-theory action
we have
(suppressing many irrelevant variables)
\eqn\invariancsmth{
Z_{M-theory}(C_{(0)m}, G_{(3)m}- p_m H_{(3)},{\cal A}_{(1)}^{11},
G_{(4)}-\omega_{(1)}H_{(3)})=} 
$$Z_{M-theory}(C_{(0)m} + p_m, G_{(3)m},{\cal A}_{(1)}^{11}+\omega_{(1)},
G_{(4)} )   $$

Now we use \invariancsmth\ to  write the sum over all fluxes 
$G_{(4)},G_{(3)m},\quad m=8,9 $ in
the kernel of $d_3$ as 
\eqn\allthesum{
Z_H=\sum_{d_3-kernel  } Z_{M-theory}(C_{(0)m},
G_{(3)m},{\cal A}_{(1)}^{11},G_{(4)})
=
\sum_{{\cal M}_{fund}}W}
where ${\cal M}_{fund}$ stands for the 
fluxes 
in the
fundamental domain for the image of $d_3$ within the kernel of $d_3$
and
\eqn\weight{
W =\sum_{p_m\in \Z^2}\sum_{\omega_{(1)}\in H^1(X,\Z)} 
 Z_{M-theory}(C_{(0)m} + p_m, G_{(3)m},{\cal A}_{(1)}^{11}+\omega_{(1)},
G_{(4)} )   }

Now, we can recognize that $Z_H$ descends naturally to
the quotient of the weak-coupling cusp.
\eqn\recognize{
\Gamma_{\infty}' \backslash \biggl[ R\times R^2 \times SL(2,R)/SO(2)\biggr] }
where $\Gamma_{\infty}'\cong \Z^2$ is the subgroup
of the parabolic group $\Gamma_{\infty}$  consisting of elements 
of the form  
\eqn\transfmth{
L_{\bf m}^{~~\bf n}=
\pmatrix{ 1& 0& p\cr
          0 & 1 & s \cr
          0 & 0 &1 \cr },\quad p,s \in Z}
Written this way,  $Z_H$ is clearly a sum over a Lagrangian sublattice
of the $K_H(X \times T^2)$ lattice. (Recall that we are working  
in the DeRham theory, with the filtration appropriate to the second
approximation.)

The interesting point that we learn from this exercise is that in
formulating
the $K_H$ theta function, the weighting factor for the contribution of a
class in $K_H$  should be given by \weight. 
 The dependence of the action on the integers $p_m$ and 
$\omega_{(1)}\in H^1(X,\Z)$  behaves like 
$\exp[- Q(p_m,\omega_{(1)}) ]$
where $Q$ is quadratic form. 
Therefore $W$  is itself already a theta function.
This follows because 
 the dependence on  $C_{(0)m}$ 
and ${\cal A}_{(1)}^{11}$ comes entirely from the 
 real part of the classical action \realsi,
 since, as we have shown,  
the phase is independent of the metric on $X\times T^3.$
The dependence on $C_{(0)m}$ 
 comes  from
$\int_X {\tilde t}^2 
{\tilde g}^{\bf mn}G_{(3)\bf m}\wdg *G_{(3)\bf n}$ and
the dependence on ${\cal A}_{(1)}^{11}$
from $\int_X Im\rho  G_{(4)}\wdg *G_{(4)},$
where we recall that 
 $[G_{(4)}]=a-\half \lambda + [{\cal A}_{(1)}^{\bf m}G_{(3)\bf m}].$

 It would be very interesting to see if the function 
 $Z_H$ defined in \allthesum\ is in accord with a mathematically 
natural definition of a theta function for  twisted K-theory. 
But we will leave this for future work. 

As an example, let us consider $X=SU(3).$  Let
$x_3$ generate $H^3(X,\Z).$ 
Then fixing $H_{(3)}=kx_3$ we find that the fundamental domain
 of the image of $d_3$ within the kernel of $d_3$ 
 is given by
\eqn\represent{ G_{(3)8}=rx_3,\quad G_{(3)9}=px_3,\quad
 0 \le r,p\le k-1}
so that  the sum over RR fluxes in \allthesum\ is finite and
it is in this sense that  RR fluxes are k-torsion.
This  example of 
 $X=SU(3)$ is especially interesting 
since it is
 well known  \refs{\Stan,\Sch,\MMS}\ 
that at weak string coupling D-brane charges on $SU(3)$
 in the presence of $H_{(3)}=kx_3$ are classified by twisted K-theory groups
of $SU(3)$, and these groups are $k$-torsion.   As argued in 
\mw, from Gauss's law
it is then  natural to expect  that RR fluxes are also $k$-torsion. This is 
indeed what we   find in \represent.
\foot{In fact, from \MMS\ we know the order of the torsion group 
is actually   $k$ or $k/2$, according to the parity of $k$. However, given 
the crude level at which we are working we do not expect to see that 
distinction. We   expect that a more accurate account of the 
phases in the partition function will reproduce this result.  } 
On the other hand, the $M$-theory sum is indeed a full sum over 
all fluxes. This is in harmony with the result of \mmsiii\ 
for brane charges. Clearly, there is much more to understand 
here.

\bigskip
\bigskip
\bigskip  
\centerline{\bf Acknowledgements:}  

GM would like to thank E. Diaconescu and E. Witten for many important 
discussions and correspondence related to these matters both  during and
since the  collaboration leading to \mw\DMW. He would also like to 
thank B. Acharya, C. Hull,   N. Lambert, and J. Raben 
 for useful discussions  and correspondence, and 
C. Hull for some remarks on the manuscript. 
 GM would also like to thank L. Baulieu and 
B. Pioline at LTPHE, Paris  and the Isaac Newton Institute 
for hospitality during the completion of this manuscript. 
G.M. is supported in part by DOE 
grant \# DE-FG02-96ER40949.  

\bigskip
\bigskip

\appendix{A}{Duality transformations as symplectic transformations}

Here we  give
the explicit expressions for representations
of $S$ and $T$ in $Sp(2N,\Z).$
Let us choose  the following
  basis  of the lattice $\Gamma $
\eqn\basisvect{{\vec {\bf x}}=\Bigl({\vec {\bf x}_1},{\vec {\bf x}_2}\Bigr)}
\eqn\basisvecti{
{\vec {\bf x}_1}=\Bigl (y_l \one,y_l\otimes \left (L(e_0)-\one \right ),\left (L(e_s)-\one \right ),
\left (L(e_s)-\one \right )\otimes \left (L(e_0)-\one \right ), 
\left (L(\c_r d\s^m)-\one \right ),}
$$x\left (f_k d \s^m\right ),x\left (\omega_i \right )
\Bigr )$$
\eqn\basisvectii{
{\vec {\bf x}_2}=\Bigl (x({ \omega_i}) \otimes\left (L(e_0)-\one \right ),
x\left (d_k d \s^m\right ),x\left (w_r d \s^m\right ),x(u_s),
x(u_s) \otimes\left (L(e_0)-\one \right ),}
$$ 
x(h_l),x(h_l)\otimes \left (L(e_0)-\one \right ) \Bigr )$$
where we introduce 
$$ y_l\in H^0(X,\Z),\quad  h_l\in H^8(X,\Z), \quad l=1,\ldots,b_0$$
$$\c_r\in H^1(X,\Z),\quad  w_r \in H^7(X,\Z), \quad r=1,\ldots b_1$$
$$e_s \in H^2(X,\Z), \quad u_s \in H^6(X,\Z) \quad  s=1, \ldots, b_2,$$
$$f_k \in H^3(X,\Z), \quad  d_k \in H^5(X,\Z),  k=1, \ldots, {\bf b_3},\quad 
{ \omega_i}\in H^4(X,\Z), i=1,\ldots,b_4,$$
where $b_p$ is the rank of $H^p(X,\Z)$ and
${\bf b_3}$ is 
 the rank of  the sublattice of $H^3(X,\Z)$ 
 which is span by classes $f$ such that $Sq^3f=0.$

In the above basis the generators $S$ and $T$ are represented by 
\eqn\ibusaapp{
\s(S) =\pmatrix{A(S)& B(S) \cr  C(S) & D(S)}, \quad 
\s(T) =\pmatrix{A(T)& B(T) \cr  C(T) & D(T)}
}
 
\eqn\matricesrep{
A(S)=\pmatrix{ & \one_{b_0} & & & & & \cr
            -\one_{b_0} &  & & & & & \cr
               &   & & \one_{b_2} & & &  \cr
               &   &-\one_{b_2} &  & & &  \cr
               &  & & &\one_{2b_1} & & \cr
             &  & & & &\one_{2{\bf b_3}} & \cr
             &  & & & & & {\bf 0}_{b_4}\cr }\quad
B(S)=\pmatrix{ &  & & & & & \cr
             &  & & & & & \cr
               &   & & & & &  \cr
               &   & &  & & &  \cr
               &  & & & & & \cr
             &  & & & & & \cr
             &  & & & & & \one_{b_4}\cr }}
\eqn\matricesrepii{
C(S)=\pmatrix{ -\one_{b_4} &  & & & & &\cr
             &  & & & & & \cr
               &   & &  & & &  \cr
               &   & &  & & &  \cr
               &  & & & & & \cr
             &  & & & & & \cr
             &  & & & & & \cr }\quad
D(S)=\pmatrix{{\bf 0}_{b_4} &  & & & & & \cr
              & \one_{2{\bf b_3}} & & & & & \cr
              &  & \one_{2b_1} & &&& \cr
              &   & & &\one_{b_2} & &   \cr
               & &  &-\one_{b_2} &  & &   \cr
               &  & & & & &\one_{b_0} \cr
               &  & & & & -\one_{b_0}& \cr }}
\eqn\matricesrepiii{
A(T)=\pmatrix{\one_{b_0} & & & & & & \cr
              -\one_{b_0} &  \one_{b_0}& & & & & \cr
              &   &  \one_{b_2}& & & &  \cr
               &   &-\one_{b_2} &\one_{b_2}&&& \cr
              &  & & &\one_{2b_1} & & \cr
             &  & & & &\one_{2{\bf b_3}} & \cr 
               &  & & & & &  \one_{b_4}\cr }\quad  B(T)={\bf 0}_N }

\eqn\matricesrepiiii{
C(T)={\bf 0}_N,\quad 
D(T)=\pmatrix{  \one_{b_4}&  & & & & & \cr
              & \one_{2{\bf b_3}} & & & & & \cr
              &  & \one_{2b_1} & &&& \cr
              &   & & \one_{b_2}& & &   \cr
               & &  &-\one_{b_2} & \one_{b_2}  & &   \cr
               &  & & & &\one_{b_0} & &\cr
               &  & & & & -\one_{b_0}&\one_{b_0} \cr }}

\appendix{B}{Supergravity conventions}

The 10D fields  that we use are related to the fields
in \Romans as:
$$ {\tG_4\over \sqrt{2\pi}}=e^{-{3\phi \over 4}}F_{4}^{Rom},\quad
{G_2\over \sqrt{2\pi}}=-e^{-{9\phi \over 4}}F_{2}^{Rom},\quad
\quad {{\hat B}_{2} \over \sqrt{2\pi}} =-e^{3\phi \over 2}B_{2}^{Rom},
\quad  m=G_0 e^{15\phi \over 4},$$
$$ {\hat \psi}_{\hA}=e^{-{\phi \over 8}}
\psi^{(Rom)}_{\hA},\quad {\hat \Lambda}=e^{-{\phi \over 8}}
\lambda^{(Rom)},  g_{\hM \hN}=e^{\half \phi} g_{\hM \hN}^{(Rom)}$$
We also remind that we set $k_{11}=\pi$ while in \Romans\
$k_{11}=\sqrt{2\pi}$  was assumed. 

\appendix{C}{4-Fermion terms}

Below we collect 4-fermionic terms in  D=10 IIA supergravity action
which are obtained from circle reduction of the D=11 action of \CJS.
\eqn\tendfermii{
S_{4-ferm}^{(10)}={\pi  \over 2} \int \sqrt{-g_{10}} e^{-2\phi}\Bigl\{
 -{1 \over 64}\Bigl[ {\bar \chi}_{\bf E}{\hat \Gamma}^{\bf ABCDEF}\chi_{\bf F}
+12{\bar \chi}^{[\bf A}{\hat \Gamma}^{\bf BC}\chi^{\bf D]} \Bigr ]
{\bar \chi}_{[\bf A}{\hat \Gamma}_{\bf BC}\chi_{\bf D]}}
$$+{1 \over 32}\left ({\bar \chi}_{\bf E}{\hat \Gamma}^{\bf ABCEF}
\chi_{\bf F}\right )
\left ({\bar \chi}_{\bf A}{\hat \Gamma}_{\bf B}\chi_{\bf C}\right )+{1 \over 4}
\left ({\bar \chi}_{\bf A}{\hat \Gamma}^{\bf A}\chi_{\bf C} \right )
\left ({\bar \chi}_{\bf B}{\hat \Gamma}^{\bf B}\chi^{\bf C} \right )$$
$$-{1 \over 8}\left ({\bar \chi}_{\bf A}{\hat \Gamma}^{\bf B}
\chi_{\bf C} \right )
\left ({\bar \chi}_{\bf B}{\hat \Gamma}^{\bf A}\chi^{\bf C} \right )
-{1 \over 16}\left ({\bar \chi}_{\bf A}{\hat \Gamma}_{\bf B}
\chi_{\bf C} \right )
\left ({\bar \chi}^{\bf A}{\hat \Gamma}^{\bf B}\chi^{\bf C} \right ) \Bigr \}$$
where 
$$\chi_{\hA}=\Bigl[
{\hat \psi}_{\hA}+{1 \over 6 \sqrt{2}}{\hat \Gamma}_{\hA}{\hat \Lambda} 
\Bigr ]$$ 
$$\chi_{11}=-{2 \sqrt{2} \over 3  }
{\hat \Gamma}^{11}{\hat \Lambda} $$
and  ${\bf A}=(\hA,11).$

Recall  that the graviton ${\bf E}_{\bf M}^{\bf A}$ and the gravitino
$\psi^{(11)}_{\bf A}$ of 11D supergravity  are 
related to 10D fields as \CJS:
$${\bf E}_{\hM}^{\hA}=e^{-{\phi \over 3}}{\hat E}_{\hM}^{\hA},\quad
{\bf E}_{11}^{11}=e^{2\phi \over 3},\quad 
{\bf E}_{\hM}^{11}=e^{2\phi \over 3}C_{\hM} $$
$$ \psi^{(11)}_{\bf A}={1 \over \sqrt{2\pi}}e^{\phi \over 6}\chi_{\bf A} $$

\appendix{B}{Quartic couplings of ghosts and fermions }

Below we collect terms in the 8D quantum action which
are bilinear in FP ghosts and bilinear in fermions:
\eqn\tendghost{
S_{bc}^{(2)2}={\pi \over 2}\int_{X}t^8e^{-2\xi}\Biggl \{
{1 \over 8}\left ({\bar \chi}_{\bf B}{\hat \Gamma}_{A}\chi_{\bf C}+
2{\bar \chi}_{A}{\hat \Gamma}_{\bf B}\chi_{\bf C} \right )
\left( {\bar {\hat b}}{\hat \Gamma}^{A}
{\hat \Gamma}^{\bf BC}{\hat c} \right )+}
$${1 \over 6}\left ({\bar \chi}_{\bf B}{\hat \Gamma}_{\bar a}\chi_{\bf C}+
2{\bar \chi}_{\bar a}{\hat \Gamma}_{\bf B}\chi_{\bf C} \right )
\left( {\bar {\hat b}}{\hat \Gamma}^{\bar a}
{\hat \Gamma}^{\bf BC}{\hat c} \right )$$
$$
+{1 \over 6}\left ({\bar \chi}_{A}{\hat \Gamma}_{\bf BC}
\chi_{\bf D} \right )
\left( {\bar {\hat b}}{\hat \Gamma}^{A {\bf BCD} }{\hat c}\right )
+{2 \over 9}\left ({\bar \chi}_{\bar a}{\hat \Gamma}_{\bf BC}
\chi_{\bf D} \right )\left( {\bar {\hat b}}
{\hat \Gamma}^{{\bar a} {\bf BCD} }{\hat c}\right )$$
$$-{1 \over 48}{\bar {\hat b}}\Bigl [{\hat \Gamma}_A 
{\hat \Gamma}^{A {\bf BCDE} }+{4 \over 3}
{\hat \Gamma}_{\bar a} 
{\hat \Gamma}^{{\bar a} {\bf BCDE} } \Bigr ]{\hat c}
\left ({\bar \chi}_{\bf B}{\hat \Gamma}_{\bf CD}
\chi_{\bf E} \right )$$
$$- \left ({\bar {\hat c}}{\hat \Gamma}^{\bf B}
\chi_{A} \right )\left ({\bar {\hat b}}{\hat \Gamma}^{A}
\chi_{\bf B} \right )
-{ 4\over 3}\left ({\bar {\hat c}}{\hat \Gamma}^{\bf B}
\chi_{\bar a} \right )\left ({\bar {\hat b}}{\hat \Gamma}^{\bar a}
\chi_{\bf B} \right )+
{1 \over 4} \left ({\bar {\hat c}}{\hat \Gamma}^{11}
\chi_{11} \right )\Bigl [{\bar {\hat b}}{\hat \Gamma}^{A}
\chi_{A} +
{4 \over 3}{\bar {\hat b}}{\hat \Gamma}^{\bar a}
\chi_{\bar a} \Bigr ]$$
$$+L_{A {\bar a}}
\left ({\bar {\hat b}}{\hat \Gamma}^{A}\chi^{\bar a} \right )
+{4 \over 3}L_{ {\bar a} {\bf D}}
\left ({\bar {\hat b}}{\hat \Gamma}^{\bar a}\chi^{\bf D} \right )
+{1 \over 4}L_{ {\bf D} {\bf E}}{\bar {\hat b}}\Bigl(
{\hat \Gamma}^{A}{\hat \Gamma}^{\bf DE}\chi_{A} +
{4 \over 3}{\hat \Gamma}^{\bar a}{\hat \Gamma}^{\bf DE}\chi_{\bar a}\Bigr )
\Biggr \}
$$
where we now split indices as 
${\bf A}=\left( A,\bar a \right),
\quad A=0,\ldots 7,\quad {\bar a}=(a,11),\quad a=8,9.$
Nonzero components of $L_{ {\bf D} {\bf E}}$ are given by:
$$L_{ {A} {\bar d}}=- {\bar {\hat c}}{\hat \Gamma}_{A}\chi_{\bar d}, \quad
L_{ a 11}=- {\bar {\hat c}}{\hat \Gamma}_{a}\chi_{11} $$ 

$S_{bc}^{(2)2}$ is obtained by relating
8D gauge field ${\hat \psi}_{(8D)}^A$( gauge parameter ${\hat \epsilon}$)
 to  11D gravitino $\psi^{(11)}_{\bf A}$ (gauge parameter
$\epsilon^{(11)}$ ) as
$${\hat \psi}_{(8D)}^A=
\sqrt{2\pi}e^{-{\phi \over 6}}\Bigl[\psi^{(11)}_A+{1 \over 6}
{\hat \Gamma}_A{\hat \Gamma}^{\bar a}\psi^{(11)}_{\bar a} \Bigr],\quad
{\hat \epsilon}= \sqrt{2\pi}e^{\phi \over 6}\epsilon^{(11)}
$$
Let us  also remind a standard fact that to keep the gauge  
$${\bf E}_{11}^{\hA}=0,\quad
{\bf E}_{m}^{A}=0$$
used in reduction from 11D 
one has to accompany supersymmetry transformations 
of \CJS\ with field dependent Lorentz transformations.

The last line in the action  $S_{bc}^{(2)2}$
originates from such  Lorentz transformations.

To write out $S_{bc}^{(2)2}$ in terms of 8D fields 
$${\hat \psi}_{(8D)}^A:=\pmatrix{\psi^A \cr \eta^A},\quad
{\hat \Lambda}_{(8D)}:=\pmatrix{ \Sigma \cr \L_R}, \quad
{\hat \theta}_{(8D)}:=\pmatrix{l \cr \mu \cr },
\quad {\hat \nu}_{(8D)}:=\pmatrix{{\tilde l} \cr {\tilde \mu}\cr }$$ 
one should substitute
$$\chi^{A}={\hat \psi}_{(8D)}^{A}+
{1 \over 12}{\hat \Gamma}^{A}{\hat \theta}_{(8D)}
+{\sqrt{2} \over 6}{\hat \Gamma}^{A}{\hat \Lambda}_{(8D)},
\quad A=0,\ldots, 7 $$
$$\chi^{8}=\half {\hat \nu}_{(8D)}+{1 \over 3}{\hat \Gamma}^{8}\Bigl (
{\hat \theta}_{(8D)}+\sqrt{2}{\hat \Lambda}_{(8D)}\Bigr ),\quad
\chi^{9}=\half {\hat \Gamma}^{89}{\hat \nu}_{(8D)}+
{1 \over 3}{\hat \Gamma}^{9}\Bigl (
{\hat \theta}_{(8D)}+\sqrt{2}{\hat \Lambda}_{(8D)}\Bigr ) 
 $$
$$\chi_{11}=-{2\sqrt{2} \over 3}{\hat \Gamma}^{11}\Bigl ({\hat \Lambda}_{(8D)}
-{\sqrt{2} \over 4}{\hat \theta}_{(8D)}\Bigr )$$

We do not present the final expression but we
have checked that $S_{bc}^{(2)2}$ is T-duality invariant.

\appendix{E}{Measures for path integrals}

Here we explain why  $det' \Delta_p$ are divided by $V_p$ in \answ\
 This is related to the integration over zeromodes.

Introducing  a basis $a_{(p )}^i, \,i=1,\ldots,b^{p}$ 
in ${\cal H}^p_{\Z}$ let us denote
\eqn\metrlat{
V_p^{ij}=\int_{X} a_{(p )}^{i}\wdg *a_{(p)}^j,\quad
V_p=det_{i,j}V_p^{ij} 
}

Note, that $V_p$ is invariant under the choice of basis
in ${\cal H}_{\bf Z}^{p}.$

To explain integration over fermionic zero modes  
let us consider the following path-integral over
fermionic p-forms $u$ and $v.$ 
\eqn\example{
\int Du Dv\left [\prod_{i=1}^{b^p}\int_{\c_i}u 
\prod_{j=1}^{b^p}\int_{\c_j}v \right] 
e^{-\bigl (v,\Delta_p u \bigr )} }
where
$\c_i, i=1, \ldots b^p$ is a basis
of $H_p(X,{\bf Z}).$

In \example\ we have inserted $\prod_{i=1}^{b^p}\int_{\c_i}u
\prod_{j=1}^{b^p}\int_{\c_j}v, $ to get non-zero answer, 
i.e.  to   saturate fermion zero modes.

To perform the integration in \example\
we expand $u$ and $v$ in an orthonormal
basis $\{\psi_n\}$ of eigen p-forms of $\Delta_p .$
\eqn\exmpli{
u=\sum_n u_n \psi_n, \quad v=\sum_n v_n \psi_n,
\quad
\bigl ( \psi_n, \psi_m \bigr )=\delta_{n,m}
}

Let us choose the  basis $a_{(p )}^i, \, i=1,\ldots,b^{p}$ 
 of the lattice ${\cal H}_{\bf Z}^{p},$  dual to
the basis $\c_i \in H_p(X,{\bf Z}),$ i.e
$$\int_{\c_i}a_{(p)}^j=\delta_{ij}$$
Then, orthonormal zero-modes   are expressed as 
\eqn\orth{
\psi_{zm}^i= a_{(p)}^j\left ( W_p^{-1}\right )_j^i} 
where $W_p^{-1}$ is the inverse of the vielbein
for the metric on ${\cal H}_{\bf Z}^{p}$: 
$(V_p)^{ij}=\left(W_p^TW_p\right)^{ij}.$

Now, we integrate \example\ and obtain
\eqn\exmplii{
\Bigl[ {det' \Delta_p \over V_p}\Bigr ]
}

In the case of bosonic  p-forms $u$ and $v$
 we do not need to insert anything to get a non-zero answer:
\eqn\bzero{
\int Du Dv
e^{-\bigl (u,\Delta_p v \bigr )}=
\Bigl[ {det' \Delta_p  \over V_p} \Bigr ]^{-1} }
where \bzero\ the
integration over bosonic zero-modes   was performed 
\eqn\bzeroii{
\int \prod_{i=1}^{b^p} Du_{zm}^i
\prod_{j=1}^{b^p} Dv_{zm}^j=
{1 \over  \Bigl (det_{i,j} \int_{\c_i}\psi_{zm}^j \Bigr )^2}=V_p
}

\appendix{F}{Super-$K$-theory theta function}

Here we explain why 
$\widehat{\Theta}(\CF,\rho)$ defined in \thetap\
is a supertheta function for a family 
 of principally polarized
 superabelian varieties.  To show this we
use  the results of \Rabin,
where   supertheta functions  were studied.

A generic complex supertorus is defined as a quotient
of the affine superspace with even coordinates 
$z_i,\quad i=1,\ldots,N_{even}$
and odd coordinates $\xi_a,\quad a=1,\ldots,N_{odd}$
by the action of the abelian group
generated by $\{\lambda_i,\lambda_{i+N_{even}}\}$
\eqn\abelian{\lambda_i : z_j\to z_j+\d_{ij}, \quad \xi_a\to \xi_a}
\eqn\abelianii{\lambda_{i+N_{even}} : z_j\to z_j+
{\left(\Omega_{even}\right)}_{ij}, \quad \xi_a\to \xi_a+
{\left(\Omega_{odd}\right)}_{ia}}
We will restrict to  the special case  
${\left(\Omega_{odd}\right)}_{ia}=0$ relevant for our
discussion. Let us also assume that the reduced torus
(obtained from the supertorus by
forgetting  all odd coordinates)
has a structure of a principally polarized abelian variety
and denote its Kahler form by
 $\omega.$

It follows from the results of  \Rabin,
that a complex line bundle $L$ on the supertorus
 with $c_1(L)=\omega$ has a unique  section (up to constant multiple) iff
$\Omega_{even}^T=\Omega_{even}$
together with the positivity of the imaginary part of the reduced
matrix. This section is a supertheta function.

Now we can find
a family of principally polarized 
superabelian varieties relevant to our case
 simply by setting $N_{even}=N$ and $N_{odd}=N_{ferm.zm}$ and by
{\it defining} symmetric $\Omega_{even}$ as
\eqn\defomega{
Re{\left(\Omega_{even}\right)}_{ij}=Re\tau_K (x_i,x_j),}
\eqn\defomegaii{
Im{\left(\Omega_{even}\right)}_{ij}=Im\tau_K (x_i,x_j)+}
$$\sum_{p=0}^{2}\int_{X_{10}} \Bigl(G_{2p}(x_i)+G_{2p}(x_j)\Bigr)
\wdg {\hat *}{\cal J}_{2p}(zm)+\d_{ij}F(zm)$$
where $x_i,i=1,\ldots,N$ is a basis of $\Gamma_1.$
In \defomega\  ${\cal J}_{2p}(zm)$ is a 2p-form on $X_{10}$
constructed as a bilinear expression in fermion(and ghosts) zeromodes
and $F(zm)$ is a functional quartic in fermion( and ghosts)
 zeromodes, both ${\cal J}_{2p}(zm)$ and $F(zm)$
can in principle be found from the 10D fermion action 
(7.10),(14.1)
as well as from the ghost action (7.35),(7.40),(15.1).
The modified  characteristics 
${\vec {\widehat \alpha}}, 
{\vec{ \widehat \beta}}$ and prefactor $\widehat{\Delta \Phi}(\CF) $
in \thetap\ all originate from the shift of the imaginary
part of the period matrix described in \defomegaii.
It would be very nice if one could formulate this superabelian
variety in a more natural way, without reference to a Lagrangian
splitting of $\Gamma_K.$

\listrefs
\bye